\documentclass[11pt,reqno]{article}
\usepackage{latexsym}
\usepackage{url,citesort}
\usepackage[thmmarks]{ntheorem}

\usepackage{amsmath,amssymb,amsfonts,amscd,graphicx}

\renewcommand{\theequation}{\thesection.\arabic{equation}}

\let\ssection=\section

\renewcommand{\section}{\setcounter{equation}{0}\ssection}

\newcounter{mnotecount}[section]

\renewcommand{\themnotecount}{\thesection.\arabic{mnotecount}}

\newcommand{\mnote}[1]
{\protect{\stepcounter{mnotecount}}$^{\mbox{\footnotesize $
\bullet$\themnotecount}}$ \marginpar{
\raggedright\tiny\em $\!\!\!\!\!\!\,\bullet$\themnotecount: #1} }

\newcommand{\Omlrt}{\Omega(\theta_l,\theta_r,-|I_0|/2)}

\newcommand{\alp}{{\alpha_p}}

\newcommand{\xpl}{x^{(\psi,\lambda)}}
\newcommand{\Xpl}{X^{(\psi,\lambda)}}
\newcommand{\Xplt}{\Xpl_t}
\newcommand{\Xplth}{\Xpl_\theta}

\newcommand{\Id}{\mbox{Id}_{\mcH_2}}
\newcommand{\avtd}{AVTD}
\newcommand{\avtdk}{AVTD$_k$}

\newcommand{\avtdrp}{AVTD$^{(\rho,\varphi)}$}
\newcommand{\avtdrpk}{AVTD$^{(\rho,\varphi)}_k$}
\newcommand{\avtdpq}{AVTD$^{(P,Q)}$}
\newcommand{\avtdpqk}{AVTD$^{(P,Q)}_k$}
\newcommand{\avtdpqthree}{AVTD$^{(P,Q)}_3$}
\newcommand{\avtdpqz}{AVTD$^{(P,Q)}_0$}
\newcommand{\avtdpqo}{AVTD$^{(P,Q)}_1$}
\newcommand{\avtdpqi}{AVTD$^{(P,Q)}_\infty$}

\newcommand{\ftone}{|tf_1|}\newcommand{\fttwo}{|tf_2|}\newcommand{\gttwo}{|tg_2|}
\newcommand{\clim}{{\textstyle \lim_{C^0_{t_0}(\psi)}}}

\newcommand{\hczpsio}{\hat C^0_{t_1}(\psi)}

\newcommand{\Omt}{\Omega_{t-\mbox{\rm \scriptsize reg}}}

\newcommand{\Omth}{\Omega_{\theta-\mbox{\rm \scriptsize reg}}}
\newcommand{\vw}{v_{\mbox{\scriptsize \rm
weak}}}
\newcommand{\supO}{\sup_{\Omega(a,b,t_0)}}

\newcommand{\myqed}{\hfill $\Box$\\ \medskip }
\newcommand{\myqedt}{\hfill $\Box$}
\newcommand{\inti}{\int_{a+t}^{b-t}}
\newcommand{\into}{\int_{\psi+t}^{\psi-t}}
\newcommand{\rmnote}[1]{}

\def \Reel{\mathbb{R}}
\def \R {\Reel}

\def \Nat{\mathbb{N}}
\def \Z{\mathbb{Z}}
\def \N {\Nat}

\newcommand{\mutwo}{{\mu_2 }}
\newcommand{\lambdatwo}{{\lambda_2}}
\newcommand{\bel}[1]{\begin{equation}\label{#1}}
\newcommand{\beal}[1]{\begin{eqnarray}\label{#1}}
\newcommand{\bea}{\begin{eqnarray}}
\newcommand{\bean}{\begin{eqnarray}\nonumber}
\newcommand{\beadl}[1]{\begin{deqarr}\label{#1}}
\newcommand{\eeadl}[1]{\arrlabel{#1}\end{deqarr}}
\newcommand{\eeal}[1]{\label{#1}\end{eqnarray}}
\newcommand{\eead}[1]{\end{deqarr}}
\newcommand{\eea}{\end{eqnarray}}
\newcommand{\beaa}{\begin{eqnarray*}}
\newcommand{\eeaa}{\end{eqnarray*}}
\newcommand{\nn}{\nonumber}

\newcommand{\be}{\begin{equation}}
\newcommand{\ee}{\end{equation}}

\newcommand{\eq}[1]{\eqref{#1}}
\newcommand{\Eq}[1]{Equation~(\ref{#1})}
\newcommand{\Eqsone}[1]{Equations~(\ref{#1})}
\newcommand{\Eqs}[2]{Equations~(\ref{#1})-\eq{#2}}

\DeclareFontFamily{OT1}{rsfs}{}
\DeclareFontShape{OT1}{rsfs}{m}{n}{ <-7> rsfs5 <7-10> rsfs7 <10->
rsfs10}{} \DeclareMathAlphabet{\mycal}{OT1}{rsfs}{m}{n}

\newcommand \al {\alpha}
\newcommand \ep {\epsilon}

\newcommand \pa {\partial}
\newcommand \mcU {{\mycal U}}
\newcommand \mcV {{\mycal V}}
\newcommand \mcB {{\mycal B}}

\newcommand \mcG {{\mycal G}}
\newcommand \CH {{\mycal H}}
\newcommand \mcH {\CH}

\newcommand \mcO {{\mycal O}}

\newcommand \CL {{\mycal  L}}
\newcommand \CP {{\mycal  P}}
\newcommand \xt {X_{t}¥}
\newcommand \xth {X_{\theta}¥}

\newcommand \dth {D_{\theta}}

\newcommand \pthth {P_{\theta\theta}}
\newcommand \ptth {P_{t\theta}}
\newcommand \qtth{Q_{t\theta}}
\newcommand \qthth {Q_{\theta\theta}}

\newtheorem{defi}{\sc Definition\rm}[section]

\newtheorem{Theorem}[defi]{\sc Theorem\rm}

\newtheorem{Definition}[defi]{\sc Definition\rm}

\newtheorem{Thm}[defi]{\sc Theorem\rm}

\newtheorem{Proposition}[defi]{\sc Proposition\rm}
\newtheorem{Prop}[defi]{\sc Proposition\rm}
\newtheorem{lem}[defi]{\sc Lemma\rm}
\newtheorem{Lem}[defi]{\sc Lemma\rm}
\newtheorem{Lemma}[defi]{\sc Lemma\rm}

\newtheorem{Corollary}[defi]{\sc Corollary\rm}
\theorembodyfont{\upshape}
\newtheorem{Remark}[defi]{\sc Remark\rm}
\newtheorem{Rem}[defi]{\sc Remark\rm}

\theoremstyle{nonumberplain}
 \theoremsymbol{$\Box$}
\theorembodyfont{\upshape}
\newtheorem{proof}{\sc Proof:\rm}

\begin{document}

\title{On  the dynamics of Gowdy space times}

\author{Myeongju Chae\thanks{Current address: School of Mathematical Sciences, Seoul National
University,
 San56-1 Shinrim-dong Kwanak-gu, Seoul 151-747, Korea.
 Supported by BK 21 project; email
  \protect\url{mjchae@math.snu.ac.kr}}, 
Piotr T. Chru\'sciel\thanks{Part of work on this paper has been
carried at the Albert Einstein Institute of the MPG, Golm.
Partially supported by a Polish Research Committee grant  KBN 2
P03B 073 24; email \protect\url{
piotr@gargan.math.univ-tours.fr}; URL \protect\url{www.phys.univ-tours.fr\~piotr}} \\  D\'epartement de math\'ematiques\\
Facult\'e des Sciences\\ Parc de Grandmont\\ F37200 Tours, France
}

\maketitle
\begin{abstract}
We study the behavior near the singularity t=0 of Gowdy metrics.
We prove existence of an open dense set of boundary points near
which the solution is smoothly ``asymptotically velocity term
dominated" (AVTD). We show that the set of AVTD solutions
satisfying a uniformity condition is open in the set of all
solutions. We analyse in detail the asymptotic behavior of ``power
law" solutions at the (hitherto unchartered) points at which the
asymptotic velocity equals zero or one. Several other related
results are established.
\end{abstract}

 \tableofcontents

\section{Introduction}\label{sec:intro}

 The Gowdy family of
space-times~\cite{GowdyANoP} constitutes an interesting toy model
to study formation of singularities in general relativity. This
family of metrics is sufficiently simple to hope to analyse the
resulting singularities in an exhaustive way. It is sufficiently
non-trivial so that the relevant dynamical behavior has not been
understood  so far. The main question of interest is the curvature
blow-up -- or lack thereof -- at the boundary $t=0$ of the
associated space-time. The reader is referred to \cite{ChLake} for
a further discussion of this issue, we simply note that the
relevant geometric information can be obtained by deriving a sharp
asymptotic expansion of the solutions near the singular set $t=0$.
The main purpose of this work is to prove a stability result for
the existence of such expansions.

 In Gowdy space times  the
essential part of the Einstein equations reduces to a nonlinear
wave-map-type system of equations~\cite{GowdyANoP} for a map $x$
from $(M,g_{\al\beta})$ to the hyperbolic plane $(\CH,h_{ab})$,
where $M=\lbrack T, 0) \times \rm{S}^1$ with the flat metric
$g=-dt^{2}+d{\theta}^2$. The solutions are critical points of the
Lagrangean
\begin{equation}\label{lag}
    \CL\lbrack x \rbrack = \frac 12 \int_{M} t g^{\al\beta }h_{ab}\pa
    _{\al}x^a\pa_{\beta }x^b \, d\theta dt\;.
    \end{equation}
This differs from the usual wave-map Lagragean by a supplementary
multiplicative factor $t$.
     It is sometimes convenient to use coordinates
  $P,Q\in \R$  on the hyperbolic plane  in which the hyperbolic metric $h_{ab}$
  takes the form
 \begin{equation}\label{pq}
 h= dP^2 + e^{2P}dQ^2.
 \end{equation}
 Let $ X_{t}=\frac {\pa x}{\pa t}, X_{\theta}= \frac {\pa x}{\pa
 \theta}$, $D$ denote the Levi-Civita connection of $h_{ab}$,
 and $D_\theta\equiv \frac {D}{D\theta}:=D_{X_{\theta}}$, $D_t\equiv\frac {D}{D  t}:=D_{X_{t}}$.
 The Euler-Lagrange equations for (\ref{lag}) take the form
 \begin{equation}\label{euler}
   \frac {DX_{t}}{D  t} - \frac {DX_{\theta}}{D\theta} =-  \frac {X_{t}}{ t}
 \end{equation}
or, in coordinates,
$$ \Box x^{a}+ \Gamma ^{a}_{bc}\circ x \pa_{\mu}x^b \pa^\mu x^{c}=
-  \frac {\pa _{t}x^a}{t},$$ where the $\Gamma$'s are the
Christoffel symbols of $h_{ab}$, and $\Box =
\partial_t^2-\partial_\theta^2$. Global existence of smooth solutions on $(-\infty,0)$
of the Cauchy problem for \eq{euler} has been established by
V.~Moncrief~\cite{Moncrief:Gowdy}.

\begin{figure}[t]
\begin{center}
  \includegraphics[scale=1]
 {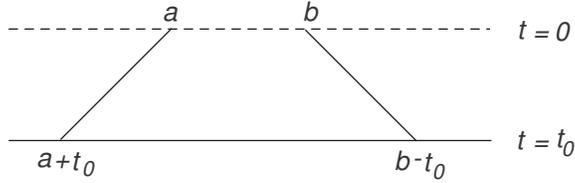}
\caption{The set $\Omega(a,b,t_0)$.}\label{F41}
\end{center}
\end{figure}

For further use we note the non-vanishing Christoffel symbols of
$h$: \bel{chri} \Gamma^P_{QQ}=-e^{2P}\;,\quad
\Gamma^Q_{PQ}=\Gamma^Q_{QP} = 1\;.\ee  In the $(P,Q)$ coordinates
one thus has \beaa &\displaystyle \partial_t^2 P -
\partial^2_\theta P = - \frac {\partial_t P }{t} + e^{2P}
\left((\partial_t Q)^2 -(\partial_\theta  Q)^2\right) \;, & \\
&\displaystyle \partial_t^2 Q - \partial^2_\theta Q = - \frac
{\partial_t Q }{t} -2 \left(\partial_t P \partial_t Q
-\partial_\theta P\partial_\theta Q\right) \;. & \eeaa  We
 consider solutions defined on sets $\Omega(a,b,t_0)$,
where \bel{intd00} t_0<0\;,\ a<b\;, \quad
\Omega(a,b,t_0):=\{t_0\le t<0\;,\ a+t\le \theta \le b-t\}\ee (see
Figure~\ref{F41}).
Thus our analysis is local, if the solution satisfies certain
properties  on an interval $[a,b]\subset S^1$, then the
conclusions hold on that interval. Throughout this work we assume
that the initial data for the map $x$ at $t_0$ are smooth
functions of $\theta$. We prove the following theorem (the Geroch
group is defined in Section~\ref{Sgeroch}; the position function
$\varphi_\infty$  is defined by \Eq{ravtd2}):

\begin{Theorem}
\label{Tstabcvd} Let $(\mathring x(t_0,\cdot),\mathring
X_t(t_0,\cdot))$ be Cauchy data for a solution of the Gowdy
equations on $\Omega(a,b,t_0)$ 
such that the associated solution $\mathring x$ has {\em uniformly
controlled blow-up}; by this we mean that
\bel{ucbuo} \sup_{\theta\in[a-|t|, b+ |t|]}
\left(\sum_{k=0}^\ell|t^{k+1}D_\theta^{k}
X_\theta|+\sum_{k=1}^\ell|t^{k+1}D_\theta^{k}
X_t|\right)(t,\theta)\to_{t\to0} 0 \;,\ee with $\ell=2$. Then:
\begin{enumerate}
\item For all  $\theta\in [a,b]$ the function $|tX_t|(t,\theta)$
converges  to a velocity function $v(\theta)$ as $t$ tends to
zero, uniformly in $\theta$. \item There exists an open dense set
on which $v$ is smooth.
\item $[a,b]$ can be covered by a finite number of intervals
$[a_i,b_i]$ with the following property: for each $i$ there exists
an element $G_i$ of the Geroch group such that $G_i\mathring x$
has a smooth velocity function $0\le v<1 $ and a smooth position
function $\varphi_\infty$, except perhaps on the boundary of the
set $\{v(\theta)=0\}$. Further $G_i\mathring x$ satisfies a power
law blow-up, \Eq{plbu}.
\item In the associated space-time the curvature scalar $R_{\alpha
\beta \gamma \delta} R^{\alpha \beta \gamma \delta}$ blows up in
finite proper time on every causal curve approaching
$$\mcB:=\{0\}\times \left([a,b]\setminus \{v(\theta)=1\}\right)\times S^1
\times S^1\;.$$ In particular the associated Gowdy space-time is
inextendible across $\mcB$. \item There exists $\eta>0$ such that
for all initial data $( x(t_0,\cdot), X_t(t_0,\cdot))$
satisfying\footnote{\label{Fdist}We equip $\Omega(a,b,t_0)$ with
the Riemannian metric $dt^2+d\theta^2$, this together with the
metric $h$ on $\mcH_2$ induces Riemannian metrics on all the
bundles involved. We use those metrics to measure the distance
between points on those bundles.}
$$\|(
x(t_0,\cdot)-\mathring{x}(t_0,\cdot),
X_t(t_0,\cdot)-\mathring{X}_t(t_0,\cdot))\|_{H^3\oplus H^2} < \eta
$$ the associated solution $x$ also satisfies \eq{ucbuo}, and hence also (i)-(iv) above.

\end{enumerate}
\end{Theorem}

\begin{Remark}
\label{Rmain} Actually it suffices to have a sequence $t_i\to 0$
along which \eq{ucbuo} holds. A recent result of
Ringstr\"om~\cite{RingstroemOberwolfach}\footnote{The results
presented in~\cite{RingstroemOberwolfach} have been made available
as a preprint~\cite{Ringstroem4} while this paper was being
prepared for publication.} can be used to lower to $\ell=1$ the
threshold $\ell$ in \eq{ucbuo}.
\end{Remark}

The proof of Theorem~\ref{Tstabcvd} can be found at the end of
Section~\ref{Sopde}.

It is of interest to enquire whether the uniform blow-up condition
\eq{ucbuo} is necessary for AVTD behavior of the solutions.
Consider, for example, an \avtdpqthree\ solution, as defined in
Section~\ref{Sspikes}, for which the error terms in
\eq{avtd1}-\eq{avtd2} and in their derivative counterparts (see
\eq{avtd1d}) are uniform in $\theta$. If $v_1$ is strictly smaller
than one (no negative lower bound assumed), then the solution
satisfies \eq{ucbuo}\footnote{This follows immediately from
Proposition~\ref{sup} together with the calculations of the proof
of Lemma~\ref{L7.2}. For points $\theta$ at which $v_1\ge 1$ one
expects $Q_\infty$ to have vanishing derivatives, compare point
(ii) of Proposition~\ref{Pnv} and Theorem~\ref{Tsvel2}. Then, if
$v_1(\theta_0)\ge 1$ and if $\partial^i_\theta
Q_\infty(\theta_0)=0$, $i=1,2,3$, then \eq{avtd1}-\eq{avtd2} and
their derivative counterparts give pointwise decay of the function
appearing under the sup at the left-hand-side of \eq{ucbuo}, but
uniformity is far from being clear.}. This shows in particular
that the set of solutions satisfying the hypotheses of
Theorem~\ref{Tstabcvd} is not empty, as existence of a large class
of \avtdpqi\ solutions satisfying $v_1<1$ follows from the results
in~\cite{Rendall:2000ih}.

The second main result of this work is the proof that for every
solution there exists an open dense set $\hat\Omega\subset S^1$
near which we have complete control of the solution:

\begin{Theorem}
\label{Tdc2} Consider a solution $x$ defined on $\Omega(a,b,t_0)$.
There exists an open dense set $\hat \Omega\subset [a,b]$ such
that $x$ is \avtdpqi\ in a neighborhood of
$\{0\}\times\hat\Omega$.
\end{Theorem}

The examples discussed in Section~\ref{Sspikes} show that the
result is sharp, with the following proviso: the known examples
have a velocity function defined everywhere, even at points where
it is not continuous, while  Theorems~\ref{Tdc2} and \ref{Tdc}
leave open the possibility of existence of points where the
velocity is not defined. Such points are characterised in point
(i) of Proposition~\ref{Pbad}.

The third main result of this paper is an exhaustive analysis of
the asymptotic behavior of power-law solutions at points at which
$v$ vanishes, or equals one. This last case is especially
important for the discussion of strong cosmic censorship, we refer
the reader to~\cite{ChLake} for applications. We note that no
results concerning those velocities were available so far in the
non-polarised case.

 The
results discussed above are established through a series of
auxiliary results which have some interest in their own. We say
that a solution  $x$ satisfies {\em a power law blow-up}, or is of
\emph{power-law type}, or is a \emph{power-law solution}, if the
norm of the theta derivatives vector $|X_\theta|$ does not blow up
faster than $ |t|^{\epsilon -1}$, for some positive constant
$\epsilon$, when approaching the singularity $t=0$. All solutions
of the smooth Cauchy problem on $T^3$ analysed in detail so far
satisfy\footnote{\label{funif}More precisely, the examples known
to us satisfy a power law decay  on all sets $C^0_{t_0}(\psi)$ as
defined in \Eq{Ctp} below. The constants are uniform in $\psi$
away from the points at which the asymptotic velocity has spikes,
or discontinuities, or crosses zero or one. In~\cite{ChLake} an
explicit self-similar solution has been given which does not
satisfy the power law decay, but this solution does not fit into a
Cauchy problem framework.} a power law decay.

It is simple  to show that every solution with a power law decay
has a continuous asymptotic velocity function $v$ (see the proof
of Theorem~\ref{Tcontv} below). The associated solutions of the
vacuum Einstein equations have curvature blowing up uniformly,
except perhaps at the set of points $\theta$ at which
$v(\theta)=1$.
 Consider the set of
initial data for solutions satisfying a power law decay and for
which $v<1$, uniformly in $\theta$. We show -- see
Theorem~\ref{Tcvs} below -- that this set is open in the set of
all initial data; this is one of the steps of the proof of
Theorem~\ref{Tstabcvd}. We further show that for those solutions
$v$ is smooth except perhaps at the boundary of the set of points
at which $v$ vanishes. Theorem~\ref{Tcvs} leads to a sharper
version of the \emph{stability of the singularity theorem} for
$(2/3,2/3,-1/3)$ Kasner metrics, see Theorem~\ref{Tstab}.

An important element of our analysis is the action of the Geroch
group, as defined in Section~\ref{Sgeroch}. In fact, the key
ingredients of our analysis are the results in~\cite{SCC} together
with the following:\begin{enumerate} \item The analysis of the
action of the Geroch group in the work of Rendall and
Weaver~\cite{RendallWeaver}; \item The reformulation of the
wave-map equations as a first-order system of scalar equations by
Christodoulou and Tahvildar-Zadeh~\cite{CT93}; \item The
small-derivatives stability result of
Ringstr\"om~\cite{Ringstroem3}.
\end{enumerate}

 We finish this introduction  by recalling
some results from~\cite{SCC} which will be useful in the sequel:

\begin{Prop}[Time-weighted pointwise estimates; Proposition 3.2.1 in \cite{SCC}]\label{sup}
Let $x(t_0, \theta) \in C^k (S^1)$, $k \ge 1$, $X_t (t_0, \theta)
\in C^{k-1}(S^1)$. For all $t \ge t_0$ we have
\begin{enumerate}
\item $(|X_t|^2 + |X_\theta|^2)(t, \theta)
\le 2 \left\{ \sup_{\psi \in [\theta - t + t_0, \theta + t - t_0]}
(|X_t|^2 + |X_\theta|^2)(t_0, \theta) \right\} \left( \frac
{t_0}{t} \right)^2$.
\item If $k \ge 2$,
then there exist constants $C$ depending only upon the arguments
listed such that, for all $1 \le |\al| \le k$,
\begin{align}\label{eqn3.2.7}
|D^\al x|(t, \theta) \le C(|\al|, t_0, \Vert X_\theta (t_0)
\Vert_{C^{|\al|-1}}, \Vert X_t (t_0) \Vert_{C^{|\al|-1}})
|t|^{-|\al|}.
\end{align}
\end{enumerate}
\end{Prop}

\begin{Remark}\label{Rnew} It has been pointed out to us by H.~Ringstr\"om
that the proof of Proposition~3.2.1 in~\cite{SCC} (compare
\cite[Equation~(3.2.5)]{SCC} together with the argument leading to
Equation~(3.2.9) there) actually gives an inequality somewhat
stronger than (i) above: \bean\lefteqn{\left( \frac {t}{t_0}
\right)^2(|X_t|^2 + |X_\theta|^2)(t, \theta)\le}&& \\ && \frac 12
\Big\{ \sup_{\psi \in [\theta - t + t_0, \theta + t - t_0]}
|X_t-X_\theta|^2(t_0, \psi) + \sup_{\psi \in [\theta - t + t_0,
\theta + t - t_0]} |X_t+X_\theta|^2(t_0, \psi) \Big\}\;. \nonumber
\\ && \eeal{impest} \Eq{impest} carries more information about the
solution than \eq{eqn3.2.7}, which can  be seen {\em e.g.\/} when
the initial data have small $\theta$-derivatives.
\end{Remark}

\begin{Prop}[Time-weighted Sobolev decay; Proposition 3.3.1 in \cite {SCC}]\label{integral}
Let $x \in C^i([t_0, 0) \times S^1)$ and let $X_\theta (t_0,
\cdot)$, $X_t (t_0, \cdot) \in H_i (S^1)$, $i \ge 1$. Then there
exist constants depending only upon the arguments listed such that
\begin{enumerate}
\item For all $1 \le |\al| \le i+1$,
\[ g^{(\al)}(t) \equiv \oint d \theta |t|^{2|\al|} |D^\al x|^2
\le C (|\al|, \Vert X_\theta (t_0) \Vert_{H_{|\al|-1}(S^1)}, \Vert
X_t (t_0) \Vert_{H_{|\al|-1}(S^1)}, t_0) .\]
\item If at least one differentiation is a $\theta$ differentiation we have
\[ \lim_{t \to 0} g^{(\al)} (t) = 0 . \]
\item If at least one differentiation is a $\theta$ differentiation
then $\frac {g^{(\al)}(t)}{|t|} \in L^1([t_0, 0])$ and
\[ \int_{t_0}^0 \frac {g^{(\al)}(s)}{|s|} ds
\le C' (|\al|, t_0, \Vert X_\theta (t_0) \Vert_{H_{|\al|-1}(S^1)},
\Vert X_t (t_0) \Vert_{H_{|\al|-1}(S^1)}) .
\]
\end{enumerate}
\end{Prop}

\section{Problems with $\theta$ derivatives, self-similar solutions}
 \label{Ssss}

\begin{figure}
\begin{center}
\begin{picture}(150,120)(30,-10)
\thinlines \put(65,65){\line(1,0){70}} \put(0,0){\line(1,0){200}}
\put(0,0){\line(1,1){100}}
\put(200,0){\line(-1,1){100}}
\put(100,115){\makebox(0,0)[t]{$(0,\psi)$}}
\put(100,10){\makebox(0,0)[t]{$t=t_0$}}
\put(100,33){\makebox(0,0)[c]{$C^{t_1}_{t_0}(\psi)$}}
\put(164,45){\makebox(0,0)[l]{$R^{t_1}_{t_0}$}}
\put(36,45){\makebox(0,0)[r]{$L^{t_1}_{t_0}$}}
\put(100,70){\makebox(0,0)[c]{$t=t_1$}}
\end{picture}
\end{center}
\caption[PDE.1]{The truncated domains of dependence
$C^t_{t_0}(\psi)$.} \label{PDE.1}
\end{figure}
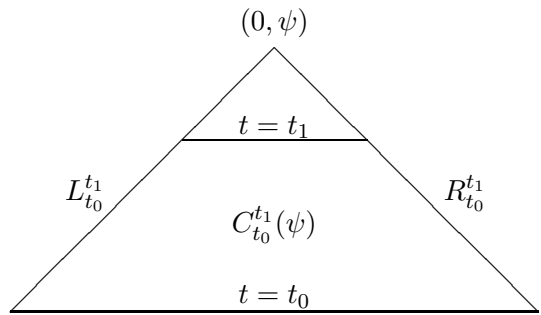

As already mentioned in the introduction, all  published solutions
of the Gowdy equations known to us, and for which the asymptotic
behavior is reasonably well
understood~\cite{CIM,SCC,Ringstroem3,isenberg90,Rendall:2000ih,RendallWeaver},
have the property that \bel{tdbo} |tX_\theta|\le C t^\epsilon\ee
for some $\epsilon>0$, with the bound being
optimal$^{\mbox{\scriptsize \ref{funif}}}$. The power law is very
useful for the control of the analytic properties of the
solutions, but it is not necessary for curvature blow-up. In any
case the bound \eq{tdbo} certainly implies, for all $\psi\in S^1$,
\bel{gdc} \int_{C^0_{t_0}(\psi)} |X_\theta|^2 \,dt\,d\theta <
\infty\;, \ee where for $t_0< 0$  the set $C^0_{t_0}(\psi)$ is
defined as (compare Figure~\ref{PDE.1})
\bel{Ctp} C^0_{t_0}(\psi)=\{t_0 \le t <0\;, \ - |t| \le
\theta - \psi \le |t|\}\;.\ee
We shall say that $\clim f = \alpha$ if
\bel{Cconv} \lim_{t\to0}\sup_{ - |t| \le \theta - \psi \le |t|}
|f(t,\theta)-\alpha| =0\;.\ee
Such limits look a little awkward
at first sight; however, they arise naturally when considering the
behavior of the geometry along causal curves with endpoints on the
boundary $t=0$. Further, such limits appear naturally in our
results below.

 We have a partial converse to \eq{tdbo}:
\begin{Proposition}
[Proposition~3.4.1 in \cite{SCC}] \label{Ptdc} At every $\psi$ at
which \eq{gdc} holds we have, for all multi-indices $\alpha$,
$$\clim |t|^{|\alpha|+1} |D^\alpha X_\theta| =0\;.$$
\end{Proposition}

\begin{Remark} In Section~\ref{SEvf} below we give further integral conditions which
ensure pointwise convergence of $|tX_t|$ to a number $v(\psi)$.
Yet another criterion for existence of $v(\psi)$ is given by
Proposition~\ref{Pbad}.
\end{Remark}

Proposition~\ref{Ptdc} begs the question of existence of solutions
for which \eq{gdc} fails. An obvious candidate is given by
self-similar solutions: \bel{sss} x(t,\theta) = y(\theta/t)\;,\ee
for some map  $y$ from $M$ to the hyperbolic space. It would be of
interest to find all solutions satisfying \eq{sss}. Here we note
the following family of such solutions: let $\alpha,\beta\in \R$
and let $\Gamma:\R\to \mcH$ be an affinely parameterised
unit-speed geodesic in $\mcH$, for $|\theta| < -t $ set \bel{sss2}
x(t,\theta) =\Gamma \Big(\alpha\arcsin\Big(\frac \theta
t\Big)+\beta\Big)\;. \ee It is easily checked that \eq{sss2}
solves the Gowdy equation~\eq{euler}. \Eq{gdc} fails for the
solution \eq{sss2} when $\alpha\ne 0$, as expected. It turns out
that  the solutions \eq{sss2} do not fit into the Cauchy problem
framework
 because they are singular on the whole light
cone $|\theta| = -t $, while the solutions we are interested in
are smooth at $t=t_0$. Singular solutions can sometimes be used to
produce smooth examples of bad behavior, but we have not managed
to exploit this solution to do that. In view of our stability
results here it would be  important to construct a solution of the
Cauchy problem where \eq{gdc} fails, or to prove that such
solutions do not exist.

\section{\protect\avtdk\ behavior, spikes and discontinuities}
\label{Sspikes} In \cite{CIM,KichenassamyRendall,Rendall:2000ih} a
large class of solutions of \eq{euler} has been constructed with
the following behavior:
\beal{avtd1old} P(t,\theta) &=& -v_1(\theta)\ln |t| +P_\infty(\theta)
+ o(1)\;,\quad \hbox{$0 < v_1(\theta)<1$}\;,
\\ Q(t,\theta) &=& Q_\infty(\theta) +
    |t|^{2v_1(\theta)}\Big(\psi_Q(\theta) + o(1)\Big)\;. \eeal{avtd2old}
    A generalisation of those formulae to arbitrary
    velocities
    requires a careful study of the field equations. For instance,
    an analysis of the indicial exponents of the linearised
    equations suggests the following behavior of $Q$ at points where
    $v(\theta)=n\in \N^*$ (compare~\cite{ChLake} for
    $n=1$)\footnote{The expected \emph{a priori} estimate $|t^{i+1}D^i_\theta X_\theta|\to 0$
    implies that $\partial^{i}_\theta Q_\infty(\theta)=0$ for $i=1,\ldots,2n-1$, which is implicit in \eq{Qexp}.}
    \beal{Qexp}Q(t,\theta) &=&
Q_\infty(\theta)
+\frac{t^{2n}}{(2n)!} \partial^{2n}_\theta Q_\infty(\theta) \ln
|t| + \psi_Q(\theta) t^{2n} +o(t^{2n})\;.\eea While there is no
existence statement for solutions with a non-zero coefficient in
the $\ln |t|$ term above, we expect that such solutions can
actually be constructed. We note that if $v\in \N^*$ on an
interval, then no log term will occur on that interval.

Applying the ``solution-generating transformation" \eq{inv}
described below to \eq{Qexp} at a point at which one further has
$Q_\infty(\theta)=0$ leads to a solution $(P',Q')$ with a negative
$P$-velocity $v'_1(\theta)=-n$ and with a logarithmically
blowing-up $Q$ function
\beal{Qexp2}Q'(t,\theta) &=&e^{2P_\infty(\theta)}\left(
\frac{1}{(2n)!} \partial^{2n}_\theta Q_\infty(\theta) \ln |t| +
\psi_Q(\theta)\right) +o(1)\;.\eea The above discussion suggests
that the following will capture the asymptotic behavior of a large
class of solutions of the Gowdy equations:
\beal{avtd1} P(t,\theta) &=& -v_1(\theta)\ln |t| +P_\infty(\theta)
+ o(1)\;,
\\ Q(t,\theta) &=& Q_\infty(\theta) + \left\{%
\begin{array}{ll}
    |t|^{2v_1(\theta)}\Big(\psi_Q(\theta) + o(1)\Big)\;, & \hbox{$0< v_1(\theta)\not\in \N$ ;} \\
    |t|^{2v_1(\theta)}\Big(Q_{\ln}(\theta)\ln |t|+\psi_Q(\theta) + o(1)\Big)\;, & \hbox{$0<v_1(\theta)\in \N$ ;} \\
    Q_{\ln}(\theta)\ln |t| + o(1)\;, & \hbox{$ v_1(\theta)\in -\N^*$ ;} \\
    o(1)\;, & \hbox{$-\N^*\not\ni v_1(\theta)\le 0$ .}  
\end{array}%
\right.\nonumber \\&& \eeal{avtd2} 
The function
\bel{vdefn} v:=|v_1|\ee will be called the \emph{velocity
function}, while $Q_\infty$ will be called the \emph{$Q$-position
function}. Those functions have the following geometric
interpretation~\cite{grubisic93}: for $v>0$ the path
$$\tau\to
\Gamma_\theta(\tau):=(P(-e^{-\tau},\theta),Q(-e^{-\tau},\theta))$$
approaches -- in a sense made precise by \eq{avtd1}-\eq{avtd2} --
the affinely parameterised $h$-geodesic
$$\tau\to \mathring\Gamma_\theta(\tau):=(v_1(\theta)\tau - \varphi(\theta),Q_\infty(\theta))\;,$$
with $v$ - the length of the velocity vector of
$\mathring\Gamma_\theta$. The point $Q_\infty(\theta)$ is then the
uniquely defined point on the conformal boundary of the hyperbolic
space at which the geodesic $\mathring\Gamma_\theta$ accumulates.
Clearly this interpretation breaks down at $v(\theta)=0$, which
suggests that solutions might display strange features, not
necessarily compatible with \eq{avtd1}-\eq{avtd2}, at the boundary
of the set $\{v(\theta)=0\}$.

We shall say that a map $x=(P,Q)$ is in the \avtdpq\ class if
there exist  functions $v_1$, $Q_\infty$, and $Q_{\ln}$ such that
\beal{avtd1b} P(t,\theta) &=& -v_1(\theta)\ln |t| + O(1)\;,
\\ Q(t,\theta) &=& \left\{%
\begin{array}{ll}
     Q_\infty(\theta) + o(1)\; , & \hbox{$v_1\not \in -\N^*$ ;} \\
    Q_{\ln}(\theta)\ln |t| + Q_\infty(\theta) + o(1)\;, & \hbox{$v_1 \in -\N^*$ .} \\
\end{array}%
\right.   \eeal{avtd2b} We shall say that a solution is in the
\avtdpqk\ class if \eq{avtd1}-\eq{avtd2} hold with functions
$v_1$, $P_\infty$, $Q_\infty$, $Q_{\ln}$ and $\psi_Q$ which are of
$C_k$ differentiability class (on closed intervals the derivatives
are understood as one-sided ones at the end points). For the
purposes of the \avtdpqk\ definition the function $Q_{\ln}$ is
assumed to be extended by $0$ to the set $v_1(\theta)\not \in-
\N^*$; we emphasise that such an extension will \emph{not} be
assumed in Definition~\ref{Davtd} below.  For $k>0$ we will assume
that the behavior \eq{avtd1}-\eq{avtd2} is preserved under
differentiation in the following way:
\bel{avtd1d} \forall\ 0\le i+j\le k \qquad
\partial_\theta^j(t\partial_t)^i\Big(P(t,\theta)
+v_1(\theta)\ln |t| -P_\infty(\theta)\Big) = o(1)\;,\ee similarly
for $Q$.

Note that the classes \avtdpq\ and \avtdpqz\ do not coincide.

Unless explicitly stated otherwise the $o(1)$ symbol denotes
\emph{pointwise convergence to zero as $t$ tends to zero at fixed
$\theta$}. Similarly $O(1)$ means \emph{uniformly bounded in $t$
at fixed $\theta$}. An alternative meaning of $o(t)$, which will
be sometimes used, is provided by uniform convergence to zero on
the set $\Omega(a,b,t_0)$ as defined by \eq{intd00}. Yet another
possibility is convergence to zero in a $\clim$ sense, as defined
by \eq{Cconv}; in any case we will make precise statements when
needed.

Using the above solutions, Rendall and Weaver~\cite{RendallWeaver}
have constructed solutions of \eq{euler} which display ``spiky
features". They separate those solutions into two classes, one
called ``false spikes" and one called ``true spikes". An
instructive example of this behavior is provided by a family of
Gowdy maps discovered by Moncrief~\cite{Vinceunpublished}, and
analysed in detail in \cite[Appendix~B]{SCC}. They are given there
in terms of the polar coordinates on the hyperbolic space,
\bel{rhophi} h = d\rho^2 + \sinh^2\rho \;d\varphi^2\;,\ee which
are related to the $(P,Q)$ coordinates by the formulae\footnote{We
are very grateful to Marsha Weaver for several enlightening
discussions concerning the issues discussed in this section, and
for providing us formulae \eq{ptP1}-\eq{ptP2}.} \bean e^P& = &
\cosh \rho + \sinh \rho \;\cos \varphi
\\ & = & \frac 12 \left( e^{\rho}(1+\cos \varphi) + e^{-\rho}(1-\cos \varphi)\right)\;,\label{ptP1}
\\ e^P Q &=& \sinh \rho \;\sin \varphi  \;.\eeal{ptP2}
Moncrief's ansatz \bel{mon1} \rho=\rho(t)\;, \quad \varphi = n
\theta\;,\quad n \in \N\;,\ee leads to the following: every
solution is uniquely determined by two numbers $v_\infty  \in
[0,1)$, $\rho_\infty\in \R$, such that \bel{mon2} \rho=-v_\infty
\ln|t| + \rho_\infty + o(1)\;.\ee Inserting \eq{mon1}-\eq{mon2}
into \eq{ptP1} one finds
\bean P(t,\theta) & = &  \left\{%
\begin{array}{ll}
    -v_\infty
\ln|t| + \rho_\infty + \ln \left( \frac{1+\cos (n\theta)}{2}\right)+o(1)\;, & \hbox{$n\theta\ne \pi \mod 2\pi$;} \\
    v_\infty
\ln|t| -\rho_\infty+o(1)\; , & \hbox{$n\theta= \pi \mod 2\pi$,} \\
\end{array}%
\right. \nonumber \\ && \label{mon3}\\ Q(t,\theta) &=&  \left\{%
\begin{array}{ll}
 \frac{\sin (n\theta)}{1+\cos (n\theta)}+o(1)\;, & \hbox{$n\theta\ne \pi \mod 2\pi$;} \\
   0\; , & \hbox{$n\theta= \pi \mod 2\pi$.} \\
\end{array}%
\right. \eeal{mon3b}
If we define $v_1:S^1\to \R$ by the equation
\bel{v1def} v_1(\theta):= \lim_{t\to0} |t|P_t\;,\ee then $v_1 =
v_\infty>0 $ except at $n$ isolated \emph{spike points} $\theta_m
= (1  +2m)\pi/n$, $m\in \N\cap[0,n]$, at which $v_1$ is equal to
$-v_\infty< 0$. \Eq{mon3} shows that the subleading term in $P$
blows up logarithmically  at the spike points, so that no
uniformity in $\theta$ for that term can be expected near those
points in norms which control pointwise behavior of $P$ and $Q$.
It is interesting that even though $Q_\infty$ blows up as one
approaches the spike points, it is finite, actually vanishing,
there. Next, even though at each fixed $\theta$ we have
$$\lim_{t\to 0} P_\theta (t,\theta)=0\;,$$
there are timelike curves reaching the boundary along which
$P_\theta$ does not go to zero: for example, for $n=1$,
$$\lim_{t\to 0} P_\theta (t,\pi+\alpha e^{-\rho(t)} )=\infty\;,\quad \lim_{t\to 0} P_\theta
(t,\pi+\alpha e^{-2\rho(t)} )= \frac{\alpha}{2}\;.$$

The above examples provide solutions with an arbitrary finite
number of spikes. Solutions with a countably infinite number of
spikes accumulating at some point $\theta_\infty\in S^1$ can be
constructed as follows: Consider an \avtdpqi\ solution such that
the function $Q_\infty$ in \eq{avtd2} has an infinite number of
isolated zeros $\theta_i$ accumulating at $\theta_\infty$, and
such that $v_1$ avoids zero in a neighborhood of $\theta_\infty$;
the existence of such solutions follows from
\cite{Rendall:2000ih}. Following~\cite{RendallWeaver}, one then
performs the following ``inversion" of the hyperbolic plane:
\bel{inv} e^{-P'} = \frac{e^{-P}}{Q^2 + e^{-2P}}\;,\quad Q' =
\frac Q {Q^2 + e^{-2P}}\;.\ee This is an isometry of $h$ and
therefore maps solutions into solutions. It is easily seen that
$(P',Q')$ will have a  spike at each of the points $\theta_i$,
yielding the desired spiky solution.

The discontinuities discussed so far consisted of isolated points
at which $v_1$ changes sign. Solutions with jumps of $v_1$ can be
constructed as follows: consider any \avtdpqi\ solution such that
the zero set of the function $Q_\infty$ in \eq{avtd2} is a closed
interval $[a,b]$, with $v_1$ strictly positive near the end
points. As before, the existence of such solutions follows from
\cite{Rendall:2000ih}. It is easily seen from \eq{inv} that the
velocity function $v'_1$ associated with the map $x'=(P',Q')$ will
jump from  $v_1(a)$ to $-v_1(a)$ at $a$, and will be continuous
from the right there. Further, as $\theta$ increases from $a$ to
$b$ the new velocity function $v'_1$ will continuously attain the
value $-v_1(b)$ when $b$ is approached from the left, and jump to
$v_1(b)$ immediately afterwards.

Clearly, the above behaviors can be combined to give infinite
sequences of pointwise jumps and/or  intervals on which $v'_1$ is
negative, with the set of discontinuities of $v_1'$ accumulating
at a given point.

 In fact, let $F\subset S^1$ or $F\subset [a,b]$ be any non-empty
closed set without interior, we claim that there exists a smooth
function $\varphi_F$ such that
$$ \varphi_F^{-1}(\{0\}) = F\;.$$
In order to see this, let $x\not\in F$, and let $(x_-,x_+)$ be the
largest open interval containing $x$ which does not meet $F$
(hence $x_\pm \in F$), we set $\sigma(x) = (x-x_-)(x_+-x)$. Define
$$ \varphi_F(x)=\left\{%
\begin{array}{ll}
    0, & \hbox{$x\in F$;} \\
    e^{-1/\sigma(x)}, & \hbox{otherwise.} \\
\end{array}%
\right.
$$
Then  $\varphi_F$ has all the required properties. Using the
function $\varphi_F$ as $Q_\infty$, with $v_1$ equal, \emph{e.g.},
to the constant function  $1/2$, after performing an inversion we
obtain a new function $v'_1$ which equals $-1/2$ on $F$, and $1/2$
on $S^1\setminus F$. If $F$ is a fat Cantor set one obtains a
rather wild set of spikes, with measure as close as desired to
that of $S^1$ or that of $[a,b]$ by choosing $F$ suitably.

 The spikes discussed so far are called
\emph{false spikes}, as they can be thought of as an artifact of
the $(P,Q)$ coordinate system above: no discontinuous behavior
occurs in the $(\rho,\varphi)$ representation of the
solutions.\footnote{Applying isometries of the hyperbolic plane to
the solution has the effect of reshuffling Killing vectors, and
can thus be considered as an irrelevant ``coordinate
transformation" from the space-time point of view. We note that
the isometry \eq{inv} changes the orientation of the hyperbolic
plane. The accompanying relabeling of the Killing vectors changes
the space-time orientation, leading thus to a non-equivalent
solution if a space-time orientation has been chosen. However, we
can always perform a second inversion about a different point,
regaining the original orientation. If the map $\theta \to
Q_\infty(\theta)$ is surjective, this will always introduce at
least one ``false spike" in the transformed solution. In other
words, in the surjective case there will be no $(P,Q)$
representation of the solution without ``false spikes". This gives
some geometric meaning to those.} However, the $(P,Q)$ coordinates
are very useful when analysing the \emph{Gowdy-to-Ernst
transformation}, because that transformation has a very simple
form precisely in the $(P,Q)$ coordinates: given a solution
$x=(P,Q)$ of the Gowdy equations, one defines a new solution $\hat
x$ by performing the ``Gowdy-to-Ernst"
transformation~\cite{RendallWeaver}:
\bel{GtE} \hat P := -P - \ln |t|\;, \quad e^{\hat P} \partial _t
\hat Q: = - e^P \partial _\theta Q\;, \quad e^{\hat P} \partial
_\theta \hat Q: = - e^P
\partial_t Q\;. \ee The new map satisfies again the Gowdy
equation~\eq{euler}. As shown by Rendall and Weaver, this has
significant consequences: By definition, a \emph{true spike} is
the image of a false spike after a Gowdy-to-Ernst transformation
has been performed; equation \eq{GtE} shows that any discontinuity
in $v_1$ leads to a discontinuity in the velocity $\hat v_1$
associated with the map $(\hat P, \hat Q)$. For AVTD solutions
this typically leads to a discontinuity in the geometric velocity
function $\hat v= |\hat v_1|$. For instance, for the solutions
\eq{mon1} the transformation \eq{GtE} leads to
\bel{mont}\hat P =  \left\{%
\begin{array}{ll}
    -(1-v_\infty)
\ln|t| - \rho_\infty - \ln \left( \frac{1+\cos (n\theta)}{2}\right)+o(1)\;, & \hbox{$n\theta\ne \pi \mod 2\pi$;} \\
    -(1+v_\infty)
\ln|t| +\rho_\infty+o(1)\; , & \hbox{$n\theta= \pi \mod 2\pi$,} \\
\end{array}%
\right. \ee which clearly results in a $|\hat v|$ which is
\emph{not\/} continuous at $n\theta= \pi \mod 2\pi$.

We refer the reader to \cite[Section~6]{RendallWeaver} for a
further discussion of iterations of the above.

It is interesting to enquire about independence of the conditions
\eq{avtd1b}-\eq{avtd2b}. It turns out that \eq{avtd1b} is the key
requirement, up to a Gowdy-to-Ernst transformation:

\begin{Proposition}
\label{Pnv} \begin{enumerate} \item At each point  $\theta$ at
which \eq{avtd1b} holds with $v_1 >0$ we also have \eq{avtd2b}. If
the term $O(1)$ in \eq{avtd1b} is uniform in $\theta$ over some
interval $I$ then $Q_\infty$ is continuous on $I$. If further
$v_1$ is
uniformly bounded away from zero, then the term $o(1)$ in
\eq{avtd1b} is uniform in $\theta\in I$.
\item If $v_1
>1$ on an interval $I$, with
the term $O(1)$ in \eq{avtd1b}  uniform in $\theta$, then there
exists a constant $A$ such that
$$\forall \;\theta \in I\quad Q(t,\theta)\to_{t\to0} A$$
(we say that $x$ is \emph{asymptotically polarised} on $I$). \item
At points $\theta$ or intervals $I$ on which \eq{avtd1b} holds
with $v_1 <1$ the conclusions of point (i) above hold for the
Gowdy-to-Ernst transformed map $\hat x$. If $v_1<0$ on $I$ then
the conclusion of point (ii) holds for $\hat Q$. (In the case of
an interval $I$ we  assume that the term $O(1)$ in \eq{avtd1b} is
uniform in $\theta$.)
\end{enumerate}
\end{Proposition}

\proof 
Point (i) of Proposition~\ref{sup} shows that \bel{Qesto} |Q_t| +
|Q_\theta| \le \frac {C(\theta)e^{-P}}{|t|}\le
C(\theta){|t|^{v_1(\theta)-1}}\;.\ee
 Integrating in $t$
one obtains \eq{avtd2b}. If $C(\theta)$ can be made
$\theta$-independent, then $Q_\infty$ is a uniform limit of
continuous functions, and therefore continuous, which gives (i).
If $v_1(\theta)
>1$, then \eq{Qesto} shows that $ |Q_\theta|$ tends to zero as $t$ goes to
zero, which easily implies that $Q_\infty$ is constant over $I$.
This proves (ii). Applying the Gowdy-to-Ernst transformation
\eq{GtE} one finds that the hatted velocity function $\hat v_1$
associated with $\hat P$ equals $1-v_1$, and point (iii) follows.
 \myqed

 Let us summarize the properties of the $(P,Q)$ coordinates
 which follow from the above:
 \begin{enumerate}
 \item They provide a simple explicit formula for the Gowdy-to-Ernst
 transformation;
 \item They describe faithfully the geometric behavior of the solutions
 except near those points at the conformal boundary with
 $\varphi=\pi\mod 2\pi$;
\item They describe faithfully the geometric behavior of those
solutions on intervals of $\theta$ on which $v$ vanishes.
 \end{enumerate}
The problem with $\varphi=\pi\mod 2\pi$ above is avoided by
turning to the $(\rho,\varphi)$-description of the solutions. So,
instead of \eq{avtd1}-\eq{avtd2} we write \beal{ravtd1}
\rho(t,\theta) &=& -v(\theta)\ln |t| +\rho_\infty(\theta) +
o(1)\;,
\\ \varphi(t,\theta) &=& \varphi_\infty(\theta) + \left\{%
\begin{array}{ll}
    |t|^{2v(\theta)}\Big(\psi_\varphi(\theta) + o(1)\Big)\;, & \hbox{$0<v(\theta)\not \in N^*$ ;} \\
    |t|^{2v(\theta)}\Big(\varphi_{\ln}(\theta)\ln |t| +\psi_\varphi(\theta) + o(1)\Big)\;,
    & \hbox{$v(\theta) \in N^*$ ;} \\
    o(1)\;, & \hbox{$v(\theta)= 0$ .}  
\end{array}%
\right.\nonumber \\&& \eeal{ravtd2} (We impose the usual
restriction that $\rho\ge 0$ so $v$ above is necessarily
non-negative, though in some situations it might be convenient not
to do this, allowing $\rho$ to be negative, but then identifying
the points  $(\rho,\varphi)$ with $(-\rho,\varphi-\pi)$.) It
follows from \eq{ptP1} that the notation $v$ for the
$\rho$-velocity in \eq{ravtd1} is comptatible with \eq{avtd1} and
\eq{vdefn}.  A map $x=(\rho,\varphi)$ will be said to be \avtdrp\
on an interval $[a,b]$ if there exist real valued functions $v$
and $\varphi_\infty$ such that for $\theta\in[a,b]$ we have
\beal{ravtd1.0} \rho(t,\theta) &=& -v(\theta)\ln |t|+ O(1)\;,
\\ \varphi(t,\theta) &=& \varphi_\infty(\theta)  + o(1)\;.
\eeal{ravtd2.0} $x$ will be said to be \avtdrpk\ on an interval
$[a,b]$ if \eq{ravtd1}-\eq{ravtd2} holds with functions $v$,
$\rho_\infty$, $\varphi_\infty$, $\varphi_{\ln}$ and
$\psi_\varphi$ which are $C^k$ on $[a,b]$. For $k>0$ the
derivatives are assumed to behave as in \eq{avtd1d}.

 Because of the usual polar
coordinate singularity at $\rho=0$ the $(\rho,\varphi)$
coordinates do not
 always reflect the geometric character of the dynamics for
solutions on intervals on which $v(\theta)=0$.

We note the following result, which follows immediately from the
calculations in the proof of Proposition~\ref{PGtE} below. It
shows that the only points $\theta$ at which the distinction
between \avtdrpk\ and \avtdpqk\ behavior matters are those at
which $v$ vanishes (where $(\rho,\varphi)$ might be singular) or
at which $\varphi_\infty=\pi \mod 2 \pi$ (where the restriction of
$Q$ to the conformal boundary of the hyperbolic space is
singular):

\begin{Proposition} \label{Pob} Let $k\ge 0$. \begin{enumerate}
\item If the map $x$ is \avtdpqk\ on $[a_1,b_1]$ with $v_1$
avoiding zero on $[a_1,b_1]$, then it is \avtdrpk. Further
$\varphi_\infty\mod 2 \pi$ avoids $\pi$ on $[a_1,b_1]$ if $v_1>0$,
while $\varphi_\infty\equiv \pi\mod 2 \pi$ if $v_1<0$. \item If
the map $x$ is \avtdrpk\ on $[a_1,b_1]$ with $\varphi_\infty\mod 2
\pi$ avoiding $\pi$ on $[a_1,b_1]$, then it is \avtdpqk.
\end{enumerate}
\end{Proposition}

\begin{Remark}
\label{RPnv} We have an obvious equivalent of points (i) and (ii)
of Proposition~\ref{Pnv} for the $(\rho,\varphi)$ representation
of the solutions, with identical proof, regardless of whether or
not  $\varphi_\infty$ meets $\pi \mod 2 \pi$.
\end{Remark}

What  has been said so far in this section leads naturally to the
following definition:

\begin{Definition}
\label{Davtd} A map $x:\Omega(a,b,t_0)\to \mcH_2$ will be said to
be \avtd, respectively \avtdk, on $[a_1,b_1]\subset [a,b]$ if
there exists a function $v:[a_1,b_1]\to \R^+$ such that:
\begin{enumerate}
\item For every interval $I\subset [a_1,b_1]$ on which $v$
vanishes the map $x$ is \avtdpq, respectively \avtdpqk, near
$\{0\}\times I$, with $|v_1|=v$. \item For every interval
$I\subset [a_1,b_1]$ on which $v$ has no zeros the map $x$ is
\avtdrp, respectively \avtdrpk, near $\{0\}\times I$.
\end{enumerate}
\end{Definition}

\section{The Geroch group and its action}\label{Sgeroch}

We consider solutions of the Gowdy equations defined on
$\Omega(a,b,t_0)$ (see \eq{intd00}), for some $a\le b$, $t_0<0$.
We fix\footnote{We emphasise that there are several coordinate
systems $(P,Q)$ in which the metric $h$ takes the form \eq{pq},
differing from each other by an isometry of $h$.}
 once and for all a $(P,Q)$ coordinate system on
$(\mcH_2,h)$. Following~\cite{Geroch:1971nt}, we define the
\emph{Geroch group} $\mcG$ as the set of finite strings of the
form $G=G_1G_2\cdots G_k$, where each of the $G_i$'s is either an
isometry of $(\mcH_2,h)$, denoted by $I_i$, or is a Gowdy-to-Ernst
transformation~\eq{GtE}, denoted by $E$.
The Geroch group acts on solutions as follows: First, $G_1\cdots
G_n$ acts on $x$ by first acting with $G_n$ on $x$, then acting
with $G_{n-1}$ on $G_n x$, \emph{etc}. Next, if $x$ is a solution
of \eq{euler}, then we start by writing it in the $(P,Q)$
coordinate system just chosen. Isometries act on solutions by
composition. This implies that the $\mcG$-group product $I_1I_2$
of two isometries $I_1$ and $I_2$ is
 the  composition $I_1\circ I_2$ of $I_1$ with $I_2$. The action of a Gowdy-to-Ernst
transformation $E$ on a solution is defined as follows: we
integrate \eq{GtE} with the integration constant chosen so that
$\hat Q(t_0,a)=Q(t_0,a)$. This leads to the group product
$E^2=\Id$, where $\Id$ is the identity isometry of the hyperbolic
plane. The $\mcG$--group products $E I$ and $I E$, are defined by
the above action on solutions.

Let $G=G_1G_2\cdots G_k$, then any two adjacent Gowdy-to-Ernst
transformations can be canceled out, leading to a shorter
presentation of $G$. Similarly any two adjacent isometries can be
replaced by a single isometry. This leads eventually to a
presentation of $G$ such that isometries and Gowdy-to-Ernst
transformations alternate. The  number of  Gowdy-to-Ernst
transformations in the resulting presentation of $G$ will be
called the \emph{order} of $G$. Thus, the order of $I$ is zero,
the order of $E$, or $IE$, or $EI$, is one, \emph{etc}.

 The behavior near $t=0 $ of all the known to us solutions of the Cauchy problem for the
 Gowdy
equations is captured in the definition of the set $\mcU_1$ below;
the set $\mcU_2$ is then a subset of $\mcU_1$ with a
\emph{genericity condition}; we expect $\mcU_2$ to be useful in
the analysis of the strong cosmic censorship problem in the class
of Gowdy space-times. The following comment is in order  here: as
emphasised in the previous section, the $(P,Q)$ variables provide
a parametrization of the hyperbolic space which does not always
correctly reflect the geometric aspects of the asymptotic behavior
of the solutions, so the reader might wonder why to invest so much
effort to characterise the asymptotics of the Gowdy solutions in
terms of those variables rather than, say, the $(\rho,\varphi)$
variables of \eq{rhophi}. The answer is that the Gowdy-to-Ernst
transformation takes a simple form in the $(P,Q)$ variables,
compare \eq{GtE2}, and this is what forces us to carry out the
analysis below.

 \begin{Definition} \label{DGtE} 1. Let $\mcU_1$ be the set of smooth solutions of the Gowdy equations
defined on $\Omega(a,b,t_0)$ for which \eq{avtd1}-\eq{avtd2} holds
with some functions $v_1$, $P_\infty$, $Q_\infty$ defined on
$[a,b]$, a function $\psi_Q$ defined on the set $\{v_1>0\}$, and a
function $Q_{\ln}$ defined on the set $\{v_1\in
\Z^*\}$,
satisfying the following:
\begin{enumerate}
\item $v_1$ is uniformly bounded. \item $v_1$ is continuous on an
open dense subset of $[a,b]$. 
\item The restrictions of $Q_\infty$ to the sets
$\{\theta\,:\,v_1(\theta)\not \in -\N\}$ and
$\{\theta\,:\,v_1(\theta)\in -\N\}$ are continuous functions on
those sets.

\item The restriction of $P_\infty$ to the set  $\{\theta: $ $Q_\infty$ and
$v_1$ are continuous at $\theta\}$ is a continuous function on
this set, similarly for $Q_{\ln}$ and $\psi_Q$.

\item At points at which $Q_{\ln}(\theta)\ne
0$ the error terms $o(1)$ in \eq{avtd1}-\eq{avtd2} have the
property that they remain $o(1)$ after multiplication by $\ln
|t|$.
\end{enumerate}
\medskip

2. We define $\mcU_2$ to be the subset of $\mcU_1$ consisting of
those solutions 
for which the sets  of discontinuities and of critical points of
$v_1$ and of $Q_\infty$ are finite.
\end{Definition}

It is an open question whether there exist solutions of the smooth
Cauchy problem for the Gowdy equations which are not in $\mcU_1$.

 In our analysis below we will
need the following:

\begin{Proposition} \label{PGtE}  $\mcU_1$ and $\mcU_2$ are invariant under the action of
isometries of $(\mcH_2,h)$.
%
%
\end{Proposition}

\begin{Remark}
The arguments of the proof of Theorem~\ref{Tdc} show that the set
of solutions which are in $\mcU_1$ 
is stable under the action of the whole Geroch group. It is not
completely clear  what happens with the action of the Geroch group
on solutions in $\mcU_2$, since the integration of the $\hat Q$
equations might introduce non-generic behavior.
\end{Remark}

\proof  Let $\phi$ be an isometry of the hyperbolic space into
itself. We will write
$$\hat x=\phi\circ x\;,$$ and denote by $(\hat P,\hat Q)$ the
associated coordinate functions.

 If $\theta_0\in [a,b]$ is such that
$v_1(\theta_0)=0$, then continuity of $\phi$ shows that
\eq{avtd1}-\eq{avtd2} does hold for $(\hat P(t,\theta),\hat
Q(t,\theta))$ at $\theta_0$, with the new velocity $\hat
v_1(\theta_0)=0=v_1(\theta)$.

 Let $\mcV$ be the open dense set on which $v_1$ is continuous, then $\mcV$ is a
nonempty countable union of open intervals $I_i$, $\mcV=\cup_i
I_i$. By definition the velocity function $v_1$ is continuous on
each $I_i$. We rewrite the $I_i$'s as
$$I_i= \underbrace{\{v_1>0 \}}_{I_{i+}}\cup\underbrace{\{v_1<0 \}}_{I_{i-}}
\cup\underbrace{\{v_1=0 \}}_{I_{i0}}\;.$$ The simplest set to
analyse is $I_{i0}$: for $(t,\theta)\in [t_0,0)\times I_{i0}$ the
map $x(t,\theta)$ stays in a compact set, so does $\phi\circ x$
for any isometry $\phi$, and the property that
$(P_\infty,Q_\infty)$ are continuous  on $I_{i0}$ is clearly
preserved under the action of isometries.

Let us use the angle $\varphi$ of \eq{rhophi} to parameterise the
conformal boundary of the hyperbolic space, then to every
 point for which $v_1(\theta)>0$ we can assign a unique
$\varphi_\infty(\theta)$ such that the trajectory $t\to
x(t,\theta)$ asymptotes to the point at infinity
$\varphi_\infty(\theta)$. From \eq{ptP1}-\eq{ptP2} we have
\bel{Qtp} v_1(\theta)> 0 \ \Longrightarrow\ Q_\infty(\theta)=
\frac {\sin(\varphi_\infty(\theta))}{1+\cos
(\varphi_\infty(\theta))}\;,\ee except  at \bel{Qtp1}
\varphi_\infty(\theta)=\pi \mod 2\pi\ee where a more careful
analysis is required. We emphasise that a possible singularity
arising here would only reflect the singular behavior of the
$Q$-parameterisation of the conformal boundary, and not a
singularity of $\varphi_\infty(\theta)$: continuity of
$\varphi_\infty$ on the set $\{v>0\}$ can be established as in
Proposition~\ref{Pnv}, working directly in the $(\rho,\varphi)$
coordinates, regardless of whether or not $\cos (\theta)=-1$. In
any case $Q_\infty$ is continuous on $I_{i+}$ by hypothesis, and
so is therefore $\varphi_\infty$.

Inverting \eq{ptP1}-\eq{ptP2} 
one finds 
\bel{ptP3} e^\rho = \frac {e^{-P} + e^P(1+Q^2)}2 +
\sqrt{ \left(\frac {e^{-P} + e^P(1+Q^2)}2\right)^2-1}\;.\ee (The
alternative solution
\bel{ptP3b} e^{-\rho} = \frac {e^{-P} + e^P(1+Q^2)}2 + \sqrt{
\left(\frac {e^{-P} + e^P(1+Q^2)}2\right)^2-1}\ee always leads  to
a negative $\rho$. \Eqs{ptP3}{ptP3b} reflect the fact that a point
$(P,Q)$ corresponds both to $(\rho,\varphi)$ and $(-\rho,
\varphi-\pi)$. In the current proof we follow the usual convention
that $\rho\ge 0$.)
It follows that \bel{ptP4} \rho(t,\theta) = -v_1(\theta)\ln |t| +
\rho_\infty(\theta) + o(1)\;,\ee for some  number
$\rho_\infty(\theta)$.  Inserting \eq{ptP4} into \eq{ptP2} one is
then led to
\bel{ptP5}\varphi(t,\theta) = \varphi_\infty(\theta) + \left\{%
\begin{array}{ll}
    |t|^{2v_1(\theta)}\Big(\psi_\varphi(\theta) + o(1)\Big)\;, & \hbox{$v_1(\theta)\not \in N^*$ ,} \\
    |t|^{2v_1(\theta)}\Big(\varphi_{\ln}(\theta)\ln |t| +\psi_\varphi(\theta) + o(1)\Big)\;,
    & \hbox{$v_1(\theta) \in N^*$ ,}
\end{array}%
\right. \ee for some numbers $\psi_\varphi(\theta)$,
$\varphi_{\ln}(\theta) $. As $\varphi_\infty$ is a continuous
function of $\theta$ on $I_{i+}$, continuity in $\theta$ of
$\rho_\infty(\theta)$ there  follows. Continuity of
$\psi_\varphi(\theta)$ and $\varphi_{\ln}(\theta)$ over the sets
$I_{i+}\cap\{v_1\in \N^*\}$ and $I_{i+}\cap\{v_1\not\in \N^*\}$
follows in an identical manner.

Recall, now, that any isometry $\phi$ of the hyperbolic space
extends to a smooth diffeomorphism of the conformal boundary of
$(\mcH_2,h)$, say $\chi$. This shows that on the set
$\{v(\theta)>0\}$ the solution $\hat x:=\phi \circ x$ will have a
$\varphi$-position function
$$\hat \varphi_\infty(\theta)=\chi(\varphi_\infty(\theta))\;.$$
The fact that isometries extend smoothly to the conformal boundary
further shows that  the asymptotic behavior \eq{ptP4}-\eq{ptP5} is
preserved under the action of isometries of the hyperbolic space,
so that the map $(\hat \rho,\hat \varphi)$ will satisfy the hatted
version of \eq{ptP4}-\eq{ptP5} at points with positive $v_1$.

Let $\theta$ be any point such that
$\chi(\varphi_\infty(\theta))\ne \pi\mod 2\pi$. A straightforward
analysis of \eq{ptP1}-\eq{ptP2} shows that one will recover
\eq{avtd1}-\eq{avtd2} for the map $(\hat P,\hat Q)$ at $\theta$,
with $\hat v_1(\theta)=v_1(\theta)>0$.

 Set \beaa\hat
I_{i+}&:=&\underbrace{\{\chi(\varphi_\infty(\theta))\ne \pi\mod
2\pi\}}_{\hat I_{i++}}\bigcup
\\ &&  \underbrace{\{\exists \ \mbox{interval
$J$ around $\theta$ such that }\chi\circ\varphi_\infty|_J\equiv
\pi\mod 2\pi\}}_{\hat I_{i+-}}\subset I_{i+}\;.\eeaa Then $\hat
I_{i+}$ is clearly open in $I_{i+}$. Suppose that $\theta\in
I_{i+}$ is such that there are no points of $\hat I_{i++}$ in a
neighborhood of $\theta$, then $Q_\infty$ is constant and equal to
$\pi\mod 2\pi$ on that neighborhood, hence $\theta\in \hat
I_{i+-}$. It follows that $\hat I_{i+}$ is dense in $I_{i+}$.

On $\hat I_{i++}$ the function  $\hat \varphi_\infty-\pi$ avoids
integer multiples of $2\pi$, and continuity of $\hat
v_1(\theta)=v_1(\theta)>0$ on  $I_{i+}$ follows. Similarly one
obtains a new continuous position function $\hat Q_\infty$ by
using \eq{Qtp} with $\varphi_\infty$ there replaced by $\hat
\varphi_\infty$.

On the other hand, at points at which $\hat \varphi_\infty=\pi\mod
2\pi$ and $v_1\not\in \N^*$ we have from \eq{ptP1}-\eq{ptP2} and
\eq{ptP4}-\eq{ptP5}
\beal{ptP6} \hat P(t,\theta) & = & v_1(\theta)
\ln|t| + \hat \rho_\infty(\theta) + \ln \left(1+\frac{(\hat
\psi_\varphi(\theta))^2}4\right) + o(1)\;,
\\ \hat Q(t,\theta) & = & - \frac 12 e^{\hat\rho(\infty) -
\hat P(\infty)}\hat\psi_\varphi(\theta)+o(1)\;. \eeal{ptP7}
Analogous equations hold with supplementary $\ln |t|$ terms for
$v_1\in \N^*$, compare \eq{Qexp2}. This shows that
\eq{avtd1}-\eq{avtd2} hold again, with a negative $v_1$. Further
$v_1$ is  continuous on $\hat I_{i+-}$. It follows that $I_{i+}$
contains an open dense set on which $v_1$ is continuous.

Consider, finally, points at which $v_1(\theta)<0$.  
\Eq{ptP3} leads to
\beaa \rho(t,\theta) &=& 
\left(-P + \frac {e^{2P}(1+3Q^2)}{4} +  O(e^{4P})\right)(t,\theta)\\
& = & -|v_1(\theta)|\,\ln |t| -P_\infty(\theta)+
o(1)
\;.\eeaa Inserting this into \eq{ptP2}
yields
$$ \sin (\varphi) = \frac{e^P}{\sinh \rho}
Q=2e^{2P_\infty}(1+o(1))|t|^{2|v_1|(\theta)}Q\;,$$ so that
$\sin(\varphi(t,\theta))$ goes to zero as $t$ does. \Eq{ptP1}
shows that we must have $\varphi(t,\theta) \to_{t\to 0} \pi \mod
2\pi$ and one obtains, again modulo $ 2 \pi$,
$$ \varphi(t,\theta) =\pi+\left\{%
\begin{array}{ll}
|t|^{2|v_1(\theta)|}2e^{2P_\infty(\theta)}\left(Q_\infty(\theta)+
o(1)\right), & \hbox{$v_1\not \in -\N^*$ ;} \\
    |t|^{2|v_1|(\theta)}2e^{2P_\infty(\theta)}\Big(Q_{\ln}(\theta)\ln |t| +Q_\infty(\theta) + o(1)\Big)\;,
    & \hbox{$v_1(\theta) \in -N^*$ .} \\
\end{array}%
\right.$$ (For $v_1(\theta) \in -N^*$  we have used the hypothesis
that $\ln |t| \times o(1)$ remains $o(1)$.) The calculations done
so far show that \eq{avtd1}-\eq{avtd2} hold for all $\theta\in
[a,b]$ for the map $\phi\circ x$. A repetition of the arguments
given on $ I_{i+}$ justifies continuity on an open dense subset of
$ I_{i-}$, and the proposition for $\mcU_1$ easily follows.

 The result for $\mcU_2$
follows immediately from the calculations above, using the fact
that for maps in $\mcU_2$ all level sets of $Q_\infty$ form a
 finite collection of points.
 \myqed

\section{A symmetric hyperbolic system} \label{Sshs}

Let us set
     \begin{align*}
     P_{t}&=f_{1}, \ \  P_{\theta}=g_{1}, \\
     e^{P}Q_{t}&=f_{2}, \ \  e^{P}Q_{\theta}=g_{2}.
     \end{align*}  Equation~\eq{euler} takes then the form of the following first order symmetric
hyperbolic system:
\begin{equation}
    \pa_{t}¥\left(
    \begin{array}{c}
      f_{1} \\   f_{2} \\  g_{1} \\  g_{2} \\
     \end{array}
     \right)=
    \left(
     \begin{array}{cccc}
      0 & 0 & 1& 0 \\  0 & 0 & 0& 1 \\  1 & 0 & 0 & 0 \\
       0 & 1 & 0 & 0 \\
     \end{array} \right)\pa_{\theta}
     \left(
     \begin{array}{c}
      f_{1} \\  f_{2} \\
      g_{1} \\   g_{2} \\
    \end{array} \right) +
    \left(
    \begin{array}{c}
    f_{2}^{2}-g_{2}^{2}-\frac {f_{1}}{t} \\
    -f_{1}f_{2}+g_{1}g_{2}-\frac{f_{2}}{t}\\
    0\\
    f_{1}g_{2}-g_{1}f_{2}\\
     \end{array}
     \right)\;,\label{shs}
     \end{equation}The new unknowns $f_{a}, g_{a}$ agree with
     the coefficient functions of $1-$forms, $\psi_{\mu}^{A}$,
     of Christodoulou and Tahvildar-Zadeh's work on spherically symmetric
     wave maps \cite{CT93}, when an appropriate trivialisation of the bundle of vectors
     tangent to the hyperbolic space
     has been chosen: Indeed, if we set \bel{bvect} e_1 = \partial_P\;,\qquad e_2 =
     e^{-P}\partial_Q\;,\ee
then, in view of \eq{pq}, the $e_a$'s form a globally defined
$h$--orthonormal frame with constant structure coefficients (and
thus constant connection coefficients), and
$$f_a = h(e_a,X_t)\;,\qquad  g_a = h(e_a,X_\theta)\;.$$

We consider solutions defined on domains of dependence
$\Omega(a,b,t_0)$, defined in \eq{intd00}. By
Proposition~\ref{sup} we have \bel{intd0} \sup_{\Omega(a,b,t_0)}
\left( |tf_1| + |tf_2| + |tg_1| + |tg_2|\right) < \infty\;. \ee

Throughout this work the value of various irrelevant constants may
change from line to line.

 Since $|X_\theta|^2 = g_1^2 +
g_2^2$, Proposition~\ref{integral} implies
\beal{intd2} &
\lim_{t\to0} \displaystyle\inti t^2 (g_1^2+ g_2^2)d\theta = 0 \;,
&\\&\displaystyle\inti t (g_1^2+ g_2^2)d\theta \in
L^1([t_0,0])\;.&\eeal{intd1} We further note that by \eq{chri} we
have \bean D_\mu X_{\nu} & = & (\partial_\mu\partial_\nu P
-e^{2P}\partial_\mu Q
\partial _\nu Q)\partial_P + ( \partial_\mu\partial_\nu Q + Q_\mu P_\nu + Q_\nu
P_\mu)\partial_Q
\\ & = & (\partial_\mu\partial_\nu P
-e^{2P}\partial_\mu Q
\partial _\nu Q)e_1 + e^{P}( \partial_\mu\partial_\nu Q + Q_\mu P_\nu + Q_\nu
P_\mu)e_2\;, \nonumber \\ && \eeal{intd2.0} so that
\bel{intd2.2} D_\theta X_\theta = (\partial_\theta g_1 -g_2^2) e_1 + (\partial_\theta g_2 + g_1 g_2) e_2
\;.\ee It then easily follows from \eq{intd0}-\eq{intd1} together
with Proposition~\ref{integral} that
\beal{intd2.6} &
\lim_{t\to0} \displaystyle\inti t^4 \left((\partial_\theta g_1)^2+
(\partial_\theta g_2)^2\right )d\theta = 0 \;,
&\\&\displaystyle\inti t \left((\partial_\theta g_1)^2+
(\partial_\theta g_2)^2\right )d\theta \in
L^1([t_0,0])\;.&\eeal{intd1.8}
 It turns out that we also have

\begin{Proposition}\label{Pf2d}
\beal{intd3} &
\lim_{t\to0} \displaystyle\inti t^2 f_2^2d\theta = 0 \;, &\\&
\displaystyle\inti t f_2^2d\theta \in L^1([t_0,0])\;.&\eeal{intd4}
\end{Proposition}

\proof Let
$$F(t)= \inti tf_1 d\theta\;,$$
then
\bean \frac {dF}{dt} & = & t(g_1-f_1)(t,b-t) - t(g_1+f_1) (t,a+t)
+ \inti t(f_2^2 -g_2^2)\;.\eea Now,  $tf_1$ is bounded, hence so
if $F$, and by integration of the last equation we obtain
$$\int_{t_0}^{t}\inti tf_2^2\, d\theta dt \le \int_{t_0}^{t}\inti tg_2^2\, d\theta dt
+C$$ for all $t_0\le t <0 $. \Eq{intd1} together with the monotone
convergence theorem imply \eq{intd4}. In order to prove \eq{intd3}
we calculate
\bean
\Bigg |\frac d {dt} \inti t^2 f_2^2\Bigg | &= & \Bigg
|-t^2f_2^2(t,b-t) -t^2f_2^2(t,a+t)
\\ && + \inti f_2\left(\partial_\theta g_2 - f_1 f_2 + g_1 g_2\right)\Bigg |
\nonumber \\ &\le & C\left(1 + \inti \left(|t|^3 (\partial_\theta
g_2)^2 +|t| (f_2^2 + g_1^2 +g_2^2)\right)\right)\;. \nonumber \\
&&\eeal{intd6} The function on the right-hand side of the last
line is in $L^1([t_0,0])$ by \eq{intd1}, \eq{intd1.8} and
\eq{intd4}. Integrating  \eq{intd6} one concludes that the limit
at the left-hand side of \eq{intd3} exists. If this limit were
different from zero \eq{intd4} couldn't hold, whence the result.
\myqed

We are ready to prove now:

\begin{Prop}\label{Pf3d} For $k\ge 0$ we have
   \beal{intd2.6h} &
    \lim_{t\to0} \displaystyle\inti \vert t \vert^{2(k+1)}\left((\pa^k_\theta
    g_1)^2
    +(\pa^k_\theta g_2)^2\right)d\theta = 0 \;,&\\&
    \displaystyle\inti \vert t \vert^{2k+1} \left((\partial^k_\theta g_1)^2+
(\partial^k_\theta g_2)^2\right )d\theta \in
L^1([t_0,0])\;,&\label{intd1.8h} \\
\label{intd3h} & \lim_{t\to0} \displaystyle\inti \vert t
\vert^{2(k+1)} (\pa^k_{\theta}f_2)^2 d\theta = 0 \;, &\\&
\displaystyle\inti \vert t \vert^{2k+1}(\pa^k_{\theta}f_2)^2
d\theta \in L^1([t_0,0])\;.&\eeal{intd4h}
\end{Prop}
\begin{proof} The cases $k=0$ have already been established, as well as \eq{intd2.6h} and \eq{intd1.8h} with $k=1$. A simple induction argument, using the
formulae~\eq{chri} for the Christoffel symbols, shows that
  \begin{equation}\label{ghi}
  D^k_{\theta}X_{\theta}=\Big(\pa^k_\theta g_{1}+
  F_{k}¥(\pa^{k-1}_{\theta}g, \ldots,g)\Big)e_{1}+
  \Big(\pa^k_\theta g_{2}+
  G_{k}¥(\pa^{k-1}_{\theta}g, \ldots,g)\Big)e_{2},
  \end{equation}
     where $ g$ denotes both  $g_{1},\;g_{2}$, while the $F_{k}(\cdot)$'s and
     $     G_{k}(\cdot)$'s
     are   polynomials in the variables
     $\pa^m_{\theta}g,\; 0\le m\le k-1$
  with the number of derivatives and factors in each of the terms $ \pa^{i_{1}}_{\theta}g\cdot\pa^{i_{2}}_{\theta}g
  \ldots\pa^{i_n}_{\theta}g$ satisfying $$n\ge 2\;,\quad \sum_{m=1}^{n} (i_{m}+1)\le
  k+1\;.$$
  This, together with Proposition \ref{integral},  proves \eq{intd2.6h} and \eq{intd1.8h}.

  Next, for $k\ge 1$ we compute
  \begin{equation}\label{fhi}
  \pa^k_{\theta} f_{2}=\sum_{i=0}^{k} C(i,k)\langle D^i_{\theta} X_{t},
  D^{k-i}_{\theta}e_{2} \rangle
  \end{equation} since $f_{2}=\langle X_{t},e_{2}\rangle.$
  By induction we obtain
  \begin{equation}\label{e_{2}}
   D^k_{\theta}e_{2}¥=\Big(\pa^{k-1}_\theta g+
  F'_{k-1}(\pa^{k-2}_{\theta}g, \ldots,g)\Big)e_{1}+
  \Big(\pa^{k-1}_\theta g +
  G'_{k-1}¥(\pa^{k-2}_{\theta}g, \ldots,g)\Big)e_{2},
   \end{equation}
     where the $F'_{k-1}¥(\cdot)$'s and $G'_{k-1}¥(\cdot)$'s have the same property as described above.
    Thus
    \begin{align*}
    \vert t \vert^{k+1} \vert \pa^k_{\theta}f_{2}\vert & \le
    \sum_{i=0}^{k} C(i,k)\vert t \vert^{i+1}\vert D^i_{\theta}X_{t} \vert
    \vert t \vert^{k-i}\vert D^{k-i}_{\theta}e_{2}\vert \\
    & \le C\sum_{i=1}^k\vert t\vert^i(\vert \pa^{i-1}_{\theta}g\vert+\vert
    F_{i-1}\vert+\vert G_{i-1}\vert)+ \vert t \vert^{k+1}\vert D^k_{\theta}X_{t}\vert.
    \end{align*}
    The proof is completed by combining Proposition~\ref{integral} with \Eqs{intd2.6h}{intd1.8h}.
\end{proof}

\section{Existence of a velocity function}\label{SEvf}

Numerical experiments (see~\cite{berger97b} and references
therein) suggest that the limit \bel{fv1} v(\theta):=\lim_{t\to
0}|tX_t|(t,\theta)\ee exists, and is a continuous function of
$\theta$ except for a ``small" exceptional set of $\theta$'s in
$S^1$. We set \bel{fv2} \Omt:= \{\theta\in S^1 \ \mbox{ such that
the limit \eq{fv1} exists }\}\;,\ee so that $v$ is a well-defined
function on $\Omt$. The existence of this limit is
 useful when analysing the geometry of the
associated space-time. Now Proposition~\ref{Pf2d} shows that
$t^2f_2^2$  goes to zero in $L^2$ as $t$ tends to zero, and since
$$|tX_t|^2 = t^2(f_1^2 + f_2^2)\;,$$
the whole information about the limit \eq{fv1} is contained in $t
f_1$, except possibly for a negligible set.

\subsection{Existence of a weak velocity function $\vw$}\label{SEwvf}

 We have the following:

\begin{Proposition}
 \label{Pwl} There exists $\vw\in L^\infty(S^1)$ such that
for any $p\in (1,\infty)$
$$ |t|f_1(t,\cdot) \stackrel{\ L^p\ }\rightharpoonup \vw\;,$$
where $\stackrel{L^p}\rightharpoonup$ denotes weak convergence in
$L^p(S^1)$.
\end{Proposition}

\begin{Remark}
In  the proof of Theorem~\ref{Tdc2} we establish existence of an
open dense set $\hat\Omega\subset S^1$ such that $\vw$ has a
smooth representative $v$ on $\hat\Omega$, with pointwise
convergence to $v$ on $\hat\Omega$.
\end{Remark}

\proof Let $v_i(\theta) = 2^{-i}f_1 (-2^{-i},\theta)$, then the
sequence $v_i$ is bounded in $L^2$ and therefore there exists
$\vw$ and a subsequence $v_{i_j}$ which converges weakly to $\vw$.
Let $\phi$ be any smooth function on $S^1$, we have
\beaa \partial_t \int_{S^1} tf_1 \phi & = & \int_{S^1} \left(
t\partial_\theta g_1 - t(f_2^2-g_2^2)\right) \phi
\\ & = &\int_{S^1} \left( -
t g_1\partial_\theta\phi  - t(f_2^2-g_2^2) \phi\right)\;. \eeaa
Integrating one finds
\beaa \lefteqn{ \left|\int_{S^1} t_1 f_1(t_1,\theta)
\phi(\theta)d\theta  -\int_{S^1} t_2 f_1(t_2,\theta)
\phi(\theta)d\theta\right| }&& \\ && = \left| \int_{t_2}^{t_1}
\int_{S^1} \left( - t g_1\partial_\theta\phi  - t(f_2^2-g_2^2)
\phi\right)d\theta dt \right|\;. \eeaa Setting $t_1 = -2^{-i_j}$
and letting $j$ go to infinity one obtains
\beaa \lefteqn{ \left|\int_{S^1} \vw(\theta)
\phi(\theta)d\theta  -\int_{S^1} |t_2| f_1(t_2,\theta)
\phi(\theta)d\theta\right| }&& \\ && \le \int_{t_2}^{0} \int_{S^1}
\left(|t g_1\partial_\theta\phi|  +|t|(f_2^2+g_2^2)
\phi\right)d\theta dt\;. \eeaa Since the integrand in the last
line is in $L^1([t_0,0]\times S^1)$ we obtain
\beaa \lim_{t\to 0} \int_{S^1} |t| f_1(t,\theta)
\phi(\theta)d\theta=\int_{S^1} \vw(\theta) \phi(\theta)d\theta
 \;, \eeaa so that $|t|f_1$ converges to $\vw$ in the sense of distributions.
Weak convergence in $L^p$ follows by elementary
 functional analysis, using the fact that smooth functions are dense in $L^{p'}$ for $p\in (1,\infty)$, with $p'$
 -- the H\"older conjugate of $p$.
 Lower semi-continuity of the norm with respect to weak
 convergence implies
 $$\| \vw\| _{L^p(S^1)} \le \lim_{j\to\infty} \|2^{-i_j}f_1(-2^{i_j},\cdot)\|_{L^p(S^1)} \le 2\pi
 \sup _{[t_0,0)\times S^1} |tf_1|\;,$$ so that
 $$\| \vw\| _{L^\infty(S^1)} \le  \frac{1}{2\pi}\sup_{p\in [2,\infty)} \|\vw\|_{L^p(S^1)} \le
 \sup _{[t_0,0)\times S^1} |tf_1|\;.$$
 \hfill$\Box$

The information contained in $\vw$ seems to be very poor. For
instance, since $\vw$ is defined only almost everywhere, one can
imagine situations in which $\vw$ has a smooth representative, but
nevertheless the dynamics has very rough features at some points.
This actually happens in the solutions with ``spikes"  discussed
in Section~\ref{Sspikes}. In those last examples one  has
pointwise convergence of $|tX_t|$ everywhere; this suggests that
even an exhaustive understanding of the properties of the velocity
function might not be enough to understand the dynamics of the
Gowdy models.

\subsection{From space-time integrals to pointwise velocity}\label{Sspcv}We introduce
\newcommand{\dgo}{\delta g_1^{\psi}}
\bel{psidef} \dgo(t) =
g_1(t,\psi-t)-g_1(t,\psi+t)\;,\ee
\bean\Omth&:=& \Big\{\psi\in S^1 \ \mbox{ such that } \  \dgo \in L^1([t_0,0)) \ \mbox { and } \\
&& \int_{C^0_{t_0}(\psi)} (f_2^2 + g_1^2 + g_2^2) \,dt\,d\theta <
\infty  \Big\}\;.\eeal{fv2b} Our aim in this section is to present
an integral criterion for pointwise existence of a velocity
function $v$ on $\Omth$; this will be used later in this work.

\begin{Theorem}
\label{Tpcv} For every $\psi\in\Omth$ there exists a number
$v(\psi)\in\R$ such that\bel{derde2c} \clim |tX_t| = v(\psi)\;,\
\mbox{  with } \ \clim |tf_2| =0\ee (recall that $\clim$ has been
defined in \eq{Cconv}). This implies in particular
$$\Omth\subset\Omt\;.$$ Moreover for
$\psi\in\Omth$ we have
\bel{derde}\clim |t^2D_\theta X_\theta|
=\clim |t^2D_\theta X_t| =\clim |tX_\theta| =0\;.\ee.
\end{Theorem}

\begin{Remark}
\label{RTpcv} We remark that for those points  $\psi\in\Omth$ for
which $v(\psi)\ne 1 $ we have curvature blow-up in the associated
space-time.
\end{Remark}

\begin{proof}We have
 \bea
\Box (tf_1) & = &
\pa_{\theta}g_{1}-(1+2tf_1)(f^2_{2}+g^2_{2})+2tf_{2}\pa_{\theta}g_{2}
-2tg_{2}\pa_{\theta}f_{2} \nonumber \\ && + 4 t f_2 g_1 g_2\;, \label{ipcv1}\\
\Box (tf_2) & = & \pa_{\theta}g_{2}+ f_{1}f_{2}+ g_{1}g_{2}
-2tf_{2}\pa_{\theta}g_{1}+2tg_{2}\pa_{\theta}f_{1}\nonumber\\
& & +tf_{2}g^2_{2} -tf_{2}g_1^{2}-tf^3_{2}+tf^2_{1}f_{2}
\;.\eeal{ipcv2} In particular
\bean 
\Box(tf_{2})^2 & = & 2tf_{2}\Box(tf_{2})+2(\pa_{t}(tf_{2}))^2-
2t^2(\pa_{\theta}f_{2})^2 \\
& = & 2tf_{2}\Box(tf_{2})-2t^2(\pa_{\theta}f_{2})^2+
2t^2(\pa_{\theta}g_{2}+ f^2_{2}-g^2_{2})^2 \eeal{ipcv4} using
(\ref{shs}). It follows from (\ref{gdc}), (\ref{gdc3}),
Propositions~\ref{Ptdc} and \cite[point b) of Lemma~3.4.1]{SCC}
that for $\psi\in\Omth$ the right-hand side of (\ref{ipcv4})
 is in
$L^1(C^0_{t_0}(\psi))$. The dominated convergence theorem applied
to the usual integral representation of solutions of the
one-dimensional wave equation,
$$u (t,x) = \mathring {u}(t,x) + \frac 12\int_{s=t_{0}}^t \int _{\theta=x-t+s}^{x+t-s}
\Box u(s,\theta) \; d\theta ds\;,$$ where $\mathring {u}$ is the
solution of the free wave equation with the same initial data,
shows that the limit $\clim (tf_2)^2$ exists. This limit has to be
zero, otherwise the integral condition on $f_2$ in  (\ref{fv2b})
wouldn't hold. If $\dgo$ is in $L^1([t_{0},0))$, then the
right-hand side of (\ref{ipcv1}) is also in
$L^1(C^0_{t_0}(\psi))$, and by a similar argument $\clim tf_1$
exists.
\end{proof}

 We close this section with the following remark:

\begin{Lemma}
\label{Lpcv} In the definition of $\Omth$, the condition
\bel{gdc3} \int_{C^0_{t_0}(\psi)} f_2^2 \,dt\,d\theta <
\infty  \ \mbox{ can be replaced by } \ \int_{C^0_{t_0}(\psi)}
|\partial _\theta f_1|  \,dt\,d\theta < \infty \;.\ee Similarly,
the condition
\bel{gdc4} \int_{t_0}^0 |\dgo| \,dt <
\infty  \ \mbox{ can be replaced by } \ \int_{C^0_{t_0}(\psi)}
|\partial _\theta g_1|  \,dt\,d\theta < \infty \;.\ee
\end{Lemma} \proof Let
$$F(t)= \into f_1(t,\theta) d\theta\;,$$
then
\beaa \frac {dF}{dt} & = & 
-f_1(t,\psi-t) - f_1 (t,\psi+t) + \into\left(\partial_\theta g_1+
\frac{f_1}{|t|} + f_2^2 -g_2^2\right)
\\  & = & 
-f_1(t,\psi-t) - f_1 (t,\psi+t) + \dgo (t) +
\into\left(\frac{f_1}{|t|} + f_2^2 -g_2^2\right)\;.\eeaa
Integration by parts gives the identity
\bel{intid} \frac 1 {|t|} \into u (\theta)d\theta = u(\theta-t) +
u(\theta+t)+ \into \frac {\psi-\theta} {|t|}
u'(\theta)d\theta\;,\ee so that
\bean \frac {dF}{dt} & = & \dgo(t) +
\into\left( \frac {\psi-\theta} {|t|} \partial_\theta f_1 +f_2^2
-g_2^2\right) \;.\eea As $F$ is bounded, integrating in $t$ gives
\eq{gdc3}. \Eq{gdc4} is obvious.
 \myqed

\section{Power law in Sobolev spaces}\label{SPlb} As discussed in the
Introduction, and as will be proved below in detail in any case, a
power law inequality \eq{tdbo} implies existence of a velocity
function. It turns out that one strategy for establishing
\eq{tdbo} is to derive a power-law for $t$-weighted Sobolev norms.
This is done in this section.

   It is useful to introduce the following
   quantities
   \bel{fq1} \mu_1  =\sup_{\Omega(a,b,t_0)}\vert tf_{1} \vert\;,\quad \mutwo = \sup_{\Omega(a,b,t_0)}
    \vert tf_{2} \vert\;,\quad \lambdatwo =
    \sup_{\Omega(a,b,t_0)}\vert tg_{2}\vert \;.\ee
    These are finite by \eq{intd0}. (Recall that $\Omega(a,b,t_0)$
    has been defined in \eq{intd00}.)

   Let $t_{0}\le t<0$, $a<b$.
    Define the $k-$th order energy $E_{k}(t)$ by
        \begin{equation}
    E_{k}(t)=\int_{a+t}^{b-t} \vert t\vert^{2k+2}\sum_{i=1,2}¥((\pa_{\theta}^k f_{i})^{2}
    + (\pa_{\theta}^k g_{i})^{2})\ d\theta.
    \end{equation}
     \begin{Proposition}\label{Ppl}
   1. If
\bel{p2new}
\beta:= \sup_{\Omega(a,b,t_0)}\left(|tf_1|+\frac  {|tf_2|}2 + \frac
{2|tg_2|^2}{\sqrt{1+4|tg_2|^2} +1}\right)<1\;,\ee
    then
    \begin{equation}\label{power}
    E_{k}(t)\le C(t_{0}, k,\alpha)\vert t \vert^{2\alpha}\;,\qquad   k\ge 1\;,
    \end{equation}
    where \bel{p1}\alpha=1-\beta
    \;.\ee
    2. Similarly if
\bel{poscond}  \hat \beta:= \sup_{\Omega(a,b,t_0)} \max(1-|t|f_1,|t|f_1) <1
\ee then \eq{power} holds with $\alpha = 1-\hat\beta$ if $\mu_1 <
1/2$, and $\alpha $ -- any number strictly smaller than $1-\mu_1$
if $\mu_1 \ge 1/2$.
    \end{Proposition}

\begin{Rem} Recall that, at fixed $a$ and $b$, the constants
$\mu_i$ and $\lambda_i$ depend upon $t_0$. In several situations
of interest $\mu_2$ and $\lambda_2$ will tend to zero as $t_0$
tends to zero, in which case the essential restriction in
\eq{p2new} is that $\mu_1$ be smaller than one sufficiently close
to the singular boundary $t=0$.
\end{Rem}
\begin{Rem} \label{Rp3} We note that \eq{p2new} will hold under
the slightly stronger but simpler condition
\bel{p3}\sup_{\Omega(a,b,t_0)}\left(|tf_1|+\frac  {|t f_2|}2 +
\frac {|t g_2|} {\sqrt{2}}\right)<1\;.\ee Proposition~\ref{sup} or
Remark~\ref{Rnew} can be used to replace \eq{p3} by a condition on
initial data using the Cauchy-Schwarz inequality  \beaa \ftone +
\frac {\fttwo}2 + \frac {\gttwo}{\sqrt{2}} &\le& \sqrt{1+ \frac 14
+ \frac 12} \sqrt{\ftone^2 + \fttwo^2 +\gttwo^2}\\
&\le& \frac{\sqrt{7}}{2} \sqrt{|tX_t|^2+|tX_\theta|^2}\\
&\le& {\sqrt{7}} \sqrt{|tX_t|^2+|tX_\theta|^2}\Big|_{t=t_0}< 1\;.
\eeaa
\end{Rem}

\begin{proof}
   Differentiating $E_{k}(t)$ in $t$, using the field equations
   and integrating by parts one has
 \begin{align}
    \frac{dE_{k}(t)}{dt}  = &
- t^{2k+2} \sum_{i=1,2}¥\left((\pa_{\theta}^k
(f_{i}-g_{i}))^{2}(t,b-t)
    + (\pa_{\theta}^k(f_{i}+ g_{i}))^{2}(t,a+t)\right)
   \nonumber \\&-(2k+2)\int_{a+t}^{b-t}\vert t \vert ^{2k+1}
     \sum_{i=1,2}¥((\pa_{\theta}^k f_{i})^{2}
    + (\pa_{\theta}^k g_{i})^{2}) \nonumber\\
    & + 2\vert t \vert^{2k+2} \int_{a+t}^{b-t}\Big(
    \pa_{\theta}^k f_{1}\cdot \pa_{\theta}^k (f_{2}^{2}-g_{2}^{2}
    -\frac{f_{1}}{t})+  \pa_{\theta}^k g_{2}\cdot \pa_{\theta}^k
    (f_{1}g_{2}-g_{1}f_{2})  \nonumber\\
    & \phantom{xxxxxxxxxxxxxx}+ \pa_{\theta}^k f_{2}\cdot\pa_{\theta}^k
    (-f_{1}f_{2}+g_{1}g_{2}-\frac{f_{2}}{t})\Big)\, d\theta \nonumber\\
     \le &-2k\int_{a+t}^{b-t}\vert t \vert ^{2k+1}\sum_{i=1,2}(\pa_{\theta}^k f_{i})^{2}
    - (2k+2)\int_{a+t}^{b-t}\vert t \vert ^{2k+1}\sum_{i=1,2}(\pa_{\theta}^k g_{i})^{2}
    \label{leading}\\
    & + 2\vert t \vert^{2k+2} \int_{a+t}^{b-t}\Big(
    f_{2}\pa_{\theta}^k f_{2}\pa_{\theta}^k f_{1}-
     g_{2}\pa_{\theta}^k f_{1}\pa_{\theta}^k g_{2}-
      f_{2}\pa_{\theta}^k g_{2}\pa_{\theta}^k g_{1}\label{middle}\\
& \phantom{xxxxxxxxxxxxxx}+  g_{2}\pa_{\theta}^k
g_{1}\pa_{\theta}^kf_{2}
        -f_{1}(\pa_{\theta}^kf_{2})^{2}+f_{1}(\pa_{\theta}^k
        g_{2})^{2}\Big)
    \label{term} \\
    & + \vert t \vert^{2k+2} \int_{a+t}^{b-t}
    \pa_{\theta}^k u\cdot \sum_{i+j=k \atop i,j>0}C(i,j,k)\pa_{\theta}^iu\cdot
    \pa_{\theta}^ju\;. \label{last}
    \end{align}
    In \eq{middle} and \eq{term} we have collected all those terms
    which contain undifferentiated functions $f_i$ or $g_i$.
     In (\ref{last}) we denote  $(f_{i}, g_{i})$ by $u$  and the $C(i,j,k)$'s
     are the coefficients  of $k-$th binomial expansions; we will ignore those coefficients
     and replace them by an overall constant from now on --- this is sufficient for estimation purposes.
     Note that when
    $k=1$, then (\ref{last}) does not appear. 
Mixed terms of the form
$\pa_{\theta}^k f_1\pa_{\theta}^k f_2$ and
   $\pa_{\theta}^k g_1\pa_{\theta}^k g_2$
   are estimated in the obvious way using    $2ab \le {a^{2}}+ b^2$. To take advantage of the different factors
   in front of the integrals appearing in \eq{leading} the mixed terms $\pa_{\theta}^k f_i\pa_{\theta}^k g_j$
   are estimated using $2ab \le \frac{a^{2}}\sigma+
   \sigma
   b^2$.
 Absorbing all the terms from \eq{middle} and \eq{term} into those appearing in (\ref{leading}) we
 obtain
   \begin{align*}
   \frac{dE_{k}(t)}{dt} \le& -\vert t \vert^{2k+1}
  \int_{a+t}^{b-t}(2k-2\ftone  -\fttwo  -\frac{2\gttwo  }{\sigma})\sum_{i=1,2} (\pa_{\theta}^k  f_{i})^2 \,d\theta \\
  &   -\vert t \vert^{2k+1}
  \int_{a+t}^{b-t}(2k+2-2\ftone  -\fttwo  -2\sigma \gttwo  )\sum_{i=1,2}(\pa_{\theta}^k g_{i})^2 \,d\theta \\
  &+ \vert t \vert^{2k+2} C(k)\int_{a+t}^{b-t}
    \pa_{\theta}^k u\cdot \sum_{i+j=k \atop i,j>0}\pa_{\theta}^iu\cdot
    \pa_{\theta}^ju\;.
    \end{align*}
For  $k=1$ this reads
  \begin{align*}
    \frac{dE_{1}(t)}{dt} \le& -\vert t \vert^{3}
  \int_{a+t}^{b-t}(2-2\ftone  -\fttwo  -\frac{2\gttwo  }{\sigma})\sum_{i=1,2} (\pa_{\theta}^k  f_{i})^2 \,d\theta \\
  &   -\vert t \vert^{3}
  \int_{a+t}^{b-t}(4-2\ftone  -\fttwo  -2\sigma \gttwo  )\sum_{i=1,2}(\pa_{\theta}^k g_{i})^2
  \,d\theta\;.
   \end{align*}
   It should be clear from what follows that the choice of
   $\sigma=\sigma(t,\theta)$ which is optimal for our purposes
   is that of equal factors in front of the sums,
   namely $\sigma = (\sqrt{1+4\gttwo^2}+1)/(2\gttwo)$.
    Choosing this
   value of $\sigma$ leads to
 $$ \frac{dE_{1}(t)}{dt} \le  -\frac{2\alpha}{|t|} E_{1}\;, $$
 with $\alpha$ as in \eq{p1}. This shows that $d((-t)^{-2\alpha}E_1)/dt \le 0$, and by
 integration one obtains
   $$E_{1}(t)\le \left \vert \frac{t}{t_{0}} \right\vert^{2\alpha }E_{1}(t_{0}).$$
   (This inequality holds whatever the sign of $\alpha$, but for $\alpha\le0$
   it does not carry any new information.)

 To cover the case \eq{poscond}
we shall need a Lemma:

\begin{Lemma}
\label{Lposcond} Let $$F(t)= \int_{a+t}^{b-t}(f_2^2+g_2^2)\;.$$
Under \eq{poscond} we have
$$F(t)\le F(t_0)\left|\frac{t}{t_0}\right|^{-2\hat \beta}\;.$$
\end{Lemma}

\proof We have
\bean \frac {d F(t)}{dt}  & = & - (f_2-g_2)^2 (t,b-t) - (f_2+ g_2)^2
(t,a+t)\nonumber\\ && + \frac 2 {|t|} \inti (1-|t|f_1)f_2^2 +
|t|f_1g_2^2 \nonumber\\ &\le&\displaystyle \frac {2\hat \beta
F(t)} { |t|}\;,\eeal{Cpl4n} and the result follows by integration
as before.
 \myqed

Returning to the estimation of $dE_1/dt$, assume that \eq{poscond}
holds and consider any of the terms of the form $f_2
\,\partial_\theta u\,
\partial_\theta  u$ or $g_2
\,\partial_\theta u\,
\partial_\theta  u$ in \eq{middle} and \eq{term}; they are
estimated as \beaa \inti \Big| |t|^4 \,f_2\, \partial_\theta u\,
\partial_\theta  u\Big| &\le& \inti \frac{\epsilon}{4C} |t|^7 |\partial _\theta u| ^4 + \frac {16C} \epsilon |t| f_2^2
\\ &\le & \inti \frac \epsilon 4  |t|^3 |\partial _\theta u| ^2 + \frac {16C} \epsilon |t| f_2^2
\\ &\le &  \frac{\epsilon E_1(t)} {4|t|} + C'(\epsilon)
|t|^{1-2\hat \beta}\;, \eeaa where in the last line we have used
Lemma~\ref{Lposcond}; similarly for $g_2$. It follows that
 $$ \frac{dE_{1}(t)}{dt} \le  -\frac{2(1-\mu_1-\epsilon)}{|t|} E_{1} + C'(\epsilon)
|t|^{1-2\hat \beta} \;, $$ Multiplying by   $\vert \frac{ t_{0}}
  { t }\vert^{2(1-\mu_1-\epsilon)}$ and integrating  in $t$, one
  obtains
  $$E_1(t)\le C(\alpha) \left|\frac t {t_0}\right|^{2\alpha}\;,$$
with $\alpha = \mu_1$ if $\mu_1 < 1/2$, and $\alpha $ -- any
number strictly smaller than $1-\mu_1$ if $\mu_1 \ge 1/2$.

 The cases $k\ge 2$ are established by
induction: Suppose, thus, that (\ref{power}) holds for $k=m-1$; we
have already shown that it holds for $k=1$. Then the terms from
line \eq{last} in  $ \frac{dE_{m}(t)}{dt} $ are estimated as
    $$\vert t \vert ^{2m+2}\int_{a+t}^{b-t} \pa_{\theta}^m u\cdot
    \sum_{i+j=m \atop i,j>0}\pa_{\theta}^iu\cdot
    \pa_{\theta}^ju\le C(t_{0})\vert t \vert^{m+1}\int_{a+t}^{b-t}
    \sum_{i+j=m \atop 0< i,j<m}\vert \pa_{\theta}^iu\cdot\pa_{\theta}^ju \vert ,$$
   letting $C(t_{0})=\sup_{(t,\theta)}\vert t \vert^{m+1} \vert \pa_{\theta}^m u\vert$, which
   is finite by \eq{intd0}. Using the induction
   hypothesis for $i,j<m$ on $\pa_{\theta}^iu\cdot \pa_{\theta}^j u$, we get
      \begin{align*}
   \vert t \vert^{m+1} \int_{a+t}^{b-t}\sum_{i+j=m \atop 0<i,j<m}
   \vert\pa_{\theta}^iu\vert\cdot \vert\pa_{\theta}^ju\vert & =  \vert t \vert^{-1}
   \int_{a+t}^{b-t}\sum_{i+j=m \atop 0<i,j<m}\vert t \vert^{i+1}\vert\pa^i_{\theta}u\vert
   \cdot \vert t \vert^{j+1}\vert\pa^j_{\theta}u\vert \nonumber \\
   & \le \vert t \vert^{-1} \int_{a+t}^{b-t}\sum_{i+j=m \atop 0<i,j<m}
   \vert t \vert^{2i+2}(\pa^i_{\theta}u)^2+
   \vert t \vert^{2j+2}(\pa^j_{\theta}u)^2\nonumber\\
   & \le C(m)\vert t \vert^{2\alpha-1}\;.
    \end{align*}
It follows that  \begin{align}\label{argu}
   \frac{dE_{m}(t)}{dt}
  & \le -\frac {2(m + \alpha-1)}{\vert t \vert}E_{m}(t)
  +C(t_{0},m)\vert t \vert^{2\alpha-1}\;.
  \end{align}
  Multiplying by $\vert\frac{ t_{0}}
  { t }\vert^{2(m+\alpha-1)}$ on both sides and integrating over $(t_{0}, t)$ in $t$,
we obtain
\begin{align*}
    E_{m}(t) & \le  \vert\frac{t}{t_{0}}\vert^{2(m+\alpha-1)} E_{k}(t_{0})+
    C(t_{0},m)\vert t \vert^{2(m+\alpha-1)}
    \int_{t_{0}}^t \vert s \vert^{1-2m} \, ds\\
    & \le  C(t_{0},m)\vert t \vert^{2\alpha }\;,
    \end{align*}
    as claimed. 
  \end{proof}

\medskip

\begin{Corollary}
\label{Cpl} Under the conditions of Proposition~\ref{Ppl}, there
exists a constant $C$ such that \bel{Cpl1} \inti t^2 |X_\theta|^2
d\theta = \inti t^2 \left((g_1)^2 + (g_2)^2\right) d\theta \le C
|t|^{2\alpha}  \;. \ee If
\bel{Cpl20} \sigma_1:= \inf_{\Omega(a,b,t_0)}|t|
f_1 
> 0\;,\ee
then we also have \bel{Cpl2} \inti t^2 f_2^2 d\theta \le C
|t|^{2\alpha'} \;, \ee
 with $\alpha' = \alpha$ if $\alpha <
\sigma_1$, or $\alpha'$ -- any number smaller than $\sigma_1$
otherwise.
\end{Corollary}

\proof We calculate
\bean \frac d {dt} \inti t^2 g_2^2 & = & -t^2 g_2^2 (t,b-t) - t^2 g_2^2
(t,a+t)\nonumber\\ && + 2t \inti g_2^2 + 2\inti t^2 g_2
\Big(\underbrace{\partial_\theta f_2}_I + \underbrace{f_1
g_2}_{II} - \underbrace{g_1 f_2}_{III}\Big) \nonumber\\ &\le & -
\frac{2}{|t|} \inti t^2 g_2^2 + \underbrace{\epsilon \inti |t|
g^2_2 + \frac 1 {\epsilon} \inti |t|^3 (\partial_\theta f_2)^2}_I
\nonumber\\& & + \underbrace{\frac{2\mu_1}{|t|}\inti t^2
g_2^2}_{II}
+\underbrace{\frac{\mu_2}{|t|}\inti t^2\left(\frac {g_1^2}{\epsilon'}+ \epsilon' g_2^2\right)}_{III}
\nonumber\\
&\le & - \frac{2-2\mu_1-\epsilon' \mu_2-\epsilon}{|t|} \inti t^2
g_2^2 + C(\epsilon,\epsilon')|t|^{2\alpha -1}\;.\nonumber\\ &&
\eeal{Cpl4} Choosing $\epsilon$, $\epsilon'$ appropriately and
arguing as in the paragraph following \eq{argu} one obtains the
bound for $\inti g_2^2$. A similar, but simpler, calculation with
$g_2$ replaced by $g_1$ proves \eq{Cpl2}. In order to establish
\eq{Cpl2} we note that
\beaa \frac d {dt} \inti t^2 f_2^2 & = & -t^2 f_2^2 (t,b-t) - t^2 f_2^2
(t,a+t)\\ && +  2\inti t^2 f_2 \Big(\underbrace{\partial_\theta
g_2}_I - \underbrace{f_1f_2}_{II} + \underbrace{g_1
g_2}_{III}\Big)
\\ &\le &  \underbrace{\epsilon \inti |t| f^2_2 + \frac 1 {\epsilon} \inti |t|^3
(\partial_\theta g_2)^2}_I
\\& & - \underbrace{\frac{2\sigma_1}{|t|}\inti t^2 f_2^2}_{II}
+\underbrace{\frac{\mu_2}{|t|}\inti t^2\left( {g_1^2}+  g_2^2\right)}_{III}\\
&\le & - \frac{2\sigma_1-\epsilon}{|t|} \inti t^2 f_2^2 +
C(\epsilon)|t|^{2\alpha -1}\;, \eeaa and we conclude as before.
 \myqed
\section{Pointwise power law}
\label{Spwd}

We start with the following observation:

\begin{Lemma}
\label{Lpwd} \begin{enumerate} \item Suppose that there exists
$\alpha>0$ such that \bel{Lpwd0a} \inti t^2 |X_\theta|^2d\theta
\le C|t|^{2\alpha}\;.\ee Then for all multi-indices $\nu$ we
have\beal{Lpwd1} &\displaystyle \inti \left(|t|^{|\nu|+1}|D^\nu
X_\theta|\right)^2d\theta \le C(\nu)|t|^{2\alpha}\;, &\eea  \item
If $|tX_\theta|\le C |t|^\alp$ then we also have \beal{Lpwd2}
&\supO |t|^{|\nu|+1}|D^\nu X_\theta| \le C'(\nu)
|t|^{\alp}\;.&\eea \item If the constant $\alpha$ in \eq{Lpwd0a}
satisfies $\alpha>1/2$ then \eq{Lpwd2} holds with
$\alp=\alpha-1/2$.
\end{enumerate}
\end{Lemma}

\proof \Eq{Lpwd1} follows by a straightforward adaptation of the
proof of \cite[Proposition~3.3.1]{SCC}. \Eq{Lpwd2} is obtained
from \cite[Remark, p.~73]{SCC}.  To establish point (iii) it
remains to show that \bel{pw4} \supO |t||X_\theta| \le C
|t|^{\alpha-1/2}\;.\ee That last inequality is obtained by
applying the interpolation inequality
$$ \|\partial_\theta u \|_{L^\infty( [a+t,b-t])} \le C\|\partial_\theta^2 u \|_{L^2(
[a+t,b-t])}^{3/4} \| u \|_{L^2( [a+t,b-t])}^{1/4}$$
%
(see,
\emph{e.g.,\/} \cite[p.~94]{Aubin}; the condition there that $u$
vanishes on the boundary is not necessary) to the functions
$tg_i$, $i=1,2$.
 \myqed

Thus, power-law blow-up in Sobolev spaces implies a pointwise one
if the decay rate is larger than $1/2$. The unpleasant feature of
the above argument is the loss of $1/2$ decay rate in point (iii)
of Lemma~\ref{Lpwd}. This can be avoided by working directly with
$L^\infty$ norms, as follows: Consider two fields $f,g$ satisfying
the symmetric hyperbolic set of equations \bean
\partial_t f -\partial_\theta g &=& S_f\;,
\\
\partial_t g -\partial_\theta f &=& S_g\;,
\eeal{sh1} set \bel{sh2} T_{tt}[f,g]=T_{\theta\theta}[f,g]: =
\frac 12 (f^2+g^2)\;,\quad T_{t\theta}[f,g]=T_{\theta t}[f,g]: =
fg\;,\ee and define \bel{sh3} j_\mu [f,g]= \partial_\nu(
T_\mu{}^\nu[f,g])\;.\ee
 Writing $T_{\mu\nu}$ for $T_{\mu\nu}[f,g]$, \emph{etc},\/ we have the identity~\cite[Equation~(3.2.5)]{SCC}
 \bean T_{tt}(t_1,\theta_1) &=& -\frac 12 \int_{t_0}^{t_1} \Big(
 (j_t+j_\theta)(t,\theta_1+t_1-t)+
 (j_t-j_\theta)(t,\theta_1+t_1+t)\Big) dt \\
 && + \frac 12\Big(
 (T_{tt}+T_{t\theta})(t,\theta_1+t_1-t_0)+
 (T_{tt}+T_{t\theta})(t,\theta_1+t_1+t_0)\Big)\;.
 \nonumber \\ &&\eeal{sh4}
For \eq{sh1} the $j$--terms appearing in \eq{sh4} read \beal{sh5}
-\frac 12 (j_t+j_\theta) &=& \frac 12 (f+g)(S_f+S_g)\;,
\\ -\frac 12 (j_t-j_\theta) &=& \frac 12
(f-g)(S_f-S_g)\;.\eeal{sh6} If we let
$(f,g)=(\partial_\theta^kf_1,\partial_\theta^kg_1)$ we obtain
\bean -\frac 12 (j_t+j_\theta) &=& \frac 12 (\partial_\theta^k
f_1+\partial_\theta^kg_1)\left( \frac{\partial_\theta^k
 f_1}{|t|}+\partial_\theta^k \Big[(f_2-g_2)(f_2+g_2)\Big]\right)\;,
\\&&\label{sh7} \\  -\frac 12 (j_t-j_\theta) &=& \frac 12 (\partial_\theta^k
f_1-\partial_\theta^kg_1)\left( \frac{\partial_\theta^k
 f_1}{|t|}+\partial_\theta^k
 \Big[(f_2-g_2)(f_2+g_2)\Big]\right)\;.\nonumber
\\&&\eeal{sh8}
Similarly for $(f,g)=(\partial_\theta^kf_2,\partial_\theta^kg_2)$
one has \bean -\frac 12 (j_t+j_\theta) &=& \frac 12
(\partial_\theta^k f_2+\partial_\theta^kg_2)\left(
\frac{\partial_\theta^k
 f_2}{|t|}-\partial_\theta^k \Big[(f_1+g_1)(f_2-g_2)\Big]\right)\;,
\\&&\label{sh9} \\  -\frac 12 (j_t-j_\theta) &=& \frac 12 (\partial_\theta^k
f_2-\partial_\theta^kg_2)\left( \frac{\partial_\theta^k
 f_2}{|t|}-\partial_\theta^k
 \Big[(f_1-g_1)(f_2+g_2)\Big]\right)\;.\nonumber
\\&&\eeal{sh10}
We define \bel{sh11} E_k(t,\psi) = \sup_{t_0\le s\le t\,,\
\psi+s-t\le \theta \le \psi -s+t}
(T_{tt}[\partial_\theta^kf_1,\partial_\theta^kg_1]
+T_{tt}[\partial_\theta^kf_2,\partial_\theta^kg_2])(s,\theta)\;.\ee
It is useful to note that \bel{sh12} T_{tt}[f,g] = \frac 14 \Big(
(f-g)^2+(f+g)^2\Big)\;.\ee

 Define
\bel{mupmdef}\mu_\pm = \sup_{\Omega(a,b,t_0)}|t(f_1\pm g_1)|\;.
\ee
 We have the following pointwise equivalent of Proposition~\ref{Ppl}:
\begin{Proposition}\label{Pplp}
 Suppose  that
\bel{plp1} \sup_{\Omega(a,b,t_0)}\left(\mu_- +|tf_1+tg_1| + \sqrt{
(\mu_- +|tf_1+tg_1|)^2+ 4|tf_2+tg_2|^2}\right) <4\;,\end{equation}
\bel{plp2}\sup_{\Omega(a,b,t_0)} \left(\mu_+ +|tf_1-tg_1| + \sqrt{
(\mu_+ +|tf_1-tg_1|)^2+ 4|tf_2-tg_2|^2}\right)<4\;.\end{equation}
Then there exist constants  $C,\alp>0$ so that we have the
inequality
\bel{plbu} |tX_\theta|\le C|t|^\alp\;.\end{equation}
\end{Proposition}
\proof We start by deriving an integral inequality for $E_1$ using
\eq{sh4}. Let $S$ denote the sum of \eq{sh7} and \eq{sh9} with
$k=1$:
\bean
S &=& \frac 12 (\partial_\theta f_1+\partial_\theta g_1)
\left(\frac{\partial_\theta
 f_1+\partial_\theta g_1+\partial_\theta
f_1-\partial_\theta g_1}{2|t|}\right. \\\nn
 && \left.\phantom{\frac12}+\partial_\theta
 \Big[(f_2-g_2)(f_2+g_2)\Big]\right)\\\nn
  && +\frac 12(\partial_\theta f_2+\partial_\theta g_2)
   \left( \frac{\partial_\theta f_2+\partial_\theta
 g_2+\partial_\theta f_2-\partial_\theta g_2}{2|t|}\right. \\\nn
  && \left.\phantom{\frac 12} -\partial_\theta
\Big[(f_2-g_2)(f_1+g_1)\Big]\right)\\\nn &=&
\frac{1}{4|t|}\Big[(\partial_\theta f_1+\partial_\theta g_1)^2
+(\partial_\theta f_2+\partial_\theta g_2)^2+ (\partial_\theta
f_1+\partial_\theta g_1) (\partial_\theta f_1-\partial_\theta
g_1)\\\nn &&\phantom{xx} +(\partial_\theta f_2+\partial_\theta
g_2) (\partial_\theta f_2-\partial_\theta g_2)\Big]
 +\frac 12 (f_2+g_2)(\partial_\theta f_1+\partial_\theta g_1)
(\partial_\theta f_2-\partial_\theta g_2)\\ && \phantom{xx}- \frac
12 (f_1+g_1)(\partial_\theta f_2+\partial_\theta g_2)
(\partial_\theta f_2-\partial_\theta g_2). \eeal{my1} A formula
for the sum of  \eq{sh8} and \eq{sh10} can be obtained
 by changing $g_a$ to $-g_a$ in \eq{my1}. We apply Young's inequality $ab\le \frac c2 a^2+\frac 1
{2c} b^2$ to estimate the mixed terms
$(\pa_{\theta}f+\pa_{\theta}g)(\pa_{\theta}f-\pa_{\theta}g)$,
using a $c$ which might possibly depend upon $t$ and $\theta$.
With a little work one finds  that the integrand in the first line
of \eq{sh4} can be estimated by the sup, over $\theta$ in the
relevant range, of the quantity \bean
 &&\!\!\!\!\!\!\!\!\!\!\!\!\! \frac 1 {2|t|}\Bigg[
 \left(1+\frac {c_1 |tf_2+tg_2|}{2 }\right)(\partial_\theta
f_1+\partial_\theta g_1)^2 \\\nn && \ +\left(1+ \frac {\mu_-}2+
\frac 12 \left ( |tf_1+tg_1 | +\frac
{|tf_2+tg_2|}{c_1}\right)\right) (\partial_\theta
f_2-\partial_\theta g_2)^2 \\ \nn &&\ +\left(1+\frac {c_2
|tf_2-tg_2|}{2 }\right)(\partial_\theta f_1-\partial_\theta g_1)^2
\\ && \ +\left(1+ \frac {\mu_+}2+ \frac 12 \left ( |tf_1- tg_1
|+\frac {|tf_2-tg_2|}{c_2}\right)\right) (\partial_\theta
f_2+\partial_\theta g_2)^2  \Bigg]\;. \eeal{my2} If all the
factors in front of the derivative squared terms are strictly
smaller than $2$, say smaller than or equal to $2-2\alp$, then
\eq{my2} is smaller than or equal to $(4-4\alp)E_1/|t|$. From
\eq{sh4} applied to $E_1$ and from Gronwall's Lemma ({\em cf.,
e.g.,\/}~\cite[Lemma~3.2.3]{SCC}) one obtains an integral
inequality for $E_1$, which translates into the inequality
\bel{ge0}|t^2\partial_\theta f_a| +|t^2\partial_\theta g_a| \le C
|t|^\alp\;.\ee  It should be clear
that an optimal estimate will be obtained in \eq{my2} 
if $c_1=c_1(t,\theta)$ is chosen so that
$$c_1 |f_2+g_2|= \mu_- +|f_1+g_1| + \frac
{|f_2+g_2|}{c_1} \;,$$ similarly for $c_2$. 
This gives
$$c_1 |f_2+g_2|\le   \frac 12 \left(\mu_- +|f_1+g_1| + \sqrt{
(\mu_- +|f_1+g_1|)^2+ 4|f_2+g_2|^2}\right)\;,$$
$$c_2 |f_2-g_2|\le \frac 12
\left(\mu_+ +|f_1-g_1| + \sqrt{ (\mu_+ +|f_1-g_1|)^2+
4|f_2-g_2|^2}\right)\;,$$ with equalities if
$(f_2+g_2)(f_2-g_2)\ne 0$, and the  condition mentioned above will
be satisfied if \eq{plp1}-\eq{plp2} holds.

 \Eq{shs} gives
\bel{ge1} \partial_t g_1 = O(|t|^{\alp-2})\;,\ee and by
integration along rays $\theta=\psi+\lambda t$, $\lambda\in
[-1,1]$ one obtains on $C^0_{t_0}(\psi)$
$$|t g_1| \le C
|t|^\alp\;.$$ Returning to \eq{shs} one finds \bel{ge2}
\partial_t g_2 -f_1g_2 = O(|t|^{\alp-2})\;,\ee which can
be integrated to give \bean 0>t_2>t_1\ge t_0 \qquad g_2(t_2,\psi)
&=&
e^{\int_{t_1}^{t_2} f_1(s,\psi)ds}g_2(t_1,\psi) \\
&& + \int_{t_1}^{t_2} e^{\int_{u}^{t_2}
f_1(s,\psi)ds}O(|u|^{\alp-2})du\;.\nonumber \\ \eeal{ge4} Since
$|f_1|\le (1-\gamma)\left|\ln|t|\right| +C$ one easily concludes
that
$$ g_2(t,\psi)=O(|t|^{\alp-1})\;.$$ Integrating in $\theta$ from
$\psi$ to $\theta$, at fixed $t$, the desired estimate for $g_2$
on $C^0_{t_0}(\psi)$ is obtained using \eq{ge0}. Since $\psi$ was
arbitrary in $[a,b]$, and since all the constants were uniform in
$\psi$, the result follows.
 \myqed

The main result of this section is the following:

  \begin{Thm} \label{Tcontv} Suppose that either
  \bel{p2newbis}
 \sup_{\Omega(a,b,t_0)}\left(|tf_1|+\frac  {|tf_2|}2 + \frac
{2|tg_2|^2}{\sqrt{1+4|tg_2|^2} +1}\right)<\frac 12\;,\ee or that
\eq{plp1}-\eq{plp2} hold. Then the velocity
  $\vw$ has a continuous representative $v<1$ on $[a,b]$, and the weak convergence in
  Proposition~\ref{Pwl} can be replaced by convergence in sup norm to $v$. In other
  words,
   $\lim_{t\to 0} \vert tX_{t} \vert (t, \theta)$ exists and is
   a continuous function on $[a,b]$.
   Moreover the solution satisfies a power law blow-up, \Eq{plbu}.
   \end{Thm}

\medskip

   \begin{Remark}\label{RTcontv}
   Further information concerning the properties of the solutions considered in Theorem~\ref{Tcontv}
   can be found in Theorem~\ref{Tcvs} below.
   \end{Remark}

\medskip

\proof Under \eq{p2newbis} Corollary~\ref{Cpl} applies, so that
the conclusion of Lemma~\ref{Lpwd} point (iii) holds. It now
follows from Proposition~\ref{Pplp} that both under \eq{p2newbis},
or under \eq{plp1}-\eq{plp2}, point (ii) of Lemma~\ref{Lpwd}
applies,  and thus there exists $\ep<1$ such that
  $\vert t D_{\theta}X_{\theta}
 \vert +\vert t D_{\theta}X_{t}
 \vert \le C\vert t \vert  ^{-\epsilon}$. Theorem~\ref{Tcontv} is now a straightforward consequence of the
following: \myqed

\begin{Lemma}\label{Lev} Suppose that there exist positive constants $C$ and
$\alp$ such that on $\Omega(a,b,t_0)$ we have \bel{secder}
|t^2D_\theta X_\theta| + |t^2D_\theta X_t| \le C |t|^\alp\;.\ee
Then there exists a continuous function $v$ such that on
$\Omega(a,b,t_0)$ it holds
 \bel{secder2}\Big| |tX_t|^2(t,\theta)- v^2(\theta)\Big|\le C'|t|^\alp\;. \ee
 Further, for every $\psi$ such that \eq{secder} holds on
$C^0_{t_0}(\psi)$ we also have \eq{secder2} on $C^0_{t_0}(\psi)$.
\end{Lemma}
   \begin{proof}
   We have
 \begin{equation}\label{velo}
 \pa_{t} \vert tX_{t} \vert^2= 2 h\left( tX_{t}, tD_{\theta}X_{\theta}
 \right)
 \;.
 \end{equation}
 By integration we obtain
 \bel{interm}  \vert tX_{t} \vert^2(t_1,\theta)=  \vert tX_{t} \vert^2(t_{2},\theta)+
  \int _{t_{2}}^{t_1} \pa_{t} \vert tX_{t} \vert^2(s,\theta)ds = \int _{t_{2}}^{t_1} O(s^{\alp -1}) ds\;.\ee
  It easily follows from this equation that  $$v(\theta):= \lim_{t\to 0} \vert tX_{t} \vert (t,
  \theta)$$ exists. By passing to the limit $t_1\to 0$ in \eq{interm} we
  obtain on $[t_0,0)\times
[a,b]$
  \bel{vdec} | v^2(\theta) - \vert tX_{t} \vert^2(t,\theta) | \le  \int _{t}^{0}
C  s^{\alp -1} ds\le C' |t|^\alp\;.\ee This shows that
$|tX_t|(t,\cdot)$ converges uniformly to $v$, and establishes
continuity thereof. The same argument applies on $C^0_{t_0}(\psi)$
by integrating along rays $\theta=\psi+\lambda t$, $\lambda\in
[-1,1]$; one easily checks, using \eq{secder}, that the number
$v(\psi)$ is $\lambda$--independent. This, together with the
result already established on $[t_0,0)\times [a,b]$, establishes
\eq{secder2} on $\Omega(a,b,t_0)$.
 \end{proof}

In terms of the $(f,g)$ variables the Gowdy-to-Ernst
transformation \eq{GtE} takes a remarkably simple form: \bean \hat
f_1 = -f_1-\frac 1t \;, && \quad \hat g_1 = -g_1\;,
\\\hat f_2 = -g_2 \;, && \quad \hat g_2 =
-f_2\;.\eeal{GtE2} This transformation immediately leads to the
following counterpart of Theorem~\ref{Tcontv}:

  \begin{Thm} \label{Tcontv2} Suppose that either
  \bel{T2newbis}
 \sup_{\Omega(a,b,t_0)}\left(|tf_1+1|+\frac  {|tg_2|}2 + \frac
{2|tf_2|^2}{\sqrt{1+4|tf_2|^2} +1}\right)<\frac 12\;,\ee
or
\bel{plp1bis} \sup_{\Omega(a,b,t_0)}\left(\hat \mu_- +|tf_1+1+tg_1| + \sqrt{
(\hat \mu_- +|tf_1+1+tg_1|)^2+ 4|tf_2+tg_2|^2}\right)
<4\;,\end{equation}
\bel{plp2bis}\sup_{\Omega(a,b,t_0)} \left(\hat \mu_++|tf_1+1-tg_1| + \sqrt{
(\mu_+ +|tf_1+1-tg_1|)^2+
4|tf_2-tg_2|^2}\right)<4\;,\end{equation} where
$$\hat \mu_\pm = \sup_{\Omega(a,b,t_0)}|tf_1+1\pm tg_1|\;.
$$
 Then there exists a continuous function $\hat v$ such that
\bel{p2a}|tf_1|^2+|tg_2|^2 \to_{t\to 0} \hat v^2\;,\ee uniformly
in $\theta$. Further there exist constants $C,\epsilon>0$ such
that we have \bel{p2b}|tf_2|^2+|tg_1|^2 \le C
|t|^{2\epsilon}\;.\ee
   \end{Thm}

In Section~\ref{Sopde} below we will see how to iterate the
Gowdy-to-Ernst transformation to obtain information on more
general solutions.

For further purposes it is convenient to restate the conclusions
of Lemma~\ref{Lpwd} as higher derivative estimates for the $g_a$'s
and $f_a$'s:

\begin{Lem}\label{L7.2}
    \begin{enumerate}
    \item There exists a constant $C$ such that \bel{hide} \vert t \vert^{k+1}\vert
\pa^k_{\theta}f_a\vert +\vert t \vert^{k+1}\vert
\pa^k_{\theta}g_a\vert + \vert t \vert^{k+1}\vert \pa^k_{t}f_a
\vert + \vert t \vert^{k+1}\vert \pa^k_{t}g_a \vert \le  C\;.
\end{equation}
    \item\label{first}
    Under the conditions of point (i) of Lemma~\ref{Lpwd} we have
    \beal{iLpwd1} &\displaystyle
    \inti\left (\vert t \vert^{l+k+1}\pa^k_{t}\pa^l_{\theta}u \right)^2 \,d\theta
    \le C(l,k)\vert t \vert^{2\al}\,, &
    \end{eqnarray} for every
    $l,\,k\ge 0$ when $u=g_{1},\, g_{2}$ or $l\ge 1,\, k\ge 0$ when
    $u=f_{1},f_{2}$.
    \item
    If $|tX_\theta|\le C |t|^\alp$ then we also have
    \beal{iLpwd2} & |t|^{l+k+1}\vert\pa^k_{t}\pa^l_{\theta}u\vert\le
    C(l,k)|t|^{\alp}\;,&\end{eqnarray}
     with  the ranges of the $(l,k)$'s 
     as in $(ii)$.
     \end{enumerate}
\end{Lem}
\begin{proof}
For the purposes of the proof let  $u$  stand for any of $g_a,
\,f_{2}$.

(i): From (\ref{ghi}) we have
$$(\pa^k_{\theta}g_{1})^2+ (\pa^k_{\theta}g_{2})^2
\le \vert D^k_{\theta}X_{\theta} \vert^2+ F^2_{k}+ G^2_{k} $$ then
 $\vert t \vert^{k+1}\vert \pa^k_{\theta}g\vert  \le  C$
 inductively  using part (ii) of Proposition~\ref{sup}.
 For $\vert t \vert^{k+1}\vert \pa^k_{t}f_a \vert$ we compute  by induction
\bean D^k_{t}X_{\theta} & = & (\pa^k_{t}g_{1}+
\tilde{F_{k}}(\pa^{k-1}u, \ldots, u))e_{1}+
(\pa^k_{t}g_{2}+ \tilde{G_{k}}(\pa^{k-1}u, \ldots, u))e_{2},\\
&&\label{dtxth}\\ D^k_{t}X_{t} & = & (\pa^k_{t}f_{1}+
\hat{F_{k}}(\pa^{k-1}f, \ldots, f))e_{1}+ (\pa^k_{t}f_{2}+
\hat{G_{k}}(\pa^{k-1}f, \ldots, f))e_{2}\;,\nonumber\\ &&
\eeal{dtxt} where $\tilde{F_{k}}$, $\tilde{G_{k}}$, $\hat{F_{k}}$,
$\hat{G_{k}}$ have the same properties as  described in the proof
of Proposition~\ref{Pf3d} for $F_{k}$ and $G_{k}$. The remaining
inequalities follow as before.

(ii): The proof is identical to that of Proposition~\ref{Pf3d}.

 (iii): We consider the $g_a$'s first.
Letting $e$ stand for the basis vectors $e_a$'s of \eq{bvect}, it
is sufficient to show \beal{ederi} &\displaystyle \supO \vert t
\vert^{i+j}\vert D^i_{t}D^j_{\theta}e \vert \le C\, ,
\end{eqnarray}
then the assertion follows from \beaa \pa^k_{t}\pa^l_{\theta}g & =
& \pa^k_{t}\pa^l_{\theta}\langle X_{\theta}, e\rangle =
\pa^k_{t}\sum_{j=0}^l C(l,j)\langle D^{j}_{\theta}X_{\theta},
D^{l-j}_{\theta} e\rangle \\
& = & \sum_{i=0}^k \sum_{j=1}^l C(l,k,i,j)\langle D^i_{t}
D^{j}_{\theta}X_{\theta}, D^{k-i}_{t} D^{l-j}_{\theta}e \rangle
\eeaa together with Lemma \ref{Lpwd}.   Now
\bel{det2}D_{\theta}e_{1}=g_{2}e_{2}\;,\quad
D_{\theta}e_{2}=-g_{2}e_{1}\;,\quad D_{t}e_{1}=f_{2}e_{2}\;,\quad
D_{t}e_{2}=-f_{2}e_{1}\;,\ee which implies that the $D^k_{\theta}
e_{a}  $, $D^k_ {t} e_{a}$ are of the form
\begin{align*}
D_{t}^ke_{1} & = \bar{F}_{k-1} e_{1}+ \left( \pa^{k-1}_{t}f_2+
\bar{G}_{k-1}\right) e_{2}\;,\; D_{t}^ke_{2}  =  \left(
-\pa^{k-1}_{\theta}f_2 +\bar{F}_{k-1}'\right) e_{1}+
\bar{G}_{k-1}' e_{2}\;,\;\\
D_{\theta}^ke_{1}& =\mathring F_{k-1} e_{1}+
\left(\pa^{k-1}_{\theta}g_2+ \mathring G_{k-1}\right) e_{2}\;,\;\;
D_{\theta}^ke_{2} = \left( -\pa^{k-1}_{\theta}g_2+\mathring
F_{k-1}'\right) e_{1}+ \mathring G_{k-1}' e_{2}\;,
\end{align*}
with the $\bar F_{k-1}$'s, $\bar
G_{k-1}$'s, \emph{etc.}, of a similar structure as in \eq{ghi}. By
(\ref{hide}) and by the above expression we see that $\vert
t\vert^k\vert D^k_{\theta} e\vert$,\, $\vert t\vert^k\vert D^k_{t}
e\vert$ are bounded. For mixed derivatives of $g$, we write
\beaa D^k_{t}D^l_{\theta}X_{\theta} & = & D^k_{t}\left(
(\pa^l_{\theta}g_{1}+F_{l} )
e_{1}+ (\pa^l_{\theta}g_{2}+G_{l})e_{2} \right) \\
& = &  \sum_{i=0}^k
\pa^i_{t}(\pa^l_{\theta}g_{1}+F_{l})D^{k-i}_{t}e_{1}+
\pa^i_{t}(\pa^l_{\theta}g_{2}+G_{l})D^{k-i}_{t}e_{2}\\
& = & (\pa^k_{t}\pa^l_{\theta}g_{1}+\pa^k_{t}F_{l})e_{1}+
(\pa^k_{t}\pa^l_{\theta}g_{2}+\pa^k_{t}G_{l})e_{2} \\
& + & \sum_{i=0}^{k-1}
(\pa^i_{t}\pa^l_{\theta}g_{1}+\pa^i_{t}F_{l}) D^{k-i}_{t}e_{1}+
(\pa^i_{t}\pa^l_{\theta}g_{2}+\pa^i_{t}G_{l})D^{k-i}_{t}e_{2}\;;
\eeaa above, and in what follows, we ignore constants arising from
binomial expansions. From above expression we get $\vert t
\vert^{k+l+1}\vert \pa^k_{t}\pa^l_{\theta} g\vert \le C$
inductively using (\ref{hide}) and boundedness of $ \vert t
\vert^i \vert D^i_{t}e\vert$. Finally (\ref{ederi}) follows from
writing
\begin{equation}\label{etth}\begin{split}
D^k_{t}D^l_{\theta}e & =  D^k_{t}\left(
(\sigma\pa^{l-1}_{\theta}g_2+\mathring F_{l-1} )
e_{1}+ (\hat\sigma \pa^{l-1}_{\theta}g+\mathring G_{l-1})e_{2})\right) \\
& =  \sum_{i=0}^k (\sigma\pa^i_{t}\pa^{l-1}_{\theta}g _2+
\pa^i_{t}\mathring F_{l-1}) D^{k-i}_{t}e_{1}+
(\hat\sigma\pa^i_{t}\pa^{l-1}_{\theta}g + \pa^i \mathring
G_{l-1})D^{k-i}_{t}e_{2}
\end{split}\end{equation} and then using boundedness of $\vert t \vert^{i+j+1}
\vert \pa^i_{t}\pa^j_{\theta}g\vert$,\, $ \vert t \vert^i\vert
D^i_{t}e\vert$.

Let us turn now to the $f_{2}$'s. According to (\ref{fhi}) we
write
\begin{equation*}
 \pa^l_{\theta}f_{2}= \sum_{j=0}^l\langle D^j_{\theta}X_{t},
 D^{l-j}_{\theta}e_{2}\rangle,
 \end{equation*} so that
 \beaa
 \pa^k_{t}\pa^l_{\theta}f_{2} & = & \sum_{i=0}^k \sum_{j=0}^l
 \langle D^i_{t}D^j_{\theta}X_{t}, D^{k-i}_{t} D^{l-j}_{\theta}e_{2}\rangle \,,\\
 & = & \sum_{i=0}^k  \langle D^i_{t}X_{t}, D^{k-i}_{t} D^l_{\theta}e_{2}
 \rangle  +\sum_{i=0}^k \sum_{j=1}^l  \langle D^i_{t}D^j_{\theta}X_{t},
  D^{k-i}_{t} D^{l-j}_{\theta}e_{2}\rangle
\eeaa when $l\ge 1$,\, $k\ge 0$. Multiplying by $\vert t
\vert^{k+l+1}$ we have
\begin{equation}
\vert t \vert^{k+l+1}\vert\pa^k_{t}\pa^l_{\theta}f_{2} \vert \le
\sum_{i=0}^k \vert t \vert^{k+l-i}\vert
D^{k-i}_{t}D^l_{\theta}e_{2} \vert+\sum_{i=0}^k \sum_{j=1}^l\vert
t \vert^{i+j+1}\vert D^i_{t}D^j_{\theta}X_{t} \vert \end{equation}
from Proposition~\ref{sup} together with (\ref{ederi}). Using the
expression (\ref{etth}) we get
\begin{align*}
\vert t \vert^{k+l-i}\vert D^{k-i}_{t}D^l_{\theta}e_{2} \vert &
\le
 \sum_{m=0}^{k-i}\vert t \vert^{k+l-i}
\left( \vert \pa^m_{t}\pa^{l-1}_{\theta} g\vert + \vert \pa^m_{t}
\mathring F_{l-1}¥\vert + \vert \pa^m_{t} \mathring G_{l-1}\vert
\right)
\vert D^{k-i-m}_{t}e\vert  \\
& \le  C \sum_{m=0}^{k-i}\vert t \vert^{l+m} \left( \vert
\pa^m_{t}\pa^{l-1}_{\theta} g\vert + \vert \pa^m_t \mathring
F_{l-1}¥\vert + \vert \pa^m_{t}\mathring  G_{l-1}\vert \right)
\end{align*} from the boundedness of $\vert t \vert^k \vert D^k_{t}e\vert$.
The claim on $f_{2}$  follows now from Lemma \ref{Lpwd} together
with the assertions on $g$. Finally, the result for $f_1$ follows
 from what has been proved so far together with the equation
$$\partial_t (tf_1) = t(\partial_\theta g_1 + f_2^2-g_2^2)\;.$$
\end{proof}

 \section{Stability of the $(\frac 23, \frac 23, -\frac 13)$ and  $(1,0,0)$ Kasner metrics}
 The following result
extends the \emph{singularity stability theorem} for the $(\frac
23, \frac 23, -\frac 13)$ Kasner metrics established
in~\cite{SCC}, by raising\footnote{A similar result has been
recently established by Ringstr\"om~\cite{RingstroemOberwolfach}.}
the stability threshold there by a
factor $6^{3/2}/2$ 

\begin{Theorem}
\label{Tstab} Suppose that \bel{stcond21}\sup_{\theta\in[a-
t_0,b+t_0]} t_0^2\left( |X_t|^2+|X_\theta|^2\right)(t_0,\theta) <
\frac 1 {2}
\;.\ee Then the solution is of power-law type. Further, the
curvature scalar $R_{\alpha
\beta \gamma \delta} R^{\alpha
\beta \gamma \delta}$ blows up on every causal curve with endpoint
on $\{0\}\times [a,b]\times S^1 \times S^1$. In particular the
associated Gowdy space-time is inextendible across the boundary
$\{0\}\times [a,b]\times S^1 \times S^1$.
\end{Theorem}


\proof We wish to apply Proposition~\ref{Pplp}. Let $\mu_\pm$ be
defined by \eq{mupmdef}, using the Cauchy-Schwarz inequality and
some rather obvious estimations we have
\bean  \lefteqn{\mu_-
+|tf_1+tg_1| + \sqrt{ (\mu_- +|tf_1+tg_1|)^2+ 4|tf_2+tg_2|^2}} && \\
&& \le \sqrt{2}\sqrt{2(\mu_- +|tf_1+tg_1|)^2+ 4|tf_2+tg_2|^2} \nn
\\ &&\le 2\sqrt{2}\sqrt{\mu_-^2 +|tf_1+tg_1|^2+ |tf_2+tg_2|^2}\;.\eeal{plp1nd}
This shows that both conditions \eq{plp1}-\eq{plp2} will hold if
$$  \sup_{\Omega(a,b,t_0)}\left(|tf_1-tg_1|^2+ |tf_2-tg_2|^2\right)<
1\;,$$ and if
$$  \sup_{\Omega(a,b,t_0)}\left(|tf_1+tg_1|^2+ |tf_2+tg_2|^2\right)< 1\;.$$
It follows from \eq{impest} that the last inequalities will hold
if
$$  \sup_{t=t_0,\theta\in [a-|t_0|,b+|t_0|]}|t|\left(|f_1-g_1|^2+|f_1+g_1|^2+ |f_2-g_2|^2+ |f_2+g_2|^2\right)< 1\;,$$
which is equivalent to \eq{stcond21}. The result follows now from
the arguments in the proofs of Theorem~3.5.1 and Proposition~3.5.2
in \cite{SCC}.
\myqed

\begin{Remark}
 We note that the Sobolev decay estimates of Section~\ref{SPlb} lead to a similar somewhat weaker
 statement, with $1/2$ replaced
 by $\sqrt{3/{19}}$ in \eq{stcond21}. This can be seen as follows: \Eq{p2newbis} will
hold under the slightly stronger but simpler condition
\bel{p3onehalf}\sup_{\Omega(a,b,t_0)}\left(|tf_1|+\frac  {|t
f_2|}2 + \frac {|t g_2|} {\sqrt{3}}\right)<\frac 12\;.\ee The
Cauchy-Schwarz inequality,
$$ \ftone + \frac {\fttwo}2 + \frac {\gttwo}{\sqrt{3}} \le
\sqrt{1+ \frac 14 + \frac 13} \sqrt{\ftone^2 + \fttwo^2
+\gttwo^2}\;,
$$  together with point
(i) of Proposition~\ref{sup} show that Theorem~\ref{Tcontv}
applies, and one concludes as before.
\end{Remark}

The  solution $$f_1=1/|t| \;,\quad f_2=g_1=g_2=0\;,$$ of \eq{shs}
corresponds to the flat Kasner metric. Theorem~\ref{Tcontv2}
similarly implies complete control of the behavior of the solution
for all data in a neighborhood of those for the flat Kasner
metric. In this case the geometric interpretation is more
complicated, because of occurrence of horizons. A further
discussion of the latter can be found
in~\cite{ChLake}.

\section{Behavior of power-law solutions at $v=0$ and $v=1$}
\label{sec:sv0} We consider solutions on $\Omega(a,b,t_0)$ such
that \bel{sv1} |tX_\theta|\le C|t|^\alp\;,\ee for some $\alp>0$.
It follows from \cite[Remark~7.3]{SCC} and Lemma~\ref{Lev} that
the velocity function $v$ exists. The aim of this section is to
study the behavior of such solutions at points, or intervals, on
which $v$ vanishes:

\begin{Theorem}
\label{Tsvel} Suppose that \eq{sv1} holds with some $\alp>0$ and
consider any point $\psi\in[a,b]$ such that $v(\psi)=0$. Then:
\begin{enumerate}
\item The restriction $\bar x:= x|_{C^0_{t_0}(\psi)}$ of $x$ to
$C^0_{t_0}(\psi)$ can be extended to an \avtdpqi\ map from $\R^2$
to $\mcH_2$. \item If\, $\clim t\partial^j_\theta P_t = 0$ for all
$j\in\N$, then for  $i,k\in\N$ we have
 \bel{sh46} \clim \partial_t^{2i+1}\partial _\theta^k P = \clim \partial_t^{2i+1}\partial _\theta^k Q =0\;.\ee
\item Further, if $v$ vanishes on an interval $[\theta_l,\theta_r]$, then the
restriction $\tilde x:= x|_{\Omega(\theta_l,\theta_r,t_0)}$ of $x$
to $\Omega(\theta_l,\theta_r,t_0)$ can be extended by continuity
to a smooth map from $\R^2$ to $\mcH_2$, with \eq{sh46} holding
for all $\psi\in[\theta_l,\theta_r]$.\end{enumerate}

\end{Theorem}

Theorem~\ref{Tsvel} says, in essence, that $x$ behaves on
$C^0_{t_0}(\psi)$ as if it arose from an \avtdpqi\ map defined on
$\R^2$.  We emphasize, however, that the extensions mentioned
above might fail to coincide with the original map $x$ away from
$C^0_{t_0}(\psi)$ (for $\bar x$), or away from
$\Omega(\theta_l,\theta_r,t_0)$ (for $\tilde x$). Such a situation
could arise when $x$ has an infinite number of smaller and smaller
spikes accumulating at a point at which $v(\psi)=0$.

\Eq{sh46} says, roughly speaking, that $P$ and $Q$ can be thought
of as smooth functions of $\theta$ and $t^2$ (rather than $t$).
This result is relevant to the question of extendibility of the
associated metric across Cauchy horizons. If $\clim
t\partial^j_\theta P_t = 0$ for a finite number of $j$'s, there
will be a certain number of $i$'s for which \eq{sh46} will hold.

Using the Gowdy-to-Ernst transformation we obtain immediately the
following counterpart of Theorem~\ref{Tsvel} at $v=1$:

\begin{Theorem}
\label{Tsvel2}  Suppose that there exist constants  $C,\alp>0$
such that \bel{svel12} |tg_1|+ |tf_2|\le C|t|^\alp\ee and consider
any point $\psi\in[a,b]$ such that $v_1(\psi)=1$. Then the
functions $(\bar P,\bar Q):= (P+\ln |t|, Q)|_{C^0_{t_0}(\psi)}$
can be extended  to an \avtdpqi\ map from $\R^2$ to $\mcH_2$. If\,
$\clim t\partial^j_\theta P_t = 0$ for all $j\in\N$, then for all
$i,k\in\N$ we have
 \bel{sh46a} \clim \partial_t^{2i+1}\partial _\theta^k (P +\ln |t|)= \clim \partial_t^{2i+1}\partial _\theta^k Q
 =\clim \partial _\theta^{k+1} Q =0\;.\ee
Further, if $v_1=1$  on an interval $[\theta_l,\theta_r]$, then
the restriction $(\tilde P,\tilde Q):= (P+\ln |t|,
Q)|_{\Omega(\theta_l,\theta_r,t_0)}$  can be extended  to a smooth
map from $\R^2$ to $\mcH_2$, with \eq{sh46a} holding for all
$\psi\in[\theta_l,\theta_r]$.
\end{Theorem}

\begin{Remark} \label{Rsvel12} The vanishing of the last term in \eq{sh46a} for all $k\ge 0$ is
somewhat surprising. We emphasise that the power-law condition
\eq{sv1} in Theorem~\ref{Tsvel} is justified for small initial
data by Theorem~\ref{Tcontv}, and that the condition~\eq{svel12}
is justified for initial data near the flat Kasner by
Theorem~\ref{Tcontv2}.
\end{Remark}

\noindent {\sc Proof of Theorem~\ref{Tsvel2}:} The Gowdy-to-Ernst
transformed map $\hat x$ satisfies the hypotheses of
Theorem~\ref{Tsvel}, and therefore \eq{sh46} holds for the
associated  functions $\hat P$ and $\hat Q$. The claim about $P$
in \eq{sh46a} is straightforward. From~\eq{GtE} we have
\bel{GtE1}
\partial _t
\partial _\theta^k Q = - |t|\partial _\theta^k\left(e^{2\hat P}
\partial _\theta \hat Q\right)\;, \quad  \partial _\theta^{k+1} Q =
- |t|\partial _\theta^k\left(e^{2\hat P}
\partial_t \hat Q\right)\;. \ee
Integrating in $t$ and using \eq{sh46} one obtains
\eq{sh46a}.\myqed

An iteration of isometries and Gowdy-to-Ernst transformations, as
in the proof of Theorem~\ref{Tstabcvd} below, allows one to
control the behavior of $x$ near points $(0,\psi)$ at which
$v(\psi)\in\Z$, the details are left to the reader.

\bigskip

\noindent {\sc Proof of Theorem~\ref{Tsvel}:} We start with a
lemma:

\begin{Lemma}\label{Lsvel}
Suppose that there exists $0<\alp$ such that \eq{sv1} holds. Then
there exists a function $v_1(\theta)$, with $|v_1|=v$, and
constants $C_a(\theta)$, $a=1,2$, such that on $C_{t_0}^0(\theta)$
we have
 \bel{sv18} |tf_2| \le
C_2(\theta)
\left\{%
\begin{array}{ll}
    |t|^{\alp}, & \hbox{$v(\theta)\le 0$;} \\
    |t|^{v(\theta)} +
|t|^{\alp}, & \hbox{$0<v_1(\theta)\ne \alp$;} \\
   \big(1+\big|\ln |t|\big|\big) |t|^{v(\theta)} , & \hbox{$v_1(\theta)=\alp$.} \\
\end{array}%
\right.
\ee \bel{sv20} \Big||t|f_1-v_1(\theta)\Big| \le C_1(\theta)
\left\{%
\begin{array}{ll}
    |t|^{\alp}, & \hbox{$v_1(\theta)\le0$;} \\
|t|^{2v_1(\theta)} + 
|t|^{\alp} , &\hbox{$0<v_1(\theta)\ne \alp$;} \\
    |t|^{\alp}, & \hbox{$v_1(\theta)=\alp$.} \\
\end{array}%
\right.
\ee 
The constants $C_a(\theta)$ are uniformly bounded on  compact
intervals on which $v_1$ is strictly positive, uniformly bounded
away from zero.
\end{Lemma}

\begin{Remark}
\label{Rsvel}The examples discussed in Section~\ref{Sspikes} show
that $v_1$ is not continuous in general. Further, the constants
$C_a$ are not uniformly bounded near  points at which $v_1$ has
discontinuities involving a change of sign. Similarly we do not
expect uniformity at points at which $v_1$ crosses $0$.
\end{Remark}

\proof As explained in the proof of Lemma~\ref{Lpwd}, \Eq{sv1}
implies\beal{sv2} & |t|^{|\nu|+1}|D^\nu X_\theta| \le C(\nu)
|t|^{\alp}\;.&\eea We can use Lemma~\ref{Lev} to conclude that
there exists a continuous function $v$ such that
  \bel{sv4} | v^2(\theta) - \vert tX_{t} \vert^2(t,\theta) | \le C |t|^\alp\;.\ee
\Eq{sv2} with $\nu=\theta$ and $\nu=t$ shows that the
$\theta$-derivatives conditions of Lemma~\ref{Lpcv} are satisfied,
and Theorem~\ref{Tpcv} implies that there exists a function $v_1$
with $|v_1|= v$ such that, for all $\psi\in [a,b]$, \bel{sv6}
\clim |t|f_1=v_1(\psi)\;,\ee together with \bel{sv8} \clim
|t|f_2=0\;.\ee \Eqsone{shs} and Lemma~\ref{L7.2} lead to
\beal{sv10} \partial_t (|t|f_1) &=& |t|f_2^2 +O (|t|^{\alp-1})\;,
\\\partial_t (|t|f_2) &=& -|t|f_1f_2 +O (|t|^{\alp-1})\;. \eeal{sv12}
Suppose, first, that $\theta$ is such that $v(\theta)=0$. It
follows from \eq{sv4}  that the right-hand sides of
\eq{sv10}-\eq{sv12} are $O (|t|^{\alp-1})$. The same is true for
$\partial_\theta (|t|f_a)$ by Lemma~\ref{L7.2}, and by integration
in $t$ along rays $\Gamma_\lambda(t)= (t,\theta + \lambda t)$,
$\lambda \in [-1,1]$, we find on $C^0_{t_0}(\theta)$ \bel{sv16}
v(\theta)=0 \ \Longrightarrow \ |t f_1| \le C|t|^\alp\;,\quad |t
f_2| \le C|t|^\alp\;.\ee
 In general, integrating \eq{sv10} shows first that $tf_2^2\in L^1([t_0,0)$,
 and then
\bel{sv15} |t|f_1(t,\theta) = v_1(\theta) - \int _t^0
|s|f_2^2(s,\theta)ds +O(|t|^{\alp})\;.\ee Similarly, \eq{sv12} can
be integrated to give \bean 0>t_2>t_1\ge t_0 \qquad
|t_2|f_2(t_2,\theta) &=&
e^{-\int_{t_1}^{t_2} f_1(s,\theta)ds}|t_1|f_2(t_1,\theta) \\
&& + \int_{t_1}^{t_2} e^{-\int_{u}^{t_2}
f_1(s,\theta)ds}O(|u|^{\alp-1})du\;.\nonumber \\ \eeal{sv14} If
$v_1(\theta)>0$ one finds first from \eq{sv14} that $|tf_2|\le
C|t|^\epsilon$, with $\epsilon$ equal to, say,  $ \min(
v_1(\theta)/2,\alp/4)$. Plugging this in \eq{sv10} one obtains
that $f_1-v_1=O(|t|^\epsilon)$. Returning to \eq{sv14} one is then
led to \eq{sv18}  at $(t,\theta)$ with $v_1(\theta)>0$. Inserting
\eq{sv18} into \eq{sv15} we arrive at \eq{sv20} at $(t,\theta)$,
again for $v_1(\theta)>0$.

 Suppose, finally, that $v_1(\theta)< 0$, then we rewrite
 \eq{sv14} as
\bean 0>t_2>t_1\ge t_0 \qquad |t_1|f_2(t_1,\theta) &=&
e^{\int_{t_1}^{t_2} f_1(s,\theta)ds}|t_2|f_2(t_2,\theta) \\
&& + \int_{t_1}^{t_2} e^{\int_{t_1}^{u}
f_1(s,\theta)ds}O(|u|^{\alp-1})du\;.\nonumber \\ \eeal{sv14a} One
readily checks that the integrand is in $L^1([t_0,0))$, and
passing to the limit $t_2\to 0$ one obtains \bean 0>t_1\ge t_0
\qquad |t_1|f_2(t_1,\theta) &=&\int_{t_1}^{0} e^{\int_{t_1}^{u}
f_1(s,\theta)ds}O(|u|^{\alp-1})du\;.\nonumber \\ \eeal{sv14b}
Arguing as before one obtains the result  at $(t,\theta)$.
Finally, the result for $(t,\theta_1)\in C_{t_0}^0(\theta)$ is
obtained from the one for $(t,\theta)$ by integrating
$\partial_\theta( tf_a)$ in $\theta$ from $(t,\theta_1)$ to
$(t,\theta)$, using point (iii) of Lemma~\ref{L7.2}. \myqed

 We return to the proof of Theorem~\ref{Tsvel}.
To proceed further we need better control of the $\theta$
derivatives. Let $E_k$ be defined by \eq{sh11} and consider a
point $\psi$ such that $v(\psi)=0$. From \eq{sv1}, \eq{sv16},
together with Equations~\eq{sh4} and \eq{sh7}-\eq{sh10} one
obtains \bel{sh14} E_1(t_1,\psi)\le E_1(t_0,\psi) +
\int_{t_0}^{t_1} \frac{2+O(|t|^\alp)}{|t|} E_1(t,\psi)\, dt\;;\ee
we have also made use of \eq{sh12}. Gronwall's Lemma ({\em cf.,
e.g.,\/}~\cite[Lemma~3.2.3]{SCC}) gives, for any $\epsilon >0$,
decreasing $|t_0|$ if necessary, \bel{sh16} E_1(t,\psi)\le
C\left|\frac {t_0}t\right|^{2+\epsilon}\;.\ee This allows us to
rewrite \eq{sh14} as \bel{sh18} E_1(t_1,\psi)\le E_1(t_0,\psi) +
\int_{t_0}^{t_1} \Big(\frac{2}{|t|} E_1(t,\psi)
+O(|t|^{\alp-2-\epsilon})\Big)\, dt\;,\ee which, together with
Gronwall's Lemma, implies \eq{sh16} with $\epsilon=0$. It follows
that on $C^0_{t_0}(\psi)$ we have \bel{sh20} |t\partial _\theta
f_a |+ |t
\partial _\theta g_a| \le C\;,\ee
and \eq{shs} gives \bel{sh22} |t
\partial _t g_1| \le C\;, \quad |t
\partial _t g_2| \le C(1+|t|^{2\alp-1})\;.\ee
Integrating in $t$ over rays $\theta=\psi+\lambda t$, $\lambda \in
[-1,1]$, one obtains \bel{sh24} g_1=O(\ln |t|)\;,\quad |
 g_2| \le C( \ln |t|+|t|^{2\alp-1})\;.\ee
 We have thus recovered \eq{sv1} with the exponent $\alp$ there replaced
 by $2\alp$ if $2\alp <1$, or by a number as close to one as
 desired otherwise. Iterating the argument leading from \eq{sh22}
 to \eq{sh24} a
 finite number of times if needed one thus has
\bel{sh28} |t f_a |+ |t
 g_a| \le C|t|^{1-\epsilon}\;,\ee
with $\epsilon$ as small as desired.

We continue by induction: suppose that for all $\epsilon>0$ and
for all $0\le i\le k-1$ there exists $ C_i(\epsilon)$ \bel{sh20n}
|t\partial^i _\theta f_a |+ |t
\partial _\theta^i g_a| \le C_i(\epsilon)|t|^{-\epsilon}\;.\ee
From \eq{sh4}, \eq{sh7}-\eq{sh10} one obtains \bean
E_k(t_1,\psi)&\le& E_k(t_0,\psi) + \int_{t_0}^{t_1}
\Big(\frac{2+\epsilon}{|t|} E_k(t,\psi)
+C|t|^{-2-2\epsilon}E_k^{1/2}(t,\psi)\Big)\, dt \\ &\le&
E_k(t_0,\psi) + \int_{t_0}^{t_1} \Big(\frac{2+2\epsilon}{|t|}
E_k(t,\psi) +\frac C {4\epsilon}|t|^{-2-2\epsilon}\Big)\,
dt\;,\eeal{sh30} leading to \eq{sh20n} with $i=k$ (and with
$\epsilon$ replaced by $2\epsilon$.).

The equation
\begin{equation}
    \pa_{t}¥\left(
    \begin{array}{c}
     \partial _\theta^k f_{1} \\    \partial _\theta^k f_{2} \\  \partial _\theta^k g_{1} \\   \partial _\theta^kg_{2} \\
     \end{array}
     \right)=
    \left(
     \begin{array}{cccc}
      0 & 0 & 1& 0 \\  0 & 0 & 0& 1 \\  1 & 0 & 0 & 0 \\
       0 & 1 & 0 & 0 \\
     \end{array} \right)\pa_{\theta}
     \left(
     \begin{array}{c}
      \partial _\theta^k f_{1} \\   \partial _\theta^kf_{2} \\
      \partial _\theta^k g_{1} \\   \partial _\theta^k g_{2} \\
    \end{array} \right) +
    \left(
    \begin{array}{c} \partial _\theta^k\Big(
    f_{2}^{2}-g_{2}^{2}-\frac {f_{1}}{t} \Big)\\
    \partial _\theta^k\Big(-f_{1}f_{2}+g_{1}g_{2}-\frac{f_{2}}{t}\Big)\\
    0\\
   \partial _\theta^k\Big( f_{1}g_{2}-g_{1}f_{2}\Big)\\
     \end{array}
     \right)\label{shsk}
     \end{equation}
     gives
\bel{sh32} |\partial _t\left(t\partial^i _\theta f_a\right) |+ |t
\partial _t
\partial _\theta^i g_a| \le C_i(\epsilon)|t|^{-\epsilon}\;.\ee
Integrating in $t$ over rays $\theta=\psi+\lambda t$, $\lambda \in
[-1,1]$, from \eq{sh20n} and \eq{sh32} one obtains \bel{sh34a}  |
\partial _\theta^i g_a| \le C_i(\epsilon)|t|^{-\epsilon}\;.\ee
Further, the same integration argument shows that there exist
constants $v_{a,i}(\psi)$ such that \bel{sh34b} ||t|\partial^i
_\theta f_a - v_{a,i}(\psi)|\le
C_i(\epsilon)|t|^{1-\epsilon}\;.\ee We note that $v_{a,0}(\psi)=0$
by \eq{sh28}. From \eq{shsk} one obtains now the equation \bean
\partial_t (\partial^i _\theta
|t|f_1) &=& |t|\partial^i _\theta (f_2^2) +O (|t|^{-\epsilon}) \\
&=& \left(\sum_{j=0}^i \begin{pmatrix}
  j \\ i
\end{pmatrix}v_{2,j}(\psi)v_{2,i-j}(\psi)\right) \frac 1 {|t|} + O
(|t|^{-\epsilon})\;, \eeal{sh34c} which is compatible with
boundedness of $\partial^i _\theta (|t|f_1)$ if and only if the
coefficient of $1/|t|$ vanishes. Suppose that we know that
$v_{2,j}$ vanishes for $j=1,\ldots,m-1$, then the condition of
vanishing of the offending term in \eq{sh34c} with $i=2m$ gives
$v_{2,m}=0$, and induction gives the vanishing of $v_{2,j}$ for
all $j$. It then follows from \eq{sh34b} that
\bel{sh36a} |
\partial _\theta^i f_2| \le C_i(\epsilon)|t|^{-\epsilon}\;.\ee
Integrating the equations for $\partial_t\partial _\theta^i g_a$
one arrives at \beal{sh36b}
\partial _\theta^i g_1 &=& -v_{1,i+1}(\psi)\ln |t| + g_{1,i}(\psi)+ O (|t|^{1-\epsilon})\;,
\\
 g_2 &=&  g_{2,0}(\psi)+ O (|t|^{1-\epsilon})\;,
\\
\partial^i _\theta g_2 &=& -v_{1,i}(\psi)g_{2,0}(\psi) \ln |t| + g_{2,1}(\psi)+ O (|t|^{1-\epsilon})\;,\eeal{sh36d}
for some constants $g_{a,i}(\psi)$. Similarly, as in \eq{sv14b},
\bean 0>t_1\ge t_0 \qquad |t_1|\partial _\theta^k f_2(t_1,\psi)
&=&\int_{t_1}^{0} e^{\int_{t_1}^{u}
f_1(s,\psi)ds}O(|u|^{1-\epsilon})du\nonumber
\\&=&O(|t_1|^{2-\epsilon})\;. \eeal{sh40}
 Integrating in $\theta$ from $\psi$ to $\theta$ and
using \eq{sh40} one obtains, 
on $C^0_{t_0}(\psi)$, 
\bel{sh42} \partial _\theta^k f_2(t,\psi) =
O(|t|^{1-\epsilon})\;.\ee Analogously to \eq{sv15} one has
\bel{sh38} |t|\partial _\theta^kf_1(t,\psi) - v_{1,i}(\psi) = \int
_t^0 O\left(\left|s\ln|s|\right|\right)ds
=O\left(\left|t^2\ln|t|\right|\right)\;.\ee An iterative
repetition of the arguments given gives a full asymptotic
expansion of the solution on $C^0_{t_0}(\psi)$. It should be clear
that the expansions for the derivatives obtained in this way
behave as though they arose from an \avtdpqi\ map, and the usual
extension arguments can be used to provide an \avtdpqi\ extension
of the restriction  $x|_{C^0_{t_0}(\psi)}$ of $x$.

Suppose finally that all the $v_{1,j}(\psi)$ vanish. We note that
it follows from the argument above that this will be the case when
$v=0$ on an interval containing $\psi$. It is then straightforward
to show that all the $
\partial _\theta^k f_a$'s and the $\partial _\theta^k g_a$'s can
be extended by continuity to the tip $(0,\psi)$ of
$C^0_{t_0}(\psi)$.  By an abuse of notation we will denote the
value at the tip of those extensions by $\partial^\alpha f_a
(0,\psi)$, {\em etc}.\/ \Eq{sh40} gives, for all $k$, \bel{sh44}
\partial _\theta^k f_2(0,\psi) =0\;.\ee
\Eq{sh38} implies \bel{sh45}
\partial _\theta^k f_1(0,\psi) =0\;.\ee  Arguing similarly one
obtains that any mixed $t$ and $\theta$ derivatives can also be
extended by continuity to the tip $(0,\psi)$ of $C^0_{t_0}(\psi)$.
For example, when $v_{1,j}(\psi)=0$ \eq{sh38} can be rewritten as
\bel{sh38n} |t|\partial _\theta^kf_1(t,\psi)  = \frac{\partial
_\theta^k
\left(g_2^2(0,\psi)\right)t^2}2+O\left(|t|^3\right)\;,\ee which
shows that $\partial _\theta^kf_1/t$ extends continuously to the
tip. It then follows from \eq{shsk} that $\partial_t \partial
_\theta^kf_1$ extends continuously to the tip. A similar analysis
shows that the same is true for $\partial_t \partial
_\theta^kf_2$. It then immediately follows from \eq{shsk} that
$\partial_t \partial _\theta^kg_a$ extends continuously to the
tip. Differentiating \eq{shsk} with respect to $t$ one can repeat
this analysis for all higher-order $t$-derivatives.

An identical extension property is  obviously true for $P$, $Q$,
and all their derivatives.

In order to prove \eq{sh46}, suppose that
 \beal{sh48a} \partial _\theta^k
f_a(t,\psi) &=&\sum_{j=0}^{i-1}\alpha_{a,j,k}t^{2j+1}+
O(|t|^{2i})\;,\\\partial _\theta^k g_a(t,\psi)
&=&\sum_{j=0}^{i}\beta_{a,j,k}t^{2j}+ O(|t|^{2i+1})\;;\eeal{sh48}
this clearly holds with $i=0$. Then \eq{shsk} gives \bea
\partial_t\left(t\partial _\theta^k f_a(t,\psi)\right)
&=&\sum_{j=0}^{i-1}\hat\alpha_{a,j,k}t^{2j+1}+
O(|t|^{2i+2})\;,\eeal{sh50} 
and by integration one recovers \eq{sh48a} with $i$ replaced by
$i+1$. Inserting this into \eq{shsk} one then finds
\bea\partial_t\left(\partial _\theta^k g_a(t,\psi)\right)
&=&\sum_{j=0}^{i}\hat\beta_{a,j,k}t^{2j+1}+
O(|t|^{2i+2})\;,\eeal{sh52} and integration in $t$ completes the
induction step. The usual extension results finish the proof on
$C^0_{t_0}(\psi)$. Clearly the same results remain valid if $v$
vanishes on an interval $I$, we simply note that all the estimates
so far were uniform in $\theta$ over $I$. \myqed

\section{Solutions with $v<1$ and with uniform power-law}
\begin{Theorem}
\label{Tcvs} Let $\mathring x$ be  a solution of the Gowdy
equations on $\Omega(a,b,t_0)$ such that
\bel{stc1b}\sup_{\Omega(a,b,t_0)}|t\mathring X_t| <1\;.\ee Suppose
moreover     that there exist positive constants $\epsilon$ and
$C_0$ such that on $\Omega(a,b,t_0)$ we have \bel{stc2a} |t
\mathring
    X_\theta|\le C_0 |t|^{\epsilon}\;.\ee Then: \begin{enumerate}
\item The solution belongs to the class $\mcU_1$ defined in
Section~\ref{Sgeroch}, with a velocity function $v=|v_1|$ and a
position function $\varphi_\infty$ which are smooth except perhaps
on the set
$$
\partial\{\theta: v(\theta)=0\}\;.$$
 \item There exists
$\eta>0$ such that for all initial data $( x(t_0,\cdot),
X_t(t_0,\cdot))$ satisfying$^{\mbox{\rm \scriptsize \ref{Fdist}}}$
$$\|(
x(t_0,\cdot)-\mathring{x}(t_0,\cdot),
X_t(t_0,\cdot)-\mathring{X}_t(t_0,\cdot))\|_{H^3\oplus H^2} < \eta
$$ the associated solution $x$ also satisfies \eq{stc1b}-\eq{stc2a}
(with perhaps  different values of $\epsilon$ and $C$). Further
the conclusions of Theorem~\ref{Tstab} apply.
\end{enumerate}
\end{Theorem}

\begin{Remark}
\label{Rcvs} The behavior of the solution near all  points
$(0,\psi)$ with $v(\psi)=0$, including $\{0\}\times
\left(\partial\{\theta: v(\theta)=0\}\right)$, is described
exhaustively in Theorem~\ref{Tsvel}. It is not clear whether or
not the function $\varphi_\infty$ of \eq{ravtd2} is continuous at
$\partial\{\theta: v(\theta)=0\}$.
\end{Remark}

\begin{Remark}
 Every \avtdpqo\  solution with $v$ strictly smaller than one,
 and for which the error terms
 in \eq{avtd1}-\eq{avtd2} and in their derivative counterparts
 (see \eq{avtd1d}) are uniform in $\theta$ satisfies
 \eq{stc2a}, as well as \eq{stc1b} for some $t_0$ close enough to $0$.
This  follows immediately from \eq{avtd1}-\eq{avtd2}. Further, all
initial data satisfying the hypotheses of Theorem~\ref{Tcontv2},
or of \cite[Theorems 1 and 2]{Ringstroem4}, satisfy the hypotheses
of Theorem~\ref{Tcvs}.
\end{Remark}

\proof  The arguments at the beginning of the proof of
Theorem~\ref{Tcontv2} show that $v$ exists on $[a,b]$ and is
continuous there. We continue with a lemma:

\begin{Lemma}
\label{Lrs} Suppose that $|tX_\theta|\le C|t|^\epsilon$ for some
$\epsilon>0$ and
\bel{stc3a}\inf_{\Omega(a,b,t_1)}|tX_t|\ge \gamma  >0\;, \qquad
\sup_{\Omega(a,b,t_1)}|tX_t| <1-\gamma\;,\ee for some $t_1$. Then
$\Omega(a,b,t_1)$ can be covered by a finite number $N$ of domains
of dependence on which $x$ is \avtdpqi\ in appropriate coordinates
on hyperbolic space. (This implies smoothness of $v$ and of
$\varphi_\infty$, as those properties are invariant under
isometries of the hyperbolic space.) The number $N$ depends only
upon $\epsilon$, $C_0$ and $\gamma$.
\end{Lemma}

\newcommand{\sumzt}{\sum_{k=0}^2}
\newcommand{\sumot}{\sum_{k=1}^2} We will actually prove a slightly stronger
statement:
\begin{Lemma}
\label{Lrs2} Under \eq{stc3a}, suppose  that there exists a
sequence $t_i$ such that the function \bel{stc2a1}
F(t_i):=\sup_{\theta\in [a+t_i,b-t_i] }\sumzt |t^{k+1}D^k_\theta
X_\theta|(t_i,\theta)+\sumot |t^{k+1}D^k_\theta
X_t|(t_i,\theta)\ee goes to zero as $t_i$ goes to zero. Then the
conclusion of Lemma~\ref{Lrs} holds, with the number $N$ there
depending upon $\gamma$ and the sequences $\{t_i\}$, $\{F(t_i)\}$.
\end{Lemma}

\proof We start by noting that under \eq{stc2a} we have $F(t_i)\le
C|t_i|^\epsilon$  by \eq{sv2}, so that Lemma~\ref{Lrs} does indeed
follow from Lemma~\ref{Lrs2} by setting $t_i = 2^{-i}t_0$.

We wish to apply a result of
Ringstr\"om~\cite[Theorem~9.1]{Ringstroem3}. Set \bel{gcond}\gamma
=\frac 18 \min (\inf_{\Omega(a,b,t_1)}|tX_t|,
1-\sup_{\Omega(a,b,t_1)}|tX_t|)\;.\ee  Let the constants
$\epsilon_k$, $k=0,1,2$ in~\cite[Theorem~9.1]{Ringstroem3} be
equal to $10$, decreasing $|t_1|$ if necessary we can suppose that
$t_1>-e^{-\tau_0}$, with $\tau_0$ given
by~\cite[Equation~(9.1)]{Ringstroem3}. Choose some $\psi\in
(a,b)$, and define $\hczpsio$ to be a triangle with the top vertex
at $(0,\psi)$ and  slopes  $\pm 1/2$: \bel{hcsd} \hczpsio:= \{t\ge
t_1\;,\ \psi - 2|t|\le \theta\le \psi+ 2|t|\}\;.\ee If $\psi = a$
we set \bel{hcsdb3} \hczpsio:= \{t\ge t_1\;,\ \psi - |t|\le
\theta\le \psi+ 2|t|\}\;,\ee while for $\psi =b$ we set
\bel{hcsdn} \hczpsio:= \{t\ge t_1\;,\ \psi - 2|t|\le \theta\le
\psi+ |t|\}\;.\ee For $0<\lambda\le 1$ let $\xpl$ be defined as
\bel{xpl} \hczpsio \ni (t,\theta) \to \xpl(t,\theta) := x(\lambda
t, \psi + \lambda \theta)\;.\ee We have \bel{Xplt}
\Xplt(t,\theta):=
\partial_t \xpl(t,\theta) = \lambda X_t (\lambda t, \psi +
\lambda\theta)\;, \ee similarly \bel{Xplth} \Xplth(t,\theta):=
\partial_\theta \xpl(t,\theta) = \lambda X_\theta (\lambda t, \psi
+ \lambda\theta)\;, \ee with analogous formulae for higher
derivatives. This, together with our hypothesis on the function
$F$, shows that for $\lambda= \lambda_i=t_i/t_1$ we have for
$\psi\in (a,b)$ \bel{xpl2} \sup_{\theta\in [2t_1,-2t_1] }\sumzt
|t^kD^k_\theta \Xpl_\theta|(t_1,\theta)+\sumot |t^kD^k_\theta
\Xpl_t|(t_1,\theta)\le F(\lambda_i
t_1=t_i)/|t_1|\to_{\lambda_i\to0} 0\;,\ee with similar results if
$\psi =a$ or $\psi =b$.  In particular at $t=t_1$ we have
\bel{xpl4} \sup_{\theta\in [-2|t_1|,2|t_1|]}\sumzt |D^k_\theta
\Xpl_\theta|(t_1,\theta)+\sumot |D^k_\theta
\Xpl_t|(t,_1\theta)\to_{\lambda\to0} 0\;.\ee  Now, it is easy to
show, using  \eq{intd2.0}, that
 \begin{equation}\label{sect}
 \ptth^{2}+e^{2P}¥\qtth^{2}¥\le \vert \dth\xt\vert^2 +C\vert \xt \vert^{2}¥
 \vert \xth\vert^2,
 \end{equation}
 \begin{equation}\label{secth}
 \pthth^{2}¥+e^{2P}¥\qthth^{2}¥\le \vert \dth\xth\vert^2 +C\vert \xth \vert^{4}.
 \end{equation}Similarly,
\begin{align}
P_{t\theta\theta}^{2}+ e^{2P}Q_{t\theta\theta}^{2} &\le \vert
\dth^{2}¥\xt\vert^{2}+C(\vert \xt\vert^2 (\vert \dth\xth\vert^{2}
+ \vert \xth\vert^{4}+ \vert \xth \vert^{2}) \nonumber \\ & \,\,+
\vert \xth\vert^{2}\vert \dth\xt \vert^{2}),\label{thirdt}\\
P_{\theta\theta\theta}^{2}+ e^{2P}Q_{\theta\theta\theta}^{2} &\le
\vert \dth^{2}¥\xth\vert^{2} +C(\vert \xth\vert^{2}\vert \dth\xt
\vert^{2}+ \vert \xth\vert^{6}+ \vert \xth\vert^{4}).
\label{thirdth}
\end{align}
\Eqs{xpl2}{thirdth} show that there exists $\lambda_j$ such that
Equations~(6.2) and (9.3) with $k=1,2$, of \cite{Ringstroem3} will
be satisfied by the initial data for $\xpl$ for all
$\lambda=\lambda_n\le\lambda_j$ with a multiplicative factor
$1/(2C_1)$ at the right-hand sides there, with a constant $C_1$ to
be made precise shortly. Decreasing $\lambda_j$ if necessary
Ringstr\"om's energy $\epsilon_2$ corresponding to the current
solution here will be smaller than $1/(2C_1)$. Again decreasing
$\lambda_j$ if necessary ~\cite[Equation~(9.4)]{Ringstroem3} will
hold with $2\gamma$ there replaced by $4\gamma$. It remains to
satisfy Ringstr\"om's equation (9.3) with $k=0$. Recall that the
group of isometries of the hyperbolic space $(\mcH_2,h)$ acts
transitively on the unit tangent bundle of $\mcH_2$. This implies
that there exists an isometry $\psi_\lambda$ of $(\mcH_2,h)$ such
that the map $ \psi_\lambda\circ \xpl$, when written in the local
coordinates $(P,Q)$, will satisfy
$$P(t_2,\psi)=Q(t_2,\psi)=\partial_t Q(t_2,\psi)=0\;, \quad \partial_t P(t_2,\psi)\ge 0.$$
Now,  Ringstr\"om's norms are not invariant under isometries.
However, the objects appearing in \eq{xpl2} and \eq{xpl4} are, and
those have been used to control Ringstr\"om's norms, so that the
inequalities which have already been fulfilled still hold (in any
case we could decrease $\lambda$ further to obtain the desired
inequalities). It then follows by integration in $\theta$, using
the fact that the $\theta$ derivatives of $P$ and $Q$ are already
known to be small, that there exists $d>0$ such that Ringstr\"om's
equation (9.3) with $k=0$ and with a multiplicative factor
$1/(2C_1)$ at the right-hand side there will hold on $[\psi-d+t_j,
\psi+d-t_j]$. Decreasing $d$ and $\lambda_j$ if necessary, we can
extend the initial data from that last interval to smooth periodic
initial data, without increasing all the relevant quantities by
more than a factor $C_1$. Applying Ringstr\"om's Theorem 9.1 the
result follows for the map obtained by the evolution of the
extended initial data. Uniqueness in domains of dependence
establishes the claim on each $\Omega(\psi-d,\psi+d,t_j)$.
Reverting to the original $t_1$, the result is obtained by
covering $\Omega(a,b,t_1)$ by a finite number of sets
$\Omega(\psi-d,\psi+d,t_1)$. \myqed

We note that an obvious modification of the argument just given
establishes the following version of Ringstr\"om's
result~\cite[Theorem~9.1]{Ringstroem3}:

\begin{Proposition}
\label{PRing} Let $a\le b$, $t_0<0$, $0<\gamma<1$, there exists
$\epsilon(\gamma)>0$ such that if
$$\gamma \le t_0P(t_0,\cdot) \le 1-\gamma\;,$$
$$ \sumzt |t^k_0D_\theta^k X_\theta|(t_0,\cdot) + \sumot |t^k_0D_\theta^k
X_t|(t_0,\cdot)<\epsilon\;,$$  {on $[a-|t_0|,b+|t_0|]$}, then the
solution is \avtdpqi\ in $\Omega(a,b,t_0)$, with velocity strictly
positive and strictly smaller than one.\myqed
\end{Proposition}

We return to the proof of Theorem~\ref{Tcvs}. The reader will note
that  the above proof has been worded to leave room for perturbing
the data at $t=t_1$, with $t_1$ as redefined in the paragraph
following \eq{gcond}, while still satisfying Ringstr\"om's
hypotheses; this is needed for the remainder of the argument.

We set now
$$\gamma =\frac 18 \min
(\frac 1{3}, 1-\sup_{\Omega(a,b,t_0)}|t\mathring X_t|)\;.$$ Since
$v$ is continuous, with $t\mathring X_t$ converging uniformly to
$v$, the interval $[a,b]$ can be covered by a finite number of
intervals $[a_i,b_i]$ on which either
$$\gamma \le\frac 18 \inf_{\Omega(a_i,b_i,t_1)}|t\mathring X_t|\;,$$
or on which
$$\sup_{\Omega(a_i,b_i,t_1)}|t\mathring X_t| <  \frac 1 {12}\;.$$

Let us call the latter intervals of type II, and the former of
type I. Each of the $[a_i,b_i]$'s of type I can further be chosen
to coincide with one of the intervals $[\psi-d, \psi+d]$ of the
proof of Lemma~\ref{Lrs} such that $\mathring x$ is \avtdpqi\ on
$\Omega(\psi-d,\psi+d,t_2),$ where $t_2$ is the time given in the
proof of Lemma~\ref{Lrs}. A sufficiently small pertubation of
$\mathring x$ at $t=t_2$ leads again to an \avtdpqi\ solution,
thus satisfying \eq{stc1b}-\eq{stc2a} (with $\mathring X$ there
replaced by $X$, with possibly different constants $C$ and
$\epsilon$). The usual continuous dependence of solutions upon
initial data on compact intervals of $t$ shows that the same will
remain true for sufficiently small perturbations of the initial
data at $t=t_0$.

 On each interval
of type II we have at $t=t_2$
$$|t\mathring X_\theta|\le C t^{\epsilon}<\frac 1{6}\;,$$ making $t_2$ smaller if necessary, then any sufficiently small
perturbation of the Cauchy data at $\{t_2\}\times
[a_i+t_2,b_i-t_2]$ leads to a solution $x$ such that
$$\sup_{\Omega(a_i,b_i,t_2)}|tX_t|^2+|tX_{\theta}|^2< 2(\frac
1{12}+\frac 16)=\frac 12,$$ where the factor $2$ comes from
Proposition~\ref{sup}. We can thus use Theorem~\ref{Tcontv} to
conclude $x$ will satisfy \eq{stc1b}-\eq{stc2a}.

%
\myqed

\section{Existence and smoothness of $v$ on an open dense
set}\label{Sopde}

 The proof of Theorem~\ref{Tdc2} will run in parallel with that of the following, more precise, statement:

\begin{Theorem}
\label{Tdc} Consider a solution $x$ defined on $\Omega(a,b,t_0)$
and set
$$n:= \lfloor \sup |tX_t|\rfloor\;.$$
There exists an open dense set $\Omega\subset [a,b]$ such that for
every $\psi\in\Omega$ there exist  a neighborhood
$\mcO_\psi\subset \Omega$ of $\psi$ and an element $G_\psi$ of the
Geroch group, with the order of $G_\psi$ less than or equal to
$n$, such that $G_\psi x$ has on $\mcO_\psi$ a smooth velocity
function $0\le v=|v_1|< 1$ and smooth position function
$Q_\infty$.
\end{Theorem}

\begin{figure}[!hbtp]
\begin{center}
\includegraphics[width=0.9\textwidth]{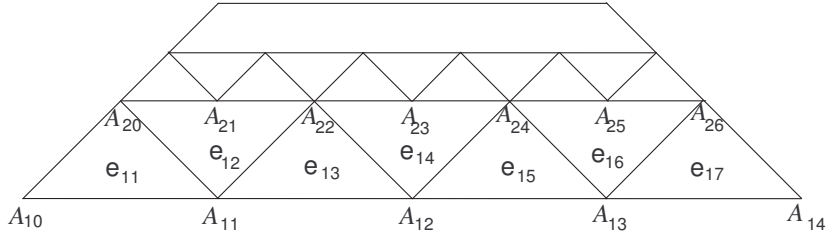}
\end{center}
\caption{\label{Fp}The points $A_{ij}$ and the triangles
$e_{ij}$.}
\end{figure}

\proof  We start with a covering argument: We choose arbitrarily
two points $a\le\theta_l<\theta_r\le b$, we set
$I_0=[\theta_l,\theta_r]$ and we decompose $\Omlrt$ into a union
of strips $\Omega_{i}$, $\Omlrt = \cup _{i=1}
^{\infty}\Omega_{i}$, where
$$\Omega_{i}= \{(t,x)\in \Omlrt \;\vert\;
-\frac{\vert I_{0}\vert }{2^{i}}\le t < -\frac {\vert I_{0}\vert}
{2^{i+1}}\}\;.$$ Note that $\vert I_{0} \vert /2$ is the height of
$\Omlrt $, and that the $\Omega_i$'s are pairwise disjoint. Let
$t_{i}=-\frac{\vert I_{0}\vert} {2^i}$ and let $B_{i}$ be the base
line of $\Omega_{i}$,
$$B_{i}=\{(t,x)\in \Omega_{i} \;\vert \;t=t_{i}\;, \ \theta_{l}-{\vert
t_i\vert} \le x\le \theta_{r}+{\vert t_{i}\vert} \}, \quad i \ge
1\;.
$$ We denote
by $A_{i0}$ the left end point of $B_{i}$, and set
$\theta_{l_{i}}=\theta_{l}-{\vert t_{i}\vert}$,
$\theta_{r_{i}}=\theta_{r}+{\vert t_{i}\vert} $. We consider a
partition $\CP_{i}$ of $B_{i}$ determined by the following
sequence of points (see Figure~\ref{Fp})
$$\CP_{i}=\{A_{i0},\ldots ,A_{ij},\ldots , A_{i,m_{i}}\},$$
where $m_{i}= 2^{i}+2$, $A_{ij}=(t_{i}, \theta_{l_{i}}+\frac
{|I_{0}|}{2^{i}}j) $. Note that the length of each sub-interval of
$\CP_{i}$ is $\frac {|I_{0}|}{2^{i}}$.

Next, we decompose $\Omega_{i}$ into a union of triangles with
$45$ degrees slopes as follows: Let $c_{ij}$  denote the triangle
with vertices at the points $ (A_{ij},A_{i,2j},A_{i,j+1})$ defined
above, where  $i\ge 1$, $j=0,1,\ldots,2^{i}+1$.  The set
$\Omega_{i}-\bigcup_{j=0}^{2^{i}+1}c_{ij}$ is then the union of
`upside down'  triangles with vertices
$(A_{i+1,2j-2},A_{ij},A_{i+1,2j})$, where $i\ge 1$, $j=1,\ldots,
2^{i}+1 $. We denote those last triangles by $d_{ij}$. It follows
that $$\Omega_{i}= (\cup_{j=0}^{2^{i}+1}
c_{ij})\bigcup(\cup_{j=1}^{2^{i}+1}d_{ij}) \;.$$ Finally we
relabel the $c_{ij}$'s and the $d_{ij}$'s  as $e_{ij}$:
\begin{align*}
e_{i,2j+1}&:= c_{ij}, \, j=0,\ldots,2^i+1\\
  e_{i,2j}&:= d_{ij}, \,
j=1,\ldots,2^i+1\\
\end{align*}

\begin{lem} \label{decom}
Let $\{\Omega_{j}\}_{j=0}^{\infty}$ be the decomposition of
$\Omlrt $ described above. Let $f$ be a nonnegative measurable
function on $\Omlrt $ with
$$\int_{\Omega} \vert t \vert f(t, \theta) \, d\theta dt <\infty\;.$$
Then for any $\ep>0 $ and $j_0\in \mathbb{N}$ there exists $j\ge
j_0$ such that $\Omega_{j}$ contains a set $\omega$ consisting of
eight \emph{consecutive} triangles $e_{ij}$ with
$$ \int_{\omega} f \, d\theta dt < \ep.$$
\end{lem}
\begin{proof}
Let $\int_{\Omega} \vert t \vert f(t, \theta) \, d\theta dt = A.$
Using the decomposition $\Omlrt =\cup_{i=1}^{\infty} \Omega_{i}$
we find
\begin{align*}
A &=
\sum_{i=1}^{\infty}\int_{\Omega_{i}} \vert t \vert f(t,\theta)\,d\theta dt\\
&\ge \vert I_{0}\vert \sum_{i=0}^{\infty}\int_{\Omega_{i}} \frac {
f(t,\theta)}{2^{i}} \, d\theta dt\;.
\end{align*}
Thus, since the last sum is finite, there exists $i(\epsilon)$
such that for all $i\ge i(\epsilon)$ we have
$$
\frac{1}{2^{i+1}}\int_{\Omega_{i}} f(t,\theta) \, d\theta dt \le
\ep$$ for any given $\ep$.
Let us show that there exist $j$ and  at least $N=8$ consecutive
triangles starting at $e_{ij}$ such that
$$ \int_{\cup_{k=0}^{N} e_{i,j+k}}f(t,\theta) \,
d\theta dt \le \ep.$$ In fact we prove this assertion with any
given $N \in \mathbb{N}$; we set
$$\begin{array}{c}
  e_{i1}\cup e_{i2}\ldots  \cup e_{iN}
=c_{1}\;,\\
  \vdots \\
  e_{i,j}\cup e_{i,j+1}\ldots \cup e_{i,j+N-1}
=c_{j}\;,\\
  \vdots  \\
  e_{i,m-N+1}\cup e_{i,m-N+2} \ldots \cup e_{im}
=c_{m-N+1}\;,\\
\end{array}$$
where $m=2^{i+1}+3$. Let $$\int_{c_{j}} f = C_{j}\;,\quad
\int_{e_{ij}} f= E_{j}\;,$$ then
\begin{align*}
 C_{1}+ \ldots
+C_{m-N+1}&=
 E_{1}+ 2E_{2}+\ldots + (N-1)E_{N-1}+ N(E_{N}+\ldots \\
 &\,\,+
E_{m-N+3})+
\ldots + 2E_{m-1}+ E_{m}\\
&\le N(E_{1}+ \ldots + E_{m})= N\int_{\Omega_{i}}
f(t,\theta)\,d\theta dt\le Nm\ep.
\end{align*}
Hence we obtain
$$\frac{C_{1}+\ldots + C_{m-N+1}}{m-N+1} \le \frac {Nm\ep}{m-N+1}\le
N\ep,$$ which implies that there exists $C_{j}$ such that $C_{j}
\le N\ep$.
\end{proof}

We apply Lemma~\ref{decom} to the function
$$f:=\sum_{k=0}^\ell
|t^{k}D_\theta^{k}X_\theta|^2+\sum_{k=1}^{\ell}|t^{k}D_\theta^{k}X_t|^2\;;$$
we are actually interested in $\ell=4$, but the argument applies
to any fixed $\ell\in\N$, $\ell\ge 2$. Point (iii) of
Proposition~\ref{integral} shows that $f$ satisfies the hypotheses
of Lemma~\ref{decom}, therefore there exists a sequence of domains
$\omega_i\subset \Omega_i$ such that $$\int_{\omega_i} f
\to_{i\to\infty}0\;.$$  The base $b_i$ of $\omega_i$ has length
$4|t_i|$, so that it can be written as
$$b_i=\{t_i\}\times [\psi_i-2|t_i|, \psi_i+2|t_i|]\;,$$ for some
$\psi_i\in [\theta_l,\theta_r]$. We scale $\omega_i$ to a union of
eight triangles with bottom edge lengths one, with the basis of
the scaled set lying at $t=-1$, and $(\psi_i, t_i)$ mapped to
$(0,-1)$; we call $\tilde \omega$ the resulting set, and we note
that the top of $\tilde \omega$ lies at $t=-1/2$. We have
\bel{fdec} \int _{\omega_i} f = \int_{\tilde
\omega}\sum_{k=0}^\ell |t^{k}D_\theta^{k}X_\theta^{(\psi_i,t_i)}
|^2+\sum_{k=1}^{\ell}|t^{k}D_\theta^{k}X_t^{(\psi_i,t_i)}|^2 \;,
\ee with $x^{(\psi_i,t_i)}$, {\emph etc.},\/  defined in \eq{xpl}.
There exists an isometry of the hyperbolic space into itself which
maps $x^{(\psi_i,t_i)}(0,-1)$ to the origin in the $(P,Q)$
coordinate system. We apply this isometry to $x^{(\psi_i,t_i)}$,
and still use the same name for the resulting map. 
Since $|t|\in [1/2,1]$ on $\tilde \omega$, isometry-invariance of
\eq{fdec} gives
$$
\| X_\theta^{(\psi_i,t_i)}\|_{H^\ell(\tilde\omega)}+ \|D_\theta
X_t^{(\psi_i,t_i)}\|_{H^{\ell-1}(\tilde\omega)} \to_{i\to\infty}
0\;,
$$ and the
Sobolev inequality implies
$$ \|X_\theta^{(\psi_i,t_i)}\|_{C^{\ell-2}(\tilde\omega)}+
\|D_\theta X_t^{(\psi_i,t_i)}\|_{C^{\ell-3}(\tilde\omega)}
\to_{i\to\infty} 0\;.
$$\newcommand{\tkone}{t^{k+1}}
Returning to the original $\omega_i$ one thus has
\bel{omdec}\sup_{\omega_i} \left(\sum_{k=0}^{\ell-2} |\tkone
D_\theta^{k}X_\theta|+\sum_{k=1}^{\ell-2}|\tkone
D_\theta^{k}X_t|\right) \to_{i\to\infty}0\;.\ee  Let $I_i=
[\psi_i-2|t_i|, \psi_i+2|t_i|]$  and set
$$F_\ell(I_i):=
\sup_{\theta\in[\psi_i-2|t_i|, \psi_i+2|t_i|]}
\left(\sum_{k=0}^\ell|\tkone
D_\theta^{k}X_\theta|+\sum_{k=1}^\ell|\tkone
D_\theta^{k}X_t|\right)(t_i,\theta)\;.$$ \Eq{omdec} shows  that
$F_2(I_i)$ approaches zero as $t_i$ tends to zero. We choose $i_0$
large enough so that for all $i\ge i_0$ we have
$$F_2(I_i)<      \frac 1 {100}\;.$$
Clearly the same bound will then also hold for $F_\ell(I_i)$ with
$0\le\ell\le 1$.
Since the sequence $|tX_t|(t_i,\psi_i)$ is bounded, passing to a
subsequence if necessary we can assume that there exists
$v_\infty$ such that $|tX_t|(t_i,\psi_i)\to v_\infty$.  Suppose,
first, that $v_\infty  < 1/200$, then $|tX_t|(t_i,\psi_i)< 1/100$
for $ i$ large enough. By integration in $\theta$ we have for
$\theta\in I_i$ \bel{vc} \left|
|tX_t|^2(t_i,\theta)-|tX_t|^2(t_i,\psi_i)\right|\le \pm
2\int_{\psi_i}^\theta t^2 |h(X_t,D_\theta X_t)| d\theta\le 4\sup
|tX_t| F_1(I_i)\;,\ee which goes to zero as $i$ goes to infinity
so that $|tX_t|(t_i,\theta)< 1/50$ for $\theta\in I_i$. Now
$|tf_1|\le |tX_t|< 1/50$, similarly for $tf_2$, while
$|tg_2|\le|tX_\theta|\le F_0(I_i)<1/100$, which proves that
Theorem~\ref{Tcontv} applies, and shows that for all $i\ge i_0$
(increasing $i_0$ if necessary) the solution satisfies a power law
decay in each of the $\Omega(\psi_i-2|t_i|,\psi_i+2|t_i|,t_i)$.
The conclusions of Theorem~\ref{Tcvs} apply to show that the
solution is \avtdpqi\ on each of the
$\Omega(\psi_i-2|t_i|,\psi_i+2|t_i|,t_i)$, except perhaps a) for
points in $\partial\{\theta: v(\theta)=0\}$ at which $Q_\infty$
may have discontinuities and/or $v_1$ might fail to be smooth
(even though it is continuous there), or b) for a set of
discontinuities of $v_1$ introduced by applying back isometries to
the isometry-transformed \avtdpqi\ solutions of Lemma~\ref{Lrs2}.
In any case Proposition~\ref{PGtE} guarantees existence of an open
dense subset of $[\theta_l,\theta_r]$ with \avtdpqi\ behavior
there.

Suppose, next, that $1/200\le v_\infty\le 1-1/200$. A scaling
argument as in the proof of Lemma~\ref{Lrs2} shows that, after
applying a suitable isometry, the solution is \avtdpqi\ in each of
the $\Omega(\psi_i-2|t_i|,\psi_i+2|t_i|,t_i)$'s, for $i$ large
enough, and hence again in an open subset of
$[\theta_l,\theta_r]$. Applying the isometry back to recover the
original solution one obtains an open subset of
$[\theta_l,\theta_r]$ with \avtdpqi\ behavior.

As the next possibility, consider the case in which $1- 1/200 <
v_\infty\le 1+1/200$. Applying an isometry $\phi_i$ we can assume
that the map $\phi_i\circ x$, still denoted by $x$, satisfies
$Q_t(t_i,\psi_i)=0$ with $P_t(t_i,\psi_i)$ -- positive, so that
$1-1/100 < |t|P_t(t_i,\psi_i) < 1+1/100$. Note that $F_\ell$ is
invariant under isometries, so that $F_2(I_i)$ remains unchanged.
Now, \bel{f2c}
\partial_\theta f_a= \partial_\theta \left(h(X_t,e_a)\right)=
h(D_\theta X_t,e_2)+h(X_t,D_\theta e_a)\;,\ee and by integration,
making use of \eq{det2}, one finds that
$$-1/100 < |t|f_2= |t|e^{P}Q_t(t_i,\theta) < 1/100$$ on $I_i$ for
$i$ large enough, similarly $1-1/50 < |t|f_1=|t|P_t(t_i,\theta) <
1+1/50$. Applying a Gowdy-to-Ernst transformation \eq{GtE}
(compare \eq{GtE2}) we obtain on $I_i$
$$-1/50 < |t|\hat P_t= 1-|t| P_t < 1/50\;, \quad  -1/100 < |t|\hat
g_1=-|t| g_1< 1/100\;,$$ as well as $$ -1/100 < |t|\hat g_2=-|t|
f_2< 1/100\;,\quad  -1/100 < |t|\hat f_2=-|t| g_2< 1/100\;.$$ This
shows that the hypotheses of Theorem~\ref{Tcontv} are satisfied by
the initial data for $\hat x$ on $I_i$, so that by
Theorem~\ref{Tcvs} the map $\hat x$ will be \avtdpqi\ on an open
dense subset of each of the
$\Omega(\psi_i-2|t_i|,\psi_i+2|t_i|,t_i)$'s for $i$ large enough.
This gives an open subset of $[\theta_l,\theta_r]$ with \avtdpqi\
behavior for $\hat x$. To analyse the behavior of $x$ we need
first to perform back the Gowdy-to-Ernst transformation. Since
$$P = -\hat P - \ln |t|\;,$$
the existence and continuity properties of $v$ are unchanged by
the Gowdy-to-Ernst map. Next, we note that \bel{Qest} |Q_t| =  |t
e^{\hat P} \hat g_2| \le C |t| ^{\epsilon-1}\ee for some $\epsilon
>0$, which by integration shows that the function $Q_\infty$ of
\eq{avtd2} exists and is a continuous function on each interval $[
\psi_i-|t_i|,\psi_i+|t_i|]$ with $i$ large enough. Hence the map
$(P,Q)$ belongs to the class $\mcU_1$ defined in
Section~\ref{Sgeroch}. Now, the original map $x$ is obtained from
the one just described by composing with an isometry of the
hyperbolic plane, and is thus again in the class $\mcU_1$ by
Proposition~\ref{PGtE}. One then obtains an \avtdpqi\ map in a
neighborhood of the set obtained by removing from $[
\psi_i-|t_i|,\psi_i+|t_i|]$ the countable set consisting of points
where $v$ has discontinuities, together with the boundary of the
set where $v=1$.

We continue by induction: suppose that we have already established
the claim for $v_\infty  \le k+ 1/200$, and suppose that there
exists $k\in \N$ such that $k+ 1/200 \le v_\infty\le k+1-1/200$.
Applying an isometry $\phi_i$ to $x$, and still denoting by $x$
the resulting map, we can assume that $f_2(t_i,\psi_i)=0$, and
that $k+1/50 < |t|P_t < k+1-1/50$ on $I_i$, while
$|tX_\theta|+|tf_2|< 1/100$ there.
    Applying a
Gowdy-to-Ernst transformation we obtain \bel{f3a} -k+1/50 <
|t|\hat P_t < -(k-1) - 1/50\ \mbox{ on $I_i$, with } \ |t\hat
f_2|< 1/100\;.\ee It follows that $|t\hat X_t|< k$ on all $I_i$'s
for $i$ large enough. We note that \eq{f2c} with $a=2$ shows, by
integration, that on $I_i$ we have \bel{f3c} |t\hat g_2|=|tf_2|
\le C F_1(I_i)\;,\quad \mbox{while} \ |t\hat g_1|=|tg_1| \le
F_0(I_i)\ \mbox{ holds trivially}.\ee Next, by \eq{intd2.2}, \bean
D_\theta \hat X_\theta &= &(\partial_\theta \hat g_1 -\hat g_2^2)
e_1 + (\partial_\theta \hat g_2 + \hat g_1 \hat g_2) e_2 \\
&=&(-\partial_\theta  g_1 -f_2^2) e_1 + (-\partial_\theta  f_2 +
g_1 f_2) e_2 \;.\eeal{f4c} Similarly,\bean D_\theta \hat X_t &=
&(\partial_\theta \hat f_1 -\hat f_2\hat g_2)
e_1 + (\partial_\theta \hat f_2 + \hat f_1 \hat g_2) e_2 \\
&=&(-\partial_\theta  f_1 -f_2g_2) e_1 + (-\partial_\theta  g_2 +
f_1 f_2 - \frac {f_2}{|t|}) e_2 \;.\eeal{f5c} It is
straightforward to check, using \eq{f3c}-\eq{f5c}, that if one
sets
$$\hat F_\ell(I_i):=
\sup_{\theta\in[\psi_i-2|t_i|, \psi_i+2|t_i|]}
\left(\sum_{k=0}^\ell|\tkone D_\theta^{k}\hat
X_\theta|+\sum_{k=1}^\ell|\tkone D_\theta^{k}\hat
X_t|\right)(t_i,\theta)\;,$$ then $\hat F_2(I_i)\to 0 $ as $i\to
\infty$. A similar calculation shows that the same is true for
$\hat F_2(I_i)$, and in fact for any higher order $\hat F_\ell$ if
true for $F_\ell$.


Returning to the proof of Theorem~\ref{Tdc}, it follows that the
map $\hat x$ satisfies all the conditions needed to apply the
result assumed to be true by the induction hypothesis. In
particular the velocity function $\hat v$ and the position
function $\hat Q_\infty$ are smooth on each interval $[
\psi_i-|t_i|,\psi_i+|t_i|]$, with $i$ large enough, except perhaps
for a countable number of points. Performing back the
Gowdy-to-Ernst transformation one obtains a map with a velocity
function $v$ which has the same properties. \Eq{Qest} becomes
\bel{Qest2} |Q_t| = |t e^{\hat P} \hat g_2| \le C e^{\hat P}\le C
|t| ^{k -1/50}\;,\ee which shows, as before, existence of a
continuous function $Q_\infty$ on each interval $[
\psi_i-|t_i|,\psi_i+|t_i|]$, with $i$ large enough.
Proposition~\ref{PGtE} implies again the \avtdpqi\ behavior in a
neighborhood of on an appropriate subset of $\{0\}\times [
\psi_i-|t_i|,\psi_i+|t_i|]$.

The possibility $k+1-1/200 < v_\infty < k+1+1/200$ is handled
similarly,
 and the induction step is completed.

Since we have a uniform bound on $|tX_t|$, the procedure stops in
a finite number of steps.

We have thus shown that every subinterval $[\theta_l,\theta_r]$ of
$[a,b]$ contains an open set so that the solution is \avtdpqi\ in
a neighborhood thereof. The union of all such sets, as
$[\theta_l,\theta_r]$ runs over all subintervals of $[a,b]$,
provides the desired  open dense set $\hat \Omega$ of
Theorem~\ref{Tdc2}. The set $\Omega$ of Theorem~\ref{Tdc} can be
taken to coincide with $\hat \Omega$. However, for the purpose of
Proposition~\ref{Pbad}, one should keep in $\Omega$ those points
at which the discontinuities in $v$ arise from the action of the
Geroch-group-inverse $G ^{-1}$ on $Gx$, when recovering  $x$ from
$Gx$ on $\Omega(\psi_i-2|t_i|,\psi_i+2|t_i|,t_i)$, by inverting
the inductive method described above.  \myqed

The arguments given above together with Remark~\ref{Rmain} can be
used to obtain the following characterisation of points which are
not in $\Omega$:

\begin{Proposition}\label{Pbad} Let $\Omega$ be the largest open
set for which the conclusions of Theorem~\ref{Tdc} hold. Then
$\psi\not\in\Omega$ if and only if \begin{enumerate}\item either
there exists $\epsilon >0$ such that for all $t$ small enough
$$ \sup_{\theta\in[\psi-|t|, \psi+|t|]}
\left(\sum_{k=0}^1|\tkone
D_\theta^{k}X_\theta|+
|t^2 D_\theta X_t|\right)(t,\theta)> \epsilon\;,$$ \item or $v$
exists and is continuous on an interval $I$ containing  $\psi$,
with $\psi$ belonging to the boundary of the set
$\{v(\theta)=0\}$. Further
$$ \sum_{k=0}^2\sup_{\theta\in I}
\left(|\tkone D_\theta^{k}X_\theta|+|\tkone
D_\theta^{k}X_t|\right)(t,\theta)\to 0\;,$$ but $Q_\infty$ is
discontinuous at $\psi$.\myqedt
\end{enumerate}
\end{Proposition}

The key open question  is that of existence of solutions for which
the set of points exhibiting the properties described in
Proposition~\ref{Pbad} is not empty. The behavior described in (i)
above seems to be a much more serious problem than the one in
(ii).


We are ready now to pass to the proof of the main result of our
paper:

\medskip

\noindent{\sc Proof of Theorem~\ref{Tstabcvd}:} Consider any point
$\psi\in [a,b]$. The argument of the proof of Theorem~\ref{Tdc}
with  $t_i = - 2^{-i}$, $\psi_i= \psi$, shows the existence of a
time $t_\psi<0$ and of an element $G_\psi$ of the Geroch group
such that $G_\psi x$ satisfies the hypotheses of
Theorem~\ref{Tcvs} on
$\Omega(\psi-|t_\psi|,\psi+|t_\psi|,t_\psi)$. (Note that the
covering argument is not needed anymore in view of the
hypothesis~\eq{ucbuo}.)
This proves (i) for
$Gx$, and what has been said concerning the action of the Geroch
group for the solutions under consideration proves (i) and (ii)
for $x$.

By compactness of $[a,b]$ a finite covering by intervals
$(a_i,b_i):=(\psi_i-|t_{\psi_i}|,\psi+|t_{\psi_i}|)$ can be
chosen, and (iii) readily follows. Point (iv) follows from the
results in \cite{SCC} (compare~\cite{ChLake}). A small change of
the initial data at $t=t_0$ will lead to a small change of initial
data at $t=t_{\psi_i}$, hence to a small change of $G_{\psi_i}
x(t_{\psi_i},\cdot)$ and its derivatives at $t_{\psi_i}$, and
point (v) follows from Theorem~\ref{Tcvs}. 
\myqed

\bigskip

\noindent{\sc Acknowledgements:} PTC wishes to thank L.~Andersson,
M.~Anderson, A.~Rendall, H.~Ringstr\"om, J.~Shatah and M.~Weaver
for useful discussions at various stages of work on this paper. We
are grateful to A.~Gerber, K.~Lake, and J.R.~Licois for help with
computer algebra calculations. The friendly hospitality of the
Albert Einstein Institute during part of work on this paper is
acknowledged.

\bibliographystyle{amsplain}
\bibliography{
../references/newbiblio,%
../references/reffile,%
../references/bibl,%
../references/Energy,%
../references/hip_bib,%
../references/netbiblio}

\end{document}